\documentclass[12pt,letterpaper]{amsart}

\usepackage{amsfonts,amssymb,amsmath,gensymb,color,epic,eepic,float,fullpage}
\usepackage[mathscr]{euscript}
\usepackage[T1]{fontenc}
\usepackage{lmodern}
\input{glyphtounicode}
\usepackage{mathtools}
\usepackage{comment}
\usepackage{bm}
\usepackage{enumerate,enumitem}
\usepackage{caption}
\usepackage[section]{placeins}
\usepackage{hyperref}
\hypersetup{
  hidelinks,
  pdftitle={Two-parameter Littlewood identities and half-space Yang-Baxter random fields for stable spin Hall-Littlewood symmetric functions},
  pdfauthor={Kailun Chen and Xiang-Mao Ding},
  pdfkeywords={Stable spin Hall-Littlewood symmetric functions, Skew Littlewood identities, half-space Yang-Baxter random field, KPZ universality class}
}
\usepackage{xcolor}
\usepackage{upgreek}
\usepackage{graphicx,color,colortbl}
\usepackage{tikz}
\usetikzlibrary{graphs,patterns,decorations.markings,arrows,matrix}
\usetikzlibrary{calc,decorations.pathmorphing,decorations.pathreplacing,shapes}
\tikzstyle{block} = [rectangle,draw,text width=10em,text centered,rounded corners,minimum height=4em]
\tikzstyle{line} = [draw, -latex']
\allowdisplaybreaks
\usepackage{array}
\usepackage{adjustbox}
\usepackage{cleveref}
\usepackage{enumitem}
\usepackage{relsize}
\usepackage{hhline}
\DeclareMathAlphabet{\mathpzc}{OT1}{pzc}{m}{it}

\newtheorem{defn}{Definition}[section]

\newtheorem{thm}[defn]{Theorem}

\newtheorem{cor}[defn]{Corollary}

\newtheorem{prop}[defn]{Proposition}

\colorlet{lgray}{white!70!black}
\colorlet{lred}{white!70!red}
\colorlet{lblue}{white!70!blue}
\colorlet{lorange}{white!50!orange}

\renewcommand{\tikz}[2]{
\begin{tikzpicture}
[scale=#1,baseline=(current bounding box.center),>=stealth]#2
\end{tikzpicture}}

\tikzset{
>=stealth',
help lines/.style={dashed, thick},
axis/.style={<->},
important line/.style={thick},
connection/.style={thick, dotted},
punkt/.style={
rectangle,
rounded corners,
draw=black, thick,
text width=4.5em,
minimum height=2em,
text centered,
},
pil/.style={
->,
thick,
gray,
shorten <=2pt,
shorten >=2pt,}
}

\newcommand{\Ufwd}{\ensuremath{\mathcal{U}^{\mathlarger{\llcorner}}}}
\newcommand{\Ubwd}{\ensuremath{\mathcal{U}^{\mathlarger{\urcorner}}}}

\numberwithin{equation}{section}

\begin{document}

\title[Littlewood identities and half-space Yang--Baxter fields]{Two-parameter Littlewood identities and half-space Yang--Baxter random fields for stable spin Hall--Littlewood symmetric functions}

\author[Kailun Chen]{Kailun Chen}
\address{Institute of Applied Mathematics, Academy of Mathematics
and Systems Science, Chinese Academy of Sciences, and University
of Chinese Academy of Sciences, Beijing, China.}
\curraddr{Department of Mathematics, National University of Singapore,
10 Lower Kent Ridge Road, Singapore 119076.}
\email[Corresponding author]{kailunxsx@gmail.com}

\author[Xiang-Mao Ding]{Xiang-Mao Ding}\address{Institute of Applied Mathematics, Academy of Mathematics and Systems Science, Chinese Academy of Sciences, Beijing, China.}\email{xmding@amss.ac.cn}

\begin{abstract}
We prove a two-parameter skew Littlewood identity for stable spin
Hall--Littlewood symmetric functions, generalizing Warnaar's identity.
This identity yields a half-space extension of the Yang--Baxter
random field of Bufetov and Petrov. Using their stochastic
Yang--Baxter move together with the skew Littlewood identity,
we construct explicit bulk and boundary sampling rules and
characterize the two boundary regimes in which these rules admit
autonomous projections onto the first $R$ column lengths for
every $R\ge1$. In these regimes, the partition-length fields agree,
after explicit coordinate and parameter changes, with the
half-space stochastic six-vertex model of Barraquand, Borodin,
Corwin and Wheeler and a subfamily of He's model.
The two-column projections retain spin dependence and converge,
under the respective continuous-time scalings, to the same
two-layer exclusion process whose first layer is open ASEP.
We also prove that the joint distributions of partition lengths
in ascending processes are independent of spin throughout the
nonnegative parameter range. Using this spin independence and
the distributional comparisons with half-space six-vertex heights,
we transfer He's asymptotic results to diagonal partition lengths,
obtaining Tracy--Widom GSE and GOE limits, Gaussian limits,
and a GSE--GOE crossover under boundary tuning.
The GOE limit also holds at the even-column specialization.
Finally, the first-layer particle count of the continuous-time
process has GOE fluctuations.
\end{abstract}
\keywords{Stable spin Hall--Littlewood symmetric functions, Skew Littlewood
identities, half-space Yang--Baxter random field, KPZ universality class}
\maketitle
\vspace{-\baselineskip}
\begingroup
\fontsize{10}{11}\selectfont
\tableofcontents
\endgroup

\section{Introduction}
\label{sec:introduction}
\subsection{Overview}
\label{ssec:overview}

The Kardar--Parisi--Zhang (KPZ) universality class describes the
large-scale fluctuations of many random growth models, interacting
particle systems and directed polymers in random environments
\cite{KardarParisiZhang1986,Corwin2012KPZ}. In one spatial dimension,
the characteristic KPZ scaling has centered height fluctuations of
order $T^{1/3}$ and a spatial correlation length of order $T^{2/3}$
at time $T$. The limiting distributions depend on the initial data
and the geometry of the model. Much of the rigorous understanding
of these predictions comes from exactly solvable models, whose
probability distributions admit formulas that can be analyzed
asymptotically.

Integrable vertex models provide a class of such solvable models.
A vertex model assigns local weights to arrow configurations on a
lattice; its partition function sums products of these weights over
configurations with prescribed boundary data. The Yang--Baxter
equation relates local configurations and permits the interchange
of rows, leading to identities between partition functions and,
with suitable boundary conditions, commuting transfer matrices
\cite{Baxter1982,BorodinPetrov1605}. When the local weights are
nonnegative and sum to one over outgoing configurations for each
fixed incoming configuration, they define stochastic updates.
For example, Borodin, Corwin and Gorin
\cite{BorodinCorwinGorin2016} proved a limit shape and one-point
Tracy--Widom GUE fluctuations for the stochastic six-vertex model.
Here GUE denotes the Gaussian unitary ensemble.

Symmetric functions arise naturally as partition functions of these
models. Borodin \cite{Borodin1410} introduced symmetric rational
functions associated with the higher-spin six-vertex model, now
called spin Hall--Littlewood functions, and proved Cauchy identities,
symmetrization formulas and orthogonality relations. Borodin and
Petrov \cite{BorodinPetrov1601} developed the fully inhomogeneous
model and derived integral formulas for its height observables.
The stochastic higher-spin vertex models of Corwin and Petrov
\cite{CorwinPetrov1502} include the stochastic six-vertex model,
the asymmetric simple exclusion process (ASEP), $q$-TASEP and
directed polymers through specializations or limits. For stable spin
Hall--Littlewood functions and their properties, see
\cite{GarbaliGierWheeler1605,BorodinWheeler1701,
BufetovMucciconiPetrov1905}. We recall their definition in
Section~\ref{ssec:fd}.

Cauchy identities for symmetric functions also give probability
measures on partitions. Macdonald processes
\cite{BorodinCorwin1111} assign weights to sequences of partitions
by products of skew Macdonald functions. In parameter ranges where
the weights are nonnegative and summable, the skew Cauchy identity
gives normalization and consistency relations. These processes
provide another way to study observables of stochastic vertex
models \cite{Borodin1608}. In particular, Borodin, Bufetov and
Wheeler \cite{BorodinBufetovWheeler1611} identified joint height
distributions of the stochastic six-vertex model along down-right
paths with joint partition lengths in Hall--Littlewood processes.

Such partition processes can be coupled into two-dimensional random
fields. Bufetov and Petrov \cite{BufetovPetrov1712} constructed a
Yang--Baxter random field by a probabilistic bijectivization of the
Yang--Baxter equation: forward and backward Markov transition
operators satisfy weighted balance relations that refine the
identity. Restrictions of the field to down-right paths have spin
Hall--Littlewood process laws, while suitable Markov projections
give stochastic vertex models and their dynamic variants.
Bufetov, Mucciconi and Petrov
\cite{BufetovMucciconiPetrov1905} formulated a skew Cauchy structure
for such constructions and treated several pairs of symmetric
functions, including stable spin Hall--Littlewood functions.
They also constructed difference operators for observables of the
associated partition measures and studied Markov projections of
the random fields.

For half-space constructions of this kind, the bulk Cauchy
relations must be supplemented by compatibility at the diagonal
boundary. Littlewood identities provide this additional relation
for the symmetric-function processes considered below.
Baik, Barraquand, Corwin and Suidan
\cite{BaikBarraquandCorwinSuidan1606} used Pfaffian Schur processes
to study last passage percolation in a half-quadrant. Barraquand,
Borodin and Corwin \cite{BarraquandBorodinCorwin1802} developed
half-space Macdonald processes, their observables and compatible
Markov dynamics. Barraquand, Borodin, Corwin and Wheeler (BBCW)
\cite{BarraquandBorodinCorwinWheeler1704} related half-space
Hall--Littlewood processes to a stochastic six-vertex model in a
half-quadrant and to open ASEP on the half-line. At each diagonal
vertex of their six-vertex model, the outgoing arrow is determined
by the incoming arrow. The corresponding Littlewood sum is
supported on partitions whose column lengths are all even.

More general boundary weights enlarge the range of models described
by these correspondences. He \cite{He2024Boundary} considered a
half-space Hall--Littlewood measure with two boundary parameters
and coefficients expressed through Rogers--Szeg\H{o} polynomials.
Its normalization follows from a specialization of Warnaar's
Littlewood identity \cite{Warnaar2007}. He identified partition
lengths under this measure with heights of a half-space stochastic
six-vertex model with two boundary parameters. He and Wheeler
\cite{HeWheeler2025FreeBoundary} subsequently related observables
of free-boundary Hall--Littlewood processes to a six-vertex model
in a strip assembled from alternating triangular domains. They
call this model the quasi-open six-vertex model.

Boundary parameters can change the limiting fluctuation law.
Early examples occur in the work of Baik and Rains
\cite{BaikRains2001} on random involutions, where varying the
number of fixed points leads to different limiting distributions
and crossover laws. For open ASEP on the half-line, BBCW
\cite{BarraquandBorodinCorwinWheeler1704} proved Tracy--Widom GOE
current fluctuations at a particular critical boundary condition.
Imamura, Mucciconi and Sasamoto
\cite{ImamuraMucciconiSasamoto2022} proved boundary phase
transitions for the half-space log-gamma polymer and KPZ equation,
using a correspondence between half-space $q$-Whittaker measures
and free-boundary Schur measures. He \cite{He2024Boundary}
established GSE, GOE and Gaussian boundary limits for half-space
ASEP and the six-vertex model, together with the GSE--GOE
crossover near the critical boundary strength. Here GOE and GSE
denote the Gaussian orthogonal and symplectic ensembles.
These results show how identities between partition measures
can be used to analyze boundary fluctuations.

Alongside these probabilistic developments, Littlewood identities
have been studied for spin Hall--Littlewood functions. Gavrilova
\cite{Gavrilova2104} proved a refined identity whose sum is
restricted to signatures with even part multiplicities and whose
right-hand side is expressed as a Pfaffian. Fischer and Gangl
\cite{FischerGangl2026} obtained product and Pfaffian Littlewood
identities with column-dependent spin parameters. Garbali,
de Gier, Mead and Wheeler \cite{GarbaliDeGierMeadWheeler2025}
constructed symmetric functions from a half-space six-vertex
model with boundary weights satisfying the reflection equation.
They proved a skew Cauchy identity using double-row transfer
operators and evaluated a triangular partition function as a
Pfaffian.

\subsection{Main results}
\label{ssec:contributions}

In this paper we generalize Warnaar's two-parameter Littlewood
identity \cite[Theorem~1.1]{Warnaar2007} for Hall--Littlewood symmetric 
functions to stable spin Hall--Littlewood symmetric functions. 
The resulting skew identity enables us to construct a half-space random field that naturally
generalizes the Yang--Baxter random field of Bufetov and Petrov
\cite{BufetovPetrov1712}. The bulk transitions use stochastic
Yang--Baxter moves, while the Littlewood identity supplies the
boundary transitions. We describe the identity, the random field
and its projected stochastic models, and the resulting
partition-length identities and boundary fluctuations below.

Fix $0<q<1$ and $-1<s\le0$. We use the stable spin
Hall--Littlewood functions $f_{\lambda/\mu}$ and $g_{\lambda/\mu}$
of \cite[Section~3]{BufetovMucciconiPetrov1905}, with
$f=\mathsf F^*$ and $g=\mathsf F$ in that paper's notation.
Their definitions are recalled in Section~\ref{ssec:fd}.
At $s=0$, they are the Hall--Littlewood functions $P_{\lambda/\mu}$
and $Q_{\lambda/\mu}$. For a partition $\lambda$, write
$\lambda'$ for its conjugate, $\ell(\lambda)=\lambda'_1$, and
$m_j(\lambda)=\lambda'_j-\lambda'_{j+1}$ for the multiplicity
of the part $j$. The notation $\mu\prec\lambda$ means
$\lambda'_j-\mu'_j\in\{0,1\}$ for every $j\ge1$.

\smallskip\noindent\textbf{The two-parameter skew Littlewood identity.}
The boundary coefficients distinguish even and odd part sizes.
For $t\in\mathbb C$ and $\zeta\ne0$, put
$U=1-t$ and $V=\zeta^{-1}-t\zeta$. Define $E_m,O_m$ by
$E_{-1}=O_{-1}=0$, $E_0=O_0=1$, and, for $m\ge0$,
\[
\begin{aligned}
(1-s^2q^m)E_{m+1}
 &=(U+sq^mV)E_m+t(1-q^m)E_{m-1},\\
(1-s^2q^m)O_{m+1}
 &=(V+sq^mU)O_m+t(1-q^m)O_{m-1}.
\end{aligned}
\]
Our boundary weight is
\[
 W_\lambda(t,\zeta)=
 \prod_{\substack{j\ge1\\j\ {\rm even}}}E_{m_j(\lambda)}
 \prod_{\substack{j\ge1\\j\ {\rm odd}}}O_{m_j(\lambda)}.
\]
In particular, $W_\varnothing=1$; dependence on $q,s$ is suppressed.

\begin{thm}[Two-parameter skew Littlewood identity]
\label{thm:intro-Littlewood}
For every partition $\mu$, $0<x<1$, $t\in\mathbb C$ and
$\zeta\in\mathbb C\setminus\{0\}$,
\begin{equation}
\label{eq:intro-Littlewood}
\sum_\lambda W_\lambda(t,\zeta)f_{\lambda/\mu}(x)
=\Phi_{t,\zeta}(x)\sum_\kappa
 W_\kappa(t,\zeta)g_{\mu/\kappa}(x),\qquad
\Phi_{t,\zeta}(x)=\frac{(1-t\zeta x)(1+x/\zeta)}{1-x^2}.
\end{equation}
The left sum converges absolutely and the right sum is finite.
For $0\le t<1$ and $\zeta>0$, the weights $W_\lambda$ are
nonnegative for every $\lambda$ if and only if
$(-s)U\le V$ and $(-s)V\le U$.
\end{thm}

Theorem~\ref{thm:fb-skew-Littlewood} proves the identity by local
$2\times2$ matrix calculations, and
Proposition~\ref{prop:fb-positivity} proves the positivity criterion.
At $s=0$ the recurrence gives
\[
 E_m=H_m(-t;q),\qquad O_m=\zeta^{-m}H_m(-t\zeta^2;q),
 \qquad
 H_m(z;q)=\sum_{k=0}^m
 \frac{(q;q)_m z^k}{(q;q)_k(q;q)_{m-k}}.
\]
Here $(q;q)_m=\prod_{k=1}^m(1-q^k)$, with $(q;q)_0=1$.
Thus the zero-spin coefficients are precisely the Rogers--Szeg\H{o}
weights in He's half-space Hall--Littlewood measures
\cite[Section~2.3]{He2024Boundary}. Our $\zeta$ is the parameter
denoted by $\nu$ there. At $(t,\zeta)=(1,1)$,
\[
 E_{2r+1}=O_{2r+1}=0,\qquad
 E_{2r}=O_{2r}=
 \prod_{k=1}^r\frac{1-q^{2k-1}}{1-s^2q^{2k-1}},
 \qquad \Phi_{1,1}(x)=1.
\]
Consequently, the identity at this specialization is a sum over
partitions whose column lengths are all even. The non-skew
multivariable identity at $(t,\zeta)=(0,1)$ is also the specialization
of \cite[Theorem~1.1]{FischerGangl2026} described in
Section~\ref{sec:boundary-family}.

\smallskip\noindent\textbf{The half-space random field.}
From now on, boundary parameters are either
$0\le t<1$, $\zeta>0$ with $(-s)U\le V$ and $(-s)V\le U$,
or $(t,\zeta)=(1,1)$. All spectral parameters belong to $(0,1)$.
Define
\[
\begin{gathered}
\Pi(x;y)=\frac{1-qxy}{1-xy},\qquad
G^{t,\zeta}_\mu(y)=\sum_{\kappa\prec\mu}
 W_\kappa(t,\zeta)g_{\mu/\kappa}(y),\\
Z_{t,\zeta}(a_1,\ldots,a_N)=
 \prod_{i=1}^N\Phi_{t,\zeta}(a_i)
 \prod_{1\le i<j\le N}\Pi(a_i;a_j).
\end{gathered}
\]
Here $G_\mu$ is a boundary function, and $Z$ is a normalizing
constant. For $A=(a_1,\ldots,a_N)$, write
$Z_{t,\zeta}(A)=Z_{t,\zeta}(a_1,\ldots,a_N)$.
The \emph{ascending process} has probability mass function
\[
 \mathbb P^{A,\uparrow}_{s;t,\zeta}
 (\lambda^{(1)},\ldots,\lambda^{(N)})
 =\frac{W_{\lambda^{(N)}}(t,\zeta)}{Z_{t,\zeta}(A)}
 \prod_{i=1}^N f_{\lambda^{(i)}/\lambda^{(i-1)}}(a_i),
 \qquad \lambda^{(0)}=\varnothing.
\]
Its partitions form an interlacing chain. Iteration of the
Littlewood identity and the skew Cauchy identity proves that
these weights sum to one.

The random field places a partition $\lambda^{(i,j)}$ at each
$(i,j)$ with $0\le i\le j$, with
$\lambda^{(0,j)}=\varnothing$ and interlacing along rightward
and upward edges. Its path distributions use the same functions.
More precisely, let
$\omega=(\omega_0,\ldots,\omega_N)$ run from $(n,n)$ to $(0,N)$
by left and up steps, and write $\omega_k=(i_k,j_k)$.
Fix parameters $x_0,x_1,\ldots\in(0,1)$. The probability of its
partitions is
\[
 \frac{G^{t,\zeta}_{\lambda^{\omega_0}}(x_n)}{Z_\omega}
 \prod_{\omega_{k+1}=\omega_k+(0,1)}
 g_{\lambda^{\omega_{k+1}}/\lambda^{\omega_k}}(x_{j_{k+1}})
 \prod_{\omega_{k+1}=\omega_k-(1,0)}
 f_{\lambda^{\omega_k}/\lambda^{\omega_{k+1}}}(x_{i_k-1}),
 \qquad \lambda^{\omega_N}=\varnothing.
\]
If $h_a$ is the height of its left step from column $a+1$ to
column $a$, the normalizer is
\[
 Z_\omega=
 \prod_{a=0}^{n-1}\Phi_{t,\zeta}(x_a)
 \prod_{a=0}^{n-1}\prod_{b=a+1}^{h_a}\Pi(x_a;x_b).
\]
Empty products equal one. Figure~\ref{fig:intro-field} illustrates
the path and the retained columns of a partition.

\begin{figure}[!htbp]
\centering
\begin{tikzpicture}[x=.75cm,y=.75cm,>=stealth,font=\small]
\begin{scope}
 \foreach \i in {0,...,5}{
  \draw[lgray,dotted] (0,\i)--(\i,\i)--(\i,5);
 }
 \draw[gray,dashed] (0,0)--(5.3,5.3);
 \draw[->] (0,0)--(5.7,0) node[right] {$i$};
 \draw[->] (0,0)--(0,5.7) node[above] {$j$};
 \draw[red,very thick,->] (3,3)--(2,3);
 \draw[red,very thick,->] (2,3)--(2,4);
 \draw[red,very thick,->] (2,4)--(1,4);
 \draw[red,very thick,->] (1,4)--(1,5);
 \draw[red,very thick,->] (1,5)--(0,5);
 \foreach \p in {(3,3),(2,3),(2,4),(1,4),(1,5),(0,5)}
  \fill[red] \p circle(2pt);
 \node[right] at (3,3) {$(n,n)$};
 \node[left] at (0,5) {$\varnothing$};
 \node at (2.5,-.8) {(a) An up-left path};
\end{scope}
\begin{scope}[shift={(9,4.8)},x=.6cm,y=.6cm]
 \foreach \r/\len in {0/4,1/3,2/1}{
  \foreach \c in {1,...,\len}
   \draw[fill=white] (\c-1,-\r) rectangle (\c,-\r-1);
 }
 \foreach \p in {(0,0),(0,-1),(0,-2),(1,0),(1,-1)}
  \draw[fill=lred] \p rectangle ++(1,-1);
 \node[above] at (2,.3) {$\lambda=(4,3,1)$};
 \node[align=center] at (2,-4.3)
  {$\pi_2(\lambda)=(\lambda'_1,\lambda'_2)=(3,2)$};
 \node at (2,-6.0) {(b) The first two columns};
\end{scope}
\end{tikzpicture}
\caption{A field is a family of partitions on the half-quadrant.
In (a), the red path has $n=3$, $N=5$; its left steps carry $f$
and its up steps carry $g$, with the index order displayed in the
path probability. The starting partition contributes
$G^{t,\zeta}_{\lambda^{(3,3)}}(x_3)$.
In (b), retaining the red columns gives the two-column projection;
the first coordinate is the partition length.}
\label{fig:intro-field}
\end{figure}

For $R\ge1$, write
$\pi_R(\lambda)=(\lambda'_1,\ldots,\lambda'_R)$.
A local projection is \emph{autonomous} if the conditional
probabilities for the projected output depend on the input
partitions only through their projections. In the bulk the inputs
are the southwest, south and west partitions of a square and
the output is its northeast partition. At the boundary the inputs
are $\lambda^{(i-1,i-1)},\lambda^{(i-1,i)}$ and the output is
$\lambda^{(i,i)}$.

\begin{thm}[Half-space random fields and autonomous projections]
\label{thm:intro-field}
For the nonnegative parameters above, there exists a field with
the specified path distributions, sampled by local Markov
transitions. The transitions can be chosen so that
every $\pi_R$, $R\ge1$, is autonomous in either of the following
regimes:
\[
\textup{(i)}\quad (t,\zeta)=(1,1),\quad 0<x_r<1;
\]
\begin{equation}
\label{eq:intro-local-curve}
\begin{gathered}
\textup{(ii)}\quad x_r=a,\quad 0<t<1,\quad t\zeta=a,\quad
r_s(a)\le\zeta\le r_s(a)^{-1},\\
r_s(a)=\frac{a-s}{1-sa},\qquad 0<a<1.
\end{gathered}
\end{equation}
The projected field laws are consistent under deleting their last
coordinates.
\end{thm}

The field theorem, the scalar identifications and the continuous-time
limit below use one common choice of local kernels.
The construction uses a stochastic Yang--Baxter update in the bulk.
At the boundary, given $\kappa=\lambda^{(i-1,i-1)}$, it first
samples an auxiliary partition with distribution
\[
 Q_x^+(\kappa,\eta)=
 \frac{W_\eta(t,\zeta)f_{\eta/\kappa}(x)}
 {\Phi_{t,\zeta}(x)G^{t,\zeta}_\kappa(x)},\qquad \kappa\prec\eta,
\]
and then uses $\eta$ as the south input of the bulk update.
All conditional laws are restricted to positive normalizers.
In regime (i), $\eta'_j=2\lceil\kappa'_j/2\rceil$ deterministically;
in regime (ii), $\eta$ is random. The explicit kernels, their
sampling rules and the classification within this construction
are proved in Sections~\ref{ssec:bulk-to}--\ref{ssec:boundary-to}.
The existence and consistency statements are
Theorems~\ref{thm:half-space-field-sampling}
and~\ref{thm:fb-local-projection}.
Autonomy concerns the retained vector at local updates; a higher
column by itself need not be Markov.

\smallskip\noindent\textbf{The scalar stochastic vertex models.}
We now take $x_r=a\in(0,1)$. An occupied edge carries one arrow.
At a bulk vertex, denote the west and south input occupations by
$(\iota_{\rm W},\iota_{\rm S})$ and the east and north outputs by
$(\iota_{\rm E},\iota_{\rm N})$.
The stochastic six-vertex model has the six probabilities in
Figure~\ref{fig:intro-scalar}, with
$c=(1-a^2)/(1-qa^2)$. Its vertices have coordinates
$1\le i\le j$, and every external west edge is occupied.
The height $H(i,j)$ counts the occupied north edges above
$(1,j),\ldots,(i,j)$; put $H(0,j)=0$.

Barraquand, Borodin, Corwin and Wheeler
\cite[Section~4.1]{BarraquandBorodinCorwinWheeler1704}
introduced the half-quadrant model whose diagonal rule is
$\iota_{\rm N}=1-\iota_{\rm W}$. We denote its height by
$H_{\rm BBCW}$. He \cite[Section~1.3]{He2024Boundary} studies
the more general diagonal probabilities
\[
 \begin{gathered}
 \mathbb P(\iota_{\rm N}=1\mid\iota_{\rm W}=0)=bp_{b,\zeta}(a),\qquad
 \mathbb P(\iota_{\rm N}=1\mid\iota_{\rm W}=1)=1-p_{b,\zeta}(a),\\
 p_{b,\zeta}(a)=\frac{1-a^2}{(1-\zeta ba)(1+a/\zeta)}.
 \end{gathered}
\]
Write $H_{\rm He}^{b,\zeta}$ for this height, whenever the
displayed numbers are probabilities. At $(b,\zeta)=(1,1)$,
$p_{1,1}(a)=1$, so this family includes the BBCW model.

\begin{figure}[!htbp]
\centering
\begin{tikzpicture}[>=stealth,font=\small,x=1cm,y=1cm]
\node at (2.1,1.05) {Bulk probabilities};
\foreach \X/\Y/\W/\S/\E/\N/\P in
 {0/0/0/0/0/0/1,2.1/0/1/1/1/1/1,4.2/0/1/0/1/0/c,
  0/-1.65/1/0/0/1/{1-c},2.1/-1.65/0/1/0/1/{qc},4.2/-1.65/0/1/1/0/{1-qc}}{
 \begin{scope}[shift={(\X,\Y)}]
  \draw[lgray] (-.5,0)--(.5,0);
  \draw[lgray] (0,-.5)--(0,.5);
  \ifnum\W=1 \draw[very thick,->] (-.55,0)--(-.05,0);\fi
  \ifnum\S=1 \draw[very thick,->] (0,-.55)--(0,-.05);\fi
  \ifnum\E=1 \draw[very thick,->] (.05,0)--(.55,0);\fi
  \ifnum\N=1 \draw[very thick,->] (0,.05)--(0,.55);\fi
  \fill (0,0) circle(1.5pt);
  \node[below] at (0,-.55) {$\P$};
 \end{scope}
}
\draw[dotted,gray] (5.35,.8)--(5.35,-2.5);
\node at (8.0,1.05) {He's diagonal probabilities};
\foreach \X/\Y/\W/\N/\P in
 {6.6/0/0/0/{1-bp},9.1/0/0/1/{bp},
  6.6/-1.65/1/0/{p},9.1/-1.65/1/1/{1-p}}{
 \begin{scope}[shift={(\X,\Y)}]
  \draw[lgray] (-.55,0)--(0,0)--(0,.55);
  \draw[lred,line width=3pt] (-.3,-.3)--(.3,.3);
  \ifnum\W=1 \draw[very thick,->] (-.55,0)--(-.05,0);\fi
  \ifnum\N=1 \draw[very thick,->] (0,.05)--(0,.55);\fi
  \fill (0,0) circle(1.5pt);
  \node[below] at (0,-.55) {$\P$};
 \end{scope}
}
\begin{scope}[shift={(.2,-5.5)},scale=.6]
 \foreach \j in {1,...,4}{
  \draw[lgray] (0,\j)--(\j,\j);
  \draw[lgray] (\j,\j)--(\j,4.5);
  \foreach \i in {1,...,\j} \fill (\i,\j) circle(2pt);
  \draw[red,thick] (\j-.15,\j-.15)--(\j+.15,\j+.15);
  \draw[red,thick] (\j-.15,\j+.15)--(\j+.15,\j-.15);
 }
 \node at (2,.25) {delete $(i,i)$};
\end{scope}
\draw[->,thick] (3.4,-4.1)--(5.3,-4.1)
 node[midway,above] {$J=j+1$};
\begin{scope}[shift={(6.1,-4.9)},scale=.6]
 \foreach \j in {1,...,3}{
  \draw[lgray] (0,\j)--(\j,\j);
  \draw[lgray] (\j,\j)--(\j,3.5);
  \foreach \i in {1,...,\j} \fill (\i,\j) circle(2pt);
 }
 \node at (2,-.2) {relabel $(i,J)\mapsto(i,J-1)$};
\end{scope}
\end{tikzpicture}
\caption{Scalar vertex probabilities and diagonal deletion.
In the upper panels, black arrows have occupation one and gray
edges have occupation zero. Here $p=p_{b,\zeta}(a)$. On the diagonal only the west input
and north output remain. The lower diagrams show the coordinate
change from an original height $H(i,J)$ to $H(i,j+1)$.
The vertices immediately above the deleted diagonal become the
new boundary; their input arrows retain the law induced by that
diagonal.}
\label{fig:intro-scalar}
\end{figure}

\begin{cor}[Identification of the scalar fields]
\label{cor:intro-scalar}
There is a choice of the local kernels in
Theorem~\ref{thm:intro-field} for which the following equalities
hold jointly over $1\le i\le j$.
In regime \textup{(i)},
\[
 \{\ell(\lambda^{(i,j)})\}_{i\le j}
 \stackrel{\mathrm d}=
 \{H_{\rm BBCW}(i,j+1)\}_{i\le j}.
\]
In regime \textup{(ii)},
\begin{equation}
\label{eq:intro-He-row-shift}
 \{\ell(\lambda^{(i,j)})\}_{i\le j}
 \stackrel{\mathrm d}=
 \{H_{\rm He}^{t,\zeta}(i,j+1)\}_{i\le j}
 \stackrel{\mathrm d}=
 \{H_{\rm He}^{qt,\zeta}(i,j)\}_{i\le j}.
\end{equation}
\end{cor}

The proofs are Propositions~\ref{prop:He-row-shift}
and~\ref{prop:even-boundary-BBCW}. In regime (ii), deleting
the diagonal is absorbed by changing the boundary parameter from
$t$ to $qt$. Thus the unshifted scalar model has $b=qt$ and
$b\zeta=qa$; these projections realize a constrained part of
He's two-parameter family. The higher-column transition
probabilities retain dependence on $s$. Their first nontrivial
continuous-time limit is the two-layer process described next.

\smallskip\noindent\textbf{A continuous-time process on two particle layers.}
The first two columns of the half-space random field give a coupled
particle process whose transition rates depend on the spin parameter.
We describe its continuous-time version directly; its derivation from
the field is given in Section~\ref{ssec:fb-local-projections}.
Fix $0<q<1$ and $-1<s\le0$. A state consists of two particle
configurations $\xi_1,\xi_2$ on the positive integers, where
$\xi_r(x)\in\{0,1\}$ and $\xi_r(x)=1$ means that site $x$ in
layer $r$ is occupied. Put
\[
 \begin{gathered}
 D_x=\sum_{y\ge x}(\xi_2(y)-\xi_1(y)),\\
 \mathcal X=\left\{(\xi_1,\xi_2):
 \sum_{x\ge1}(\xi_1(x)+\xi_2(x))<\infty,\quad
 D_x\ge0\ \text{for all }x\ge1
 \right\}.
 \end{gathered}
\]
Thus layer 2 has at least as many particles as layer 1 at or to the
right of every site. Every jump below requires an occupied departure
site and a vacant destination site; every entry or exit requires,
respectively, a vacant or occupied site $1$.

At a bond $(x,x+1)$, write $d=D_{x+1}$. A layer-1 particle jumps
right at rate $1$ and left at rate $q$. A right jump also moves
the layer-2 particle from $x$ to $x+1$ when $d=0$; the state-space
constraint ensures that the layer-2 pair is then $10$.
For $d>0$, a right jump leaves layer 2 unchanged; a left jump
always leaves layer 2 unchanged.
Layer-2 particles also make jumps that leave layer 1 unchanged. Their right
and left rates, $u_d(\omega)$ and $v_d(\omega)$, depend on the
first-layer occupations $\omega=(\xi_1(x),\xi_1(x+1))$:
\begin{equation}
\begin{gathered}
 A_d=1-sq^d,\qquad B_d=(1-q^{d+1})(1-s^2q^d),\\[3pt]
\begin{array}{c|c|c}
 \omega&u_d(\omega):\ 10\to01&v_d(\omega):\ 01\to10,\quad d\ge1\\ \hline
 10&B_d/A_d^2&qA_{d-1}^2/B_{d-1}\\[2pt]
 00\text{ or }11&A_{d+1}/A_d&qA_{d-1}/A_d\\[2pt]
 01&A_{d+1}^2/B_d&qB_{d-1}/A_d^2
\end{array}
\end{gathered}
\label{eq:fb-two-column-bulk-rates}
\end{equation}
Set $v_0(\omega)=0$, so a layer-2 left jump is blocked when $d=0$.

At site $1$, write $D=D_1$, $k=\xi_1(1)$, and put
\begin{equation}
 \rho_D(k)=
 \begin{cases}
 B_D/A_D^2,&D\text{ even},\ k=0,\\
 A_{D+1}^2/B_D,&D\text{ odd},\ k=1,\\
 A_{D+1}/A_D,&\text{otherwise}.
 \end{cases}
\label{eq:fb-two-column-boundary-rho}
\end{equation}
Particles enter layer 1 at rate $1/2$ and leave at rate $q/2$.
When $D=0$, a layer-1 entry inserts a particle into layer 2
simultaneously; otherwise layer 2 is unchanged. A layer-1 exit
always leaves layer 2 unchanged. In addition, particles enter
layer 2 alone at rate $\rho_D(k)/2$ and leave layer 2 alone at
rate $q/(2\rho_{D-1}(k))$ when $D\ge1$. The latter exit is blocked
when $D=0$. All rates are evaluated before the transition, and
all occupations not specified by a move remain unchanged.
Figure~\ref{fig:intro-two-layer-moves} illustrates the interaction
between the layers and the separate second-layer transitions.

\begin{figure}[!htbp]
\centering
\begingroup
\newcommand{\introsite}[4]{%
 \ifnum#3=2
  \node[text=#4,inner sep=0pt] at (#1,#2) {$\ast$};
 \else
  \draw[draw=#4,fill=white] (#1,#2) circle(2.7pt);
  \ifnum#3=1\fill[#4] (#1,#2) circle(2.7pt);\fi
 \fi}
\newcommand{\introbond}[5]{%
 \begin{scope}[xshift=#1cm]
  \draw[gray!50] (0,.65)--(1.3,.65);
  \draw[gray!50] (0,-.2)--(1.3,-.2);
  \introsite{.3}{.65}{#2}{black}\introsite{1}{.65}{#3}{black}
  \introsite{.3}{-.2}{#4}{red}\introsite{1}{-.2}{#5}{red}
  \node[below,font=\scriptsize] at (.3,-.43) {$x$};
  \node[below,font=\scriptsize] at (1,-.43) {$x+1$};
 \end{scope}}
\newcommand{\introboundary}[3]{%
 \begin{scope}[xshift=#1cm]
  \draw[gray!50] (.1,.65)--(1.2,.65);
  \draw[gray!50] (.1,-.2)--(1.2,-.2);
  \introsite{.65}{.65}{#2}{black}\introsite{.65}{-.2}{#3}{red}
  \node[below,font=\scriptsize] at (.65,-.43) {site $1$};
 \end{scope}}
\newcommand{\intromove}[2]{%
 \draw[->,thick] (1.5,.23)--(3.05,.23)
  node[midway,above=4pt] {$#1$}
  node[midway,below=4pt] {$#2$};}
\begin{tikzpicture}[>=stealth,font=\small,x=1cm,y=1cm]
 \begin{scope}[shift={(0,0)}]
  \node at (2.4,1.25) {(a) Layer-1 right jump};
  \introbond{0}{1}{0}{2}{2}
  \introbond{3.3}{0}{1}{2}{2}
  \intromove{1}{d>0}
 \end{scope}
 \begin{scope}[shift={(6.05,0)}]
  \node at (2.4,1.25) {(b) Simultaneous right jump};
  \introbond{0}{1}{0}{1}{0}
  \introbond{3.3}{0}{1}{0}{1}
  \intromove{1}{d=0}
 \end{scope}
 \begin{scope}[shift={(0,-2.4)}]
  \node at (2.4,1.25) {(c) Layer-2 right jump};
  \introbond{0}{2}{2}{1}{0}
  \introbond{3.3}{2}{2}{0}{1}
  \intromove{u_d(\omega)}{d\ge0}
 \end{scope}
 \begin{scope}[shift={(6.05,-2.4)}]
  \node at (2.4,1.25) {(d) Layer-1 entry};
  \introboundary{0}{0}{2}
  \introboundary{3.3}{1}{2}
  \intromove{1/2}{D>0}
 \end{scope}
 \begin{scope}[shift={(0,-4.8)}]
  \node at (2.4,1.25) {(e) Simultaneous entry};
  \introboundary{0}{0}{0}
  \introboundary{3.3}{1}{1}
  \intromove{1/2}{D=0}
 \end{scope}
 \begin{scope}[shift={(6.05,-4.8)}]
  \node at (2.4,1.25) {(f) Layer-2 entry};
  \introboundary{0}{2}{0}
  \introboundary{3.3}{2}{1}
  \intromove{\rho_D(k)/2}{D\ge0}
 \end{scope}
\end{tikzpicture}
\endgroup
\caption{Transitions of the two-layer process. The upper black row
is layer 1 and the lower red row is layer 2. Filled and open circles
denote particles and vacancies. A star denotes a fixed occupation
that remains unchanged. Each panel shows one transition, with its
rate above the arrow and its condition below; only admissible input
configurations are considered. In panel (c), $\omega$ is the pair
of first-layer occupations; in panel (f), $k$ is its occupation at
site $1$. Panels (b),(e) are simultaneous transitions, in addition
to the separate layer-2 transitions in panels (c),(f). Left jumps
and exits have the rates given in the text.}
\label{fig:intro-two-layer-moves}
\end{figure}

\begin{thm}[A two-layer extension of open ASEP]
\label{thm:intro-continuous-time}
The preceding rules define a nonexplosive continuous-time Markov
chain on $\mathcal X$, started from the empty configuration.
Its first layer is open ASEP with right and left jump rates
$1,q$ and boundary entry and exit rates $1/2,q/2$.
This two-layer chain can be obtained as a continuous-time limit
of the projection onto the first two columns of the half-space
random field in each of the boundary regimes \textup{(i)} and \textup{(ii)}.
\end{thm}
The second layer retains the spin dependence: from the empty state,
its separate entry rate is $(1-q)(1+s)/(2(1-s))$, whereas the
simultaneous entry rate is $1/2$. The scaling and the proof are
given in Section~\ref{ssec:fb-local-projections}.

\smallskip\noindent\textbf{Joint partition lengths and boundary fluctuations.}
The next result applies throughout the nonnegative boundary range,
including $(t,\zeta)=(1,1)$, without the restriction $t\zeta=a$.

\begin{thm}[Equality of joint partition-length distributions]
\label{thm:intro-joint-length}
For $A=(a_1,\ldots,a_N)\in(0,1)^N$, the joint law of
$(\ell(\lambda^{(1)}),\ldots,\ell(\lambda^{(N)}))$ under
$\mathbb P^{A,\uparrow}_{s;t,\zeta}$ is independent of $s$ among
the admissible spin values. It equals the corresponding law of
the half-space Hall--Littlewood process
$\mathbb P^{A,\uparrow}_{0;t,\zeta}$. This holds both for
$0\le t<1$ in the nonnegative range and for $(t,\zeta)=(1,1)$.
\end{thm}

The proof uses difference operators
(Theorem~\ref{thm:fb-joint-length}). Together with the scalar
identifications and He's asymptotic results \cite{He2024Boundary},
it gives the following boundary limits.
Put $v(a)=2a/(1+a)$ and $\sigma(a)=[a(1-a)]^{1/3}/(1+a)$.
For $\zeta<1$, put
\[
 m_{\rm G}=
 \frac{2a^2+a(\zeta+\zeta^{-1})}{(1+a\zeta)(1+a\zeta^{-1})},
 \qquad
 \sigma_{\rm G}^2=
 \frac{a(1-a^2)(\zeta^{-1}-\zeta)}
 {(1+a\zeta)^2(1+a\zeta^{-1})^2}.
\]
Write $F_{\rm GOE},F_{\rm GSE}$ for the Tracy--Widom distribution
functions and $F_{\rm cross}(z;\xi)$ for the Baik--Rains GSE--GOE
crossover distribution, normalized as in
\cite[Definition~5.1]{He2024Boundary}.

\begin{cor}[Boundary fluctuations, including $(t,\zeta)=(1,1)$]
\label{cor:intro-fluctuations}
The following limits hold for every $z\in\mathbb R$.
\nopagebreak[4]
\begin{enumerate}[label=\textup{(\roman*)},leftmargin=*]
\item
For the homogeneous field with spectral parameter $a\in(0,1)$
and fixed nonnegative boundary parameters, set
$J_N=\ell(\lambda^{(N,N)})$. Then
\[
 \lim_{N\to\infty}\mathbb P\left\{
 \frac{J_N-v(a)N}{\sigma(a)N^{1/3}}\le z\right\}
 =
 \begin{cases}
 F_{\rm GSE}(z),&\zeta>1,\quad t<1,\\
 F_{\rm GOE}(z),&\zeta=1,\quad 0\le t\le1.
 \end{cases}
\]
In particular, $(t,\zeta)=(1,1)$ gives the GOE limit for the
row-shifted BBCW field. For $\zeta<1$ and $t<1$,
$(J_N-m_{\rm G}N)/(\sigma_{\rm G}\sqrt N)$ converges to a
standard Gaussian. For $t<1$, the same limits hold for the
terminal ascending length $\ell(\lambda^{(N)})$.

\item
For the two-layer process above, started empty, let $N_1(T)$
be the number of particles in its first layer at time $T$. Then
\[
 \lim_{T\to\infty}\mathbb P\left\{
 \frac{(1-q)T/4-N_1(T)}
 {2^{-4/3}((1-q)T)^{1/3}}\le z\right\}=F_{\rm GOE}(z).
\]

\item
Fix $a\in(0,1)$ and $\xi>0$, and choose regime \textup{(ii)} with
$\zeta_N^{-1}=1-\xi/(\sigma(a)N^{1/3})$ and $t_N=a/\zeta_N$
for sufficiently large $N$. Then
\[
 \lim_{N\to\infty}\mathbb P\left\{
 \frac{J_N-v(a)N}{\sigma(a)N^{1/3}}\le z\right\}
 =F_{\rm cross}(z;\xi).
\]
This family approaches GOE as $\xi\downarrow0$ and GSE as
$\xi\to\infty$.
\end{enumerate}
\end{cor}

The proofs, including the separate argument at $(t,\zeta)=(1,1)$,
are in Corollaries~\ref{cor:fb-boundary-limits}--%
\ref{cor:fb-boundary-crossover}. These limits concern the first
column and the first particle layer.
\FloatBarrier

\subsection{Outline}
\label{ssec:outline}

Section~\ref{sec:ssHL} defines the stable spin Hall--Littlewood
symmetric functions, proves the two-parameter skew Littlewood identity,
and states the skew Cauchy--Littlewood structure needed for the
probabilistic construction. Section~\ref{sec:HSRF} constructs the
half-space random field and its bulk and boundary transition
operators. Its final subsection describes the Markov projections onto
the first $R$ column lengths and identifies their relations to
existing stochastic models. Section~\ref{sec:boundary-KPZ} proves the
equality of joint partition-length distributions and derives the
boundary fluctuation limits.

\subsection*{Acknowledgments}

The first version of this paper was written during the first author's doctoral studies. The first author also thanks Alexey Bufetov, Leonid Petrov and Matteo Mucciconi for helpful discussions. X.-M.~Ding was supported by the National Natural Science Foundation of China (Grant No.~11775299) and the National Key Research and
Development Program of China (Grant No.~2018YFB0704304).

\section{Stable spin Hall--Littlewood symmetric functions}
\label{sec:ssHL}

We first recall the vertex weights and the definition of the stable
spin Hall--Littlewood symmetric functions used in
Section~\ref{ssec:contributions}.  Section~\ref{sec:boundary-family}
proves the two-parameter skew Littlewood identity and the positivity
criterion in Theorem~\ref{thm:intro-Littlewood}.
Section~\ref{ssec:scls} collects the skew Cauchy--Littlewood structure
and derives the boundary functions and normalizing constants used in
the half-space random field.

\subsection{Higher spin six vertex model and vertex weights}
\label{ssec:mw}

The higher spin six vertex model is a model of directed paths on the
square lattice.  A horizontal edge carries at most one path, whereas
a vertical edge may carry any nonnegative number of paths.  The
\emph{occupation} of an edge is the number of paths on that edge.
Paths follow the orientation of each line, and the total occupation
is conserved at every vertex.  We use $I,K\in\mathbb Z_{\geq0}$
for vertical occupations and $j,l\in\{0,1\}$ for horizontal
occupations.  As in Section~\ref{ssec:contributions}, fix the
deformation parameter $0<q<1$ and the spin parameter $-1<s\leq0$.
A spectral parameter $x$ is assigned to each horizontal line.

The arguments of every vertex weight are ordered as
$(\text{lower},\text{left};\text{upper},\text{right})=(I,j;K,l)$.
At an upward-oriented vertex $L_{x,s}(I,j;K,l)$, the incoming
occupations are $I,j$ and the outgoing occupations are $K,l$;
conservation reads $I+j=K+l$.  The nonzero weights are
\begin{equation}
\label{eq:vertex-L}
\begin{aligned}
L_{x,s}(I,0;I,0)&=\frac{1-sxq^I}{1-sx},&
L_{x,s}(I,1;I,1)&=\frac{x-sq^I}{1-sx},\\
L_{x,s}(I,1;I+1,0)&=\frac{1-q^{I+1}}{1-sx},&
L_{x,s}(I+1,0;I,1)&=\frac{x(1-s^2q^I)}{1-sx},
\qquad I\geq0.
\end{aligned}
\end{equation}
These are the higher spin six vertex weights in the normalization of
\cite{BorodinWheeler1701}.  For a downward-oriented vertex, the lower
and upper labels remain $I,K$.  Its incoming occupations are therefore
$K,j$ and its outgoing occupations are $I,l$, so conservation becomes
$K+j=I+l$.  Write
$(z;q)_m=\prod_{r=0}^{m-1}(1-zq^r)$, with $(z;q)_0=1$.
Define its dual weight by
\begin{equation}
\label{eq:vertex-Mdual}
M^*_{x,s}(I,j;K,l)
=\frac{c_I}{c_K}L_{x,s}(K,j;I,l),
\qquad c_m=\frac{(s^2;q)_m}{(q;q)_m}.
\end{equation}
Here $c_0=1$ and
$c_{m+1}/c_m=(1-s^2q^m)/(1-q^{m+1})$ for $m\geq0$.
Using this ratio in \eqref{eq:vertex-Mdual} gives the explicit weights,
with $K$ the incoming vertical occupation:
\[
\begin{aligned}
M^*_{x,s}(K,0;K,0)&=\frac{1-sxq^K}{1-sx},&
M^*_{x,s}(K,1;K,1)&=\frac{x-sq^K}{1-sx},\\
M^*_{x,s}(K+1,1;K,0)&=\frac{1-s^2q^K}{1-sx},&
M^*_{x,s}(K-1,0;K,1)&=\frac{x(1-q^K)}{1-sx}.
\end{aligned}
\]
The last configuration requires $K\geq1$; all unlisted or inadmissible
weights are zero.  Figure~\ref{fig:vertex-orientations} fixes the
orientation conventions.  A configuration is an assignment of
occupations to edges satisfying the conservation rule at every
vertex.  Its weight is the product of the vertex weights and any
specified boundary weights.  The \emph{partition function} is the sum
of these weights over configurations with prescribed external edge
occupations.

For $0<x<1$, all these weights are nonnegative.
These are Boltzmann weights: their sum over outgoing
occupations need not equal one.  Probability measures are obtained
by normalization, and the compatible Markov transition operators are
constructed in Section~\ref{sec:HSRF}.  Finite row identities are
rational identities in their spectral variables wherever the
denominators are nonzero.

\begin{figure}[htbp]
\centering
\begin{tabular}{cc}
\tikz{0.6}{
\draw[lgray,line width=1.5pt,->] (-1,0) -- (1.2,0);
\draw[lgray,line width=6pt,->] (0,-1) -- (0,1.2);
\node[left] at (-1,0) {\tiny $j$};\node[right] at (1,0) {\tiny $l$};
\node[below] at (0,-1) {\tiny $I$};\node[above] at (0,1.2) {\tiny $K$};
}
\qquad\qquad
&
\tikz{0.6}{
\draw[lred,line width=1.5pt,->] (-1,0) -- (1.2,0);
\draw[lred,line width=6pt,->] (0,1) -- (0,-1.2);
\node[left] at (-1,0) {\tiny $j$};\node[right] at (1,0) {\tiny $l$};
\node[below] at (0,-1.2) {\tiny $I$};\node[above] at (0,1) {\tiny $K$};
}\\[4pt]
$L_{x,s}(I,j;K,l)$ & $M^*_{x,s}(I,j;K,l)$
\end{tabular}
\caption{The $L$- and $M^*$-vertices.  Arrows indicate the orientations
of the lattice lines, including edges with occupation zero.  The
labels $I,K$ and $j,l$ are edge occupations; the thick vertical lines
distinguish the edge type and do not specify its occupation.  The
lower and upper labels remain $I,K$ when the vertical orientation
is reversed.}
\label{fig:vertex-orientations}
\end{figure}

\subsection{The definition of stable spin Hall--Littlewood symmetric functions}
\label{ssec:fd}

We recall the functions of
\cite[Section~3]{BufetovMucciconiPetrov1905}, writing
$f=\mathsf F^*$ and $g=\mathsf F$ as in
Section~\ref{ssec:contributions}; their dependence on $q,s$ is
suppressed.  A partition $\lambda$ is a weakly decreasing finite
sequence of positive integers; $\varnothing$ denotes the empty
partition.  We append zero coordinates when needed, including
$\lambda_1=0$ for $\lambda=\varnothing$.

Recall that $\lambda'_j$ is the number of parts of $\lambda$ at least
$j$, or equivalently the length of its $j$th column, and
$\ell(\lambda)=\lambda'_1$.  The vertical occupation in vertex-model
column $j$ is instead the multiplicity
$m_j(\lambda)=\lambda'_j-\lambda'_{j+1}$ of the part $j$.
For $\mu\prec\lambda$, put
$\delta_j=\lambda'_j-\mu'_j\in\{0,1\}$.
This is the horizontal-strip condition from
Section~\ref{ssec:contributions}, and includes $\mu=\lambda$.

Consider a row of $M^*$-vertices with upper occupations $m_j(\mu)$,
lower occupations $m_j(\lambda)$, and zero horizontal occupation
sufficiently far to the right.  Conservation determines the horizontal
occupation entering column $j$ uniquely as
\[
\delta_j=\sum_{r\geq j}\bigl(m_r(\lambda)-m_r(\mu)\bigr)
=\lambda'_j-\mu'_j.
\]
The outgoing horizontal occupation is $\delta_{j+1}$, so an admissible
row exists exactly when $\mu\prec\lambda$.  In particular, the left
occupation is $d=\delta_1=\ell(\lambda)-\ell(\mu)\in\{0,1\}$.

\begin{defn}[Stable spin Hall--Littlewood symmetric functions]
\label{defn:ssHL}
For $\mu\prec\lambda$, put $d=\ell(\lambda)-\ell(\mu)=\delta_1$ and
define
\begin{equation}
\label{eq:row-definition}
f_{\lambda/\mu}(x)
=x^d\prod_{j\geq1}
M^*_{x,s}\bigl(m_j(\lambda),\delta_j;
                       m_j(\mu),\delta_{j+1}\bigr).
\end{equation}
Set this function to zero otherwise, and define its dual by
\begin{equation}
\label{eq:f-g-gauge}
g_{\lambda/\mu}(x)
=\frac{\mathsf c(\mu)}{\mathsf c(\lambda)}f_{\lambda/\mu}(x),
\qquad
\mathsf c(\rho)=\prod_{j\geq1}
\frac{(s^2;q)_{m_j(\rho)}}{(q;q)_{m_j(\rho)}}.
\end{equation}
An \emph{alphabet} is a finite list of spectral parameters.
For $A=(x_1,\ldots,x_n)$, define
\begin{equation}
\label{eq:multivariable-definition}
f_{\lambda/\mu}(A)
=\sum_{\mu=\lambda^{(0)}\prec\cdots\prec\lambda^{(n)}=\lambda}
\prod_{i=1}^n f_{\lambda^{(i)}/\lambda^{(i-1)}}(x_i),
\qquad
g_{\lambda/\mu}(A)
=\frac{\mathsf c(\mu)}{\mathsf c(\lambda)}f_{\lambda/\mu}(A).
\end{equation}
For the empty alphabet these functions equal
$\mathbf1_{\lambda=\mu}$, where $\mathbf1_E$ is one if $E$ holds
and zero otherwise.  We abbreviate
$f_\lambda=f_{\lambda/\varnothing}$ and
$g_\lambda=g_{\lambda/\varnothing}$.
\end{defn}

The factor $x^d$ assigns weight $1$ to an empty left horizontal edge
and weight $x$ to an occupied one.  These are prescribed boundary
weights.  For the columnwise constructions below, we regard this
factor as the weight of an auxiliary vertex in column zero, with one
outgoing horizontal edge and no vertical occupation.  There is no
multiplicity $m_0$; this vertex only represents the boundary factor.
Figure~\ref{fig:ssHL-one-row} shows the resulting one-row partition
function.

\begin{figure}[htbp]
\centering
\begin{tikzpicture}[x=1cm,y=1cm,>=stealth,font=\small]
\draw[lred,line width=1.5pt,->] (-0.5,0) -- (7.8,0);
\foreach \xx/\jj in {1/1,3/2,6.5/R}{
  \draw[lred,line width=5pt,->] (\xx,0.85) -- (\xx,-0.85);
  \node[above] at (\xx,0.85) {$m_{\jj}(\mu)$};
  \node[below] at (\xx,-0.85) {$m_{\jj}(\lambda)$};
  \node[below] at (\xx,-1.45) {column $\jj$};
}
\node[above] at (0.12,0) {$\delta_1$};
\node[above] at (2,0) {$\delta_2$};
\node[above] at (3.8,0) {$\delta_3$};
\node[fill=white,inner sep=3pt] at (4.8,0) {$\cdots$};
\node[above] at (5.75,0) {$\delta_R$};
\node[above] at (7.45,0) {$0$};
\node[left] at (-0.7,0) {$x$};
\node[align=center] at (-0.8,-1.05)
  {boundary weight\\$x^{\delta_1}$};
\end{tikzpicture}
\caption{The one-variable function $f_{\lambda/\mu}(x)$.
Choose $R\geq\max\{3,\lambda_1\}$.  Every intersection has weight
$M^*_{x,s}$, with upper occupations prescribed by $\mu$ and lower
occupations prescribed by $\lambda$.  The left horizontal
occupation is $\delta_1=\ell(\lambda)-\ell(\mu)$ and receives the
additional weight $x^{\delta_1}$.  The right horizontal occupation
is zero.  Conservation determines all the displayed horizontal
occupations, so this one-row partition function is the single
product in \eqref{eq:row-definition}.}
\label{fig:ssHL-one-row}
\end{figure}

The product in \eqref{eq:row-definition} has finitely many nontrivial
factors: for $j>\lambda_1$ all occupations vanish and
$M^*_{x,s}(0,0;0,0)=1$.
The sum in \eqref{eq:multivariable-definition} is also finite, since
all intermediate partitions lie between the fixed endpoints $\mu$
and $\lambda$.

The multivariable definition is the partition function of $n$ rows
placed one below another, with spectral parameters $x_1,\ldots,x_n$
from top to bottom; see Figure~\ref{fig:ssHL-multiple-rows}.
For $1\leq i<n$, the partition $\lambda^{(i)}$ encodes the vertical
occupations between rows $i$ and $i+1$, which are summed over.

\begin{figure}[htbp]
\centering
\begin{tikzpicture}[x=1cm,y=1cm,>=stealth,font=\small]
\foreach \xx/\jj in {1.2/1,3.5/2,7/R}{
  \draw[lred,line width=5pt,->] (\xx,1.1) -- (\xx,-3.3);
  \node[above] at (\xx,1.1) {$m_{\jj}(\mu)$};
  \node[below] at (\xx,-3.3) {$m_{\jj}(\lambda)$};
}
\foreach \yy/\ii in {0.5/1,-0.8/2,-2.6/n}{
  \draw[lred,line width=1.5pt,->] (-0.5,\yy) -- (8.0,\yy);
  \node[left] at (-0.65,\yy) {$x_{\ii}$};
  \node[above] at (-0.12,\yy) {$\delta_1^{(\ii)}$};
  \node[above] at (7.65,\yy) {$0$};
  \node[fill=white,inner sep=3pt] at (5.25,\yy) {$\cdots$};
}
\foreach \xx in {1.2,3.5,7}{
  \node[fill=white,inner sep=3pt] at (\xx,-1.7) {$\vdots$};
}
\node[fill=white,inner sep=2pt] at (1.2,-0.17)
  {$m_1(\lambda^{(1)})$};
\node[fill=white,inner sep=2pt] at (3.5,-0.17)
  {$m_2(\lambda^{(1)})$};
\node[fill=white,inner sep=2pt] at (7,-0.17)
  {$m_R(\lambda^{(1)})$};
\end{tikzpicture}
\caption{The partition function
$f_{\lambda/\mu}(x_1,\ldots,x_n)$.
Choose $R\geq\max\{3,\lambda_1\}$ as in Figure~\ref{fig:ssHL-one-row}.
The spectral parameters are ordered from top to bottom.  The
vertical occupations between successive rows are summed over and
are encoded by $\lambda^{(1)},\ldots,\lambda^{(n-1)}$.
In row $i$, the horizontal occupation entering column $j$ is
$\delta_j^{(i)}=(\lambda^{(i)})'_j-(\lambda^{(i-1)})'_j$,
and the left boundary contributes $x_i^{\delta_1^{(i)}}$.
All right boundary occupations are zero.}
\label{fig:ssHL-multiple-rows}
\end{figure}

For $\mu\prec\lambda$, the local relation \eqref{eq:vertex-Mdual}
gives a useful representation of the dual function.  The factors
$c_{m_j(\lambda)}/c_{m_j(\mu)}$ multiply to
$\mathsf c(\lambda)/\mathsf c(\mu)$ and cancel the ratio in
\eqref{eq:f-g-gauge}.  Hence
\begin{equation}
\label{eq:dual-row}
g_{\lambda/\mu}(x)
=x^d\prod_{j\geq1}
L_{x,s}\bigl(m_j(\mu),\delta_j;
                     m_j(\lambda),\delta_{j+1}\bigr).
\end{equation}
This is an upward-oriented $L$-row, with $\mu$ below and $\lambda$
above, and with the same left boundary weight $x^d$.
The representation follows from the definition and will be used in
Section~\ref{sec:boundary-family}.

\subsection{Two-parameter skew Littlewood identity}
\label{sec:boundary-family}

We prove Theorem~\ref{thm:intro-Littlewood}, the identity that
normalizes the boundary transitions of our half-space field.
We first derive explicit boundary coefficients and then compare
the two partition functions by local identities of $2\times2$
matrices.

Fix $0<q<1$ and $-1<s\leq0$.  The two boundary parameters are
$t\in\mathbb C$ and $\zeta\in\mathbb C\setminus\{0\}$; their
nonnegative range will be determined separately.  The zero-spin
specialization is compared with
\cite[Section~2.3]{He2024Boundary} below.  Write
$U=1-t$ and $V=\zeta^{-1}-t\zeta$.  Define the sequences
$E_m=E_m(t,\zeta)$ and $O_m=O_m(t,\zeta)$ by
$E_{-1}=O_{-1}=0$, $E_0=O_0=1$, and
\begin{align}
 (1-s^2q^m)E_{m+1}
 &=(U+sq^mV)E_m+t(1-q^m)E_{m-1},
 \label{eq:fb-even-recurrence}\\
 (1-s^2q^m)O_{m+1}
 &=(V+sq^mU)O_m+t(1-q^m)O_{m-1},\qquad m\geq0.
 \label{eq:fb-odd-recurrence}
\end{align}
The positive denominators determine both sequences uniquely, with
$E_1=(U+sV)/(1-s^2)$ and $O_1=(V+sU)/(1-s^2)$.

Recall that $m_j(\lambda)$ is the number of parts of $\lambda$
equal to $j$.  Set
\begin{equation}
 W_\lambda(t,\zeta)=
 \prod_{\substack{j\geq1\\j\ {\rm even}}}E_{m_j(\lambda)}(t,\zeta)
 \prod_{\substack{j\geq1\\j\ {\rm odd}}}O_{m_j(\lambda)}(t,\zeta).
 \label{eq:fb-boundary-weights}
\end{equation}
Here parity refers to the part size and the subscript to its
multiplicity: for example, $W_{(4,4,3,1,1)}=E_2O_1O_2$.
Only positive part sizes enter, all but finitely many factors are
one, and $W_\varnothing=1$.  We suppress the dependence on $q,s$.

\smallskip
\noindent\emph{Explicit coefficients and their generating functions.}
In addition to the finite products defined in Section~\ref{ssec:mw},
write $(z;q)_\infty=\prod_{r\geq0}(1-zq^r)$ and
$\genfrac{[}{]}{0pt}{}{m}{k}_q=(q;q)_m/((q;q)_k(q;q)_{m-k})$.
The solutions of the recurrences are
\begin{align}
 E_m&=\frac{1}{(s^2;q)_m}\sum_{k=0}^m
 \genfrac{[}{]}{0pt}{}{m}{k}_q(-t)^{m-k}
 (-s/\zeta;q)_k(-s\zeta;q)_{m-k},
 \label{eq:fb-even-finite}\\
 O_m&=\frac{1}{(s^2;q)_m}\sum_{k=0}^m
 \genfrac{[}{]}{0pt}{}{m}{k}_q\zeta^{-k}(-t\zeta)^{m-k}
 (-s\zeta;q)_k(-s/\zeta;q)_{m-k}.
 \label{eq:fb-odd-finite}
\end{align}
The corresponding generating functions, used again in
Section~\ref{ssec:boundary-to}, are
\begin{align}
 F_E(z):=\sum_{m\geq0}\frac{(s^2;q)_mE_mz^m}{(q;q)_m}
 &=\frac{(-sz/\zeta;q)_\infty(st\zeta z;q)_\infty}
 {(z;q)_\infty(-tz;q)_\infty},
 \label{eq:fb-even-generating}\\
 F_O(z):=\sum_{m\geq0}\frac{(s^2;q)_mO_mz^m}{(q;q)_m}
 &=\frac{(-sz;q)_\infty(stz;q)_\infty}
 {(z/\zeta;q)_\infty(-t\zeta z;q)_\infty}.
 \label{eq:fb-odd-generating}
\end{align}
These identities hold as power series about $z=0$.
To verify them, use the $q$-binomial expansion
\[
 \frac{(az;q)_\infty}{(z;q)_\infty}
 =\sum_{m\geq0}\frac{(a;q)_m}{(q;q)_m}z^m.
\]
Indeed the ratio $R(z)=\sum r_mz^m$ is analytic near zero and
satisfies $(1-z)R(z)=(1-az)R(qz)$.  Coefficient comparison gives
$r_0=1$ and $(1-q^m)r_m=(1-aq^{m-1})r_{m-1}$, proving the
expansion.  For $F_E$, apply it with $(a,z)$ replaced by
$(-s/\zeta,z)$ and $(-s\zeta,-tz)$; for $F_O$, use
$(-s\zeta,z/\zeta)$ and $(-s/\zeta,-t\zeta z)$.
In each product the coefficient of $z^m$ is the convolution over
$k+(m-k)=m$.  Multiplication by $(q;q)_m/(s^2;q)_m$ gives
\eqref{eq:fb-even-finite} and \eqref{eq:fb-odd-finite}, including
$t=0$ without division by $t$.

To identify these coefficients with the recurrence-defined sequences,
cancel the first factors in the proposed infinite products.  This gives
\begin{align*}
 \frac{F_E(qz)}{F_E(z)}
 &=\frac{(1-z)(1+tz)}{(1+sz/\zeta)(1-st\zeta z)}
 =\frac{1-Uz-tz^2}{1+sVz-ts^2z^2},\\
 \frac{F_O(qz)}{F_O(z)}
 &=\frac{(1-z/\zeta)(1+t\zeta z)}{(1+sz)(1-stz)}
 =\frac{1-Vz-tz^2}{1+sUz-ts^2z^2}.
\end{align*}
Put $C_m=(s^2;q)_mE_m/(q;q)_m$.  The first quotient gives
$(1+sVz-ts^2z^2)F_E(qz)=(1-Uz-tz^2)F_E(z)$.
Comparing coefficients of $z^{m+1}$, for $m\geq1$, yields
\[
 (1-q^{m+1})C_{m+1}
 =(U+sVq^m)C_m+t(1-s^2q^{m-1})C_{m-1}.
\]
Dividing by $(s^2;q)_m/(q;q)_m$ and canceling the
$q$-Pochhammer factors gives
\eqref{eq:fb-even-recurrence}.  At $m=0$, comparison of the
coefficient of $z$ gives $(1-q)C_1=U+sV$, hence the same
recurrence.  The second quotient interchanges $U,V$ and proves
\eqref{eq:fb-odd-recurrence}.  The constant coefficients are one,
so uniqueness establishes all four formulas.

\begin{prop}[Nonnegativity of the boundary weights]
\label{prop:fb-positivity}
Let $0\leq t<1$, $\zeta>0$, and $\sigma=-s$.  The weights
$W_\lambda(t,\zeta)$ are nonnegative for every partition if and only if
\begin{equation}
 \sigma U\leq V,\qquad \sigma V\leq U.
 \label{eq:fb-positive-domain}
\end{equation}
If both inequalities are strict, every boundary weight is positive.
\end{prop}
\begin{proof}
The weights $W_{(2)}=(U-\sigma V)/(1-\sigma^2)$ and
$W_{(1)}=(V-\sigma U)/(1-\sigma^2)$ imply the necessity of
\eqref{eq:fb-positive-domain}.
Conversely, these inequalities give $U>0$, $V\geq\sigma U\geq0$,
and, for $m\geq0$,
\[
 U-\sigma q^mV\geq U-\sigma V\geq0,\qquad
 V-\sigma q^mU\geq V-\sigma U\geq0.
\]
The remaining recurrence coefficients $t(1-q^m)$ are nonnegative,
and the denominators are positive.  Induction from
$E_{-1}=O_{-1}=0$, $E_0=O_0=1$ proves $E_m,O_m\geq0$ and hence
$W_\lambda\geq0$.  Under strict inequalities, the coefficients
of $E_m,O_m$ in the respective recurrences are positive, so the
same induction gives strict positivity.
\end{proof}

At $s=0$, the criterion becomes $t\zeta^2\leq1$.
For every fixed $s\in(-1,0]$ and $0\leq t<1$, its inequalities
are strict at $\zeta=1$, where $U=V=1-t$, and hence on an open
interval about one.  The algebraic identity below also permits
complex boundary parameters.

Define the scalar factor
\begin{equation}
 \Phi_{t,\zeta}(x)=\frac{1+Vx-tx^2}{1-x^2}
 =\frac{(1-t\zeta x)(1+x/\zeta)}{1-x^2}.
 \label{eq:fb-boundary-factor}
\end{equation}
\begin{thm}[Two-parameter skew Littlewood identity]
\label{thm:fb-skew-Littlewood}
Fix $0<q<1$ and $-1<s\leq0$.  For every partition $\mu$,
$0<x<1$, $t\in\mathbb C$, and
$\zeta\in\mathbb C\setminus\{0\}$,
\begin{equation}
 \sum_\lambda W_\lambda(t,\zeta)f_{\lambda/\mu}(x)
 =\Phi_{t,\zeta}(x)
 \sum_\kappa W_\kappa(t,\zeta)g_{\mu/\kappa}(x).
 \label{eq:fb-skew-one-row}
\end{equation}
The left sum converges absolutely; the right sum is finite.
\end{thm}

Only $\mu\prec\lambda$ contributes on the left, and only
$\kappa\prec\mu$ on the right.  Thus the left sum permits an
arbitrarily large first part, whereas
every partition on the right lies inside $\mu$.
Figure~\ref{fig:two-parameter-skew-Littlewood} shows the two sums
using the row definitions from Section~\ref{ssec:fd}.

\begin{figure}[htbp]
\centering
\begin{tikzpicture}[x=.95cm,y=.8cm,>=stealth,font=\scriptsize]
\node at (2.55,2.65) {$f_{\lambda/\mu}(x)$: downward $M^*$ vertices};
\node at (10.05,2.65) {$g_{\mu/\kappa}(x)$: upward $L$ vertices};
\node at (2.55,2.08) {fixed partition $\mu$};
\node at (10.05,2.08) {fixed partition $\mu$};
\draw[lred,line width=1.4pt,->] (-.45,0)--(5.35,0);
\draw[lgray,line width=1.4pt,->] (7.05,0)--(12.85,0);
\foreach \p/\j in {.7/1,2.25/2,3.8/3}{
 \draw[lred,line width=3pt,->] (\p,1.08)--(\p,-1.56);
 \node[above] at (\p,1.08) {$m_{\j}(\mu)$};
 \node[fill=white,inner sep=1pt] at (\p,-.97) {$m_{\j}(\lambda)$};
 \fill[lred] (\p,0) circle (2pt);
}
\foreach \p/\j in {8.2/1,9.75/2,11.3/3}{
 \draw[lgray,line width=3pt,->] (\p,-1.56)--(\p,1.08);
 \node[above] at (\p,1.08) {$m_{\j}(\mu)$};
 \node[fill=white,inner sep=1pt] at (\p,-.97) {$m_{\j}(\kappa)$};
 \fill[lgray] (\p,0) circle (2pt);
}
\node[above] at (0,0) {$\delta_1$};
\node[above] at (1.46,0) {$\delta_2$};
\node[above] at (3.02,0) {$\delta_3$};
\node[fill=white,inner sep=1pt] at (4.67,0) {$\cdots$};
\node[right] at (5.35,0) {$0$};
\node[below] at (-.12,-.15) {$x^{\delta_1}$};
\node[above] at (7.5,0) {$\varepsilon_1$};
\node[above] at (8.96,0) {$\varepsilon_2$};
\node[above] at (10.52,0) {$\varepsilon_3$};
\node[fill=white,inner sep=1pt] at (12.17,0) {$\cdots$};
\node[right] at (12.85,0) {$0$};
\node[below] at (7.38,-.15) {$x^{\varepsilon_1}$};
\draw[lred,line width=.8pt] (-.1,-2.46) rectangle (5.2,-1.56);
\node at (2.55,-2.01) {$\textstyle\sum_{\lambda:\,\mu\prec\lambda}W_\lambda(t,\zeta)$};
\draw[lgray,line width=.8pt] (7.4,-2.46) rectangle (12.7,-1.56);
\node at (10.05,-2.01) {$\textstyle\sum_{\kappa:\,\kappa\prec\mu}W_\kappa(t,\zeta)$};
\node at (6.34,-1.12) {$=\ \Phi_{t,\zeta}(x)$};
\end{tikzpicture}
\caption{The two partition functions in the skew Littlewood identity.
The upper occupations are fixed by $\mu$; the lower occupations are
summed with weight $W_\lambda$ on the left and $W_\kappa$ on the right.
Horizontal occupations are $\delta_j=\lambda'_j-\mu'_j$ and
$\varepsilon_j=\mu'_j-\kappa'_j$, respectively.  The factors at the
left ends are $x^{\delta_1}$ and $x^{\varepsilon_1}$.  The dots denote
the remaining columns; sufficiently far to the right all occupations
are zero.  Vertical arrows fix the $M^*$ and $L$ conventions of
Section~\ref{ssec:mw}; the boxes denote summation over partitions,
not additional vertex weights.}
\label{fig:two-parameter-skew-Littlewood}
\end{figure}

\begin{proof}
Fix $\mu$.  Each shared horizontal occupation is binary;
conservation then fixes the multiplicity in the partition being
summed.  Thus the two rows have $2\times2$ transfer matrices.

\smallskip
\noindent\emph{Step 1: the matrix for the $f$-row.}
Put $\delta_j=\lambda'_j-\mu'_j$.  For an allowed horizontal strip,
all $\delta_j$ are zero or one, and are eventually zero.  Since
$\lambda'_1=\ell(\lambda)$, the boundary factor in
\eqref{eq:row-definition} is $x^{\delta_1}$.  The telescoping sum
$\delta_1=\sum_{j\geq1}(\delta_j-\delta_{j+1})$ gives
\[
 x^{\delta_1}=\prod_{j\geq1}x^{\delta_j-\delta_{j+1}}.
\]
Assign the $j$th factor to column $j$; since $x>0$, this
redistribution is valid even when its exponent is negative.

Fix $K=m_j(\mu)$ and write $\theta_m$ for the boundary factor
assigned to multiplicity $m$ at this column, with $\theta_{-1}=0$.
Using $m_j(\rho)=\rho'_j-\rho'_{j+1}$ gives
$m_j(\lambda)=K+\delta_j-\delta_{j+1}$.  The column weight is
\[
 x^{\delta_j-\delta_{j+1}}
 \theta_{K+\delta_j-\delta_{j+1}}
 M^*_{x,s}(K+\delta_j-\delta_{j+1},\delta_j;
                    K,\delta_{j+1}).
\]
Use $\delta_{j+1}$ as the row index and $\delta_j$ as the column
index, both ordered $0,1$.  Substitution of the four $M^*$ weights
from Section~\ref{ssec:mw} gives the following numerators; every
entry has the common denominator $1-sx$:
\[
\begin{array}{c|c|c|c|c}
\delta_{j+1}&\delta_j&m_j(\lambda)&
x^{\delta_j-\delta_{j+1}}&\text{weighted numerator}\\ \hline
0&0&K&1&\theta_K(1-sxq^K)\\
0&1&K+1&x&x\theta_{K+1}(1-s^2q^K)\\
1&0&K-1&x^{-1}&\theta_{K-1}(1-q^K)\\
1&1&K&1&\theta_K(x-sq^K)
\end{array}
\]
Thus the matrix is
\begin{equation}
 A_K[\theta]=\frac1{1-sx}\begin{pmatrix}
 \theta_K(1-sxq^K)&x\theta_{K+1}(1-s^2q^K)\\
 \theta_{K-1}(1-q^K)&\theta_K(x-sq^K)
 \end{pmatrix}.
 \label{eq:fb-transfer-A}
\end{equation}
When $K=0$, the entry that would give multiplicity $-1$ is zero;
our convention $\theta_{-1}=0$ includes this case.

\smallskip
\noindent\emph{Step 2: the matrix for the $g$-row.}
For $\kappa\prec\mu$, put
$\varepsilon_j=\mu'_j-\kappa'_j\in\{0,1\}$.  Then
$m_j(\kappa)=K-\varepsilon_j+\varepsilon_{j+1}$.
Using \eqref{eq:dual-row} and distributing
$x^{\varepsilon_1}=\prod_jx^{\varepsilon_j-\varepsilon_{j+1}}$
in the same way gives column weight
\[
 x^{\varepsilon_j-\varepsilon_{j+1}}
 \theta_{K-\varepsilon_j+\varepsilon_{j+1}}
 L_{x,s}(K-\varepsilon_j+\varepsilon_{j+1},\varepsilon_j;
                    K,\varepsilon_{j+1}).
\]
Use $1-\varepsilon_{j+1}$ as row index and
$1-\varepsilon_j$ as column index.  The four entries, with
denominator $1-sx$, are
\[
\begin{array}{c|c|c|c|c}
\text{row}&\text{column}&(\varepsilon_{j+1},\varepsilon_j)&
m_j(\kappa)&\text{weighted numerator}\\ \hline
0&0&(1,1)&K&\theta_K(x-sq^K)\\
0&1&(1,0)&K+1&\theta_{K+1}(1-s^2q^K)\\
1&0&(0,1)&K-1&x\theta_{K-1}(1-q^K)\\
1&1&(0,0)&K&\theta_K(1-sxq^K)
\end{array}
\]
We obtain
\begin{equation}
 B_K[\theta]=\frac1{1-sx}\begin{pmatrix}
 \theta_K(x-sq^K)&\theta_{K+1}(1-s^2q^K)\\
 x\theta_{K-1}(1-q^K)&\theta_K(1-sxq^K)
 \end{pmatrix}.
 \label{eq:fb-transfer-B}
\end{equation}

\begin{samepage}
\smallskip
\noindent\emph{Step 3: the local matrix calculation.}
Let $\theta_j$ denote the sequence $(E_m)_{m\geq-1}$ for even
$j$ and $(O_m)_{m\geq-1}$ for odd $j$.  Define
\begin{equation}
 S_{\rm even}=\begin{pmatrix}
 x(V+xU)&t(x^2-1)\\x^2-1&U+xV
 \end{pmatrix},\qquad
 S_{\rm odd}=\begin{pmatrix}
 x(U+xV)&t(x^2-1)\\x^2-1&V+xU
 \end{pmatrix}.
 \label{eq:fb-intertwiners}
\end{equation}
\end{samepage}
Write $S_j=S_{\rm even}$ for even $j$ and $S_j=S_{\rm odd}$ for
odd $j$, including $S_0=S_{\rm even}$.  We claim that
\begin{equation}
 A_K[\theta_j]S_{j-1}=S_jB_K[\theta_j].
 \label{eq:fb-local-intertwining}
\end{equation}

We verify all four entries at once for either parity.  In this
calculation write $\theta=\theta_j$ and abbreviate
\[
 p=q^K,\quad h=\theta_K,\quad
 D=(1-s^2p)\theta_{K+1},\quad T=(1-p)\theta_{K-1},
 \quad d=1-sx,\quad z=x^2-1.
\]
Set $(\alpha,\beta)=(U,V)$ when $j$ is even and
$(\alpha,\beta)=(V,U)$ when $j$ is odd.  The recurrence at this
column is exactly $D=(\alpha+sp\beta)h+tT$.
With $a=h(1-sxp)$ and $c=h(x-sp)$, the matrices take the form
\begin{align*}
 dA_K[\theta_j]&=\begin{pmatrix}a&xD\\T&c\end{pmatrix},&
 dB_K[\theta_j]&=\begin{pmatrix}c&D\\xT&a\end{pmatrix},\\
 S_j&=\begin{pmatrix}x(\beta+x\alpha)&tz\\z&\alpha+x\beta\end{pmatrix},&
 S_{j-1}&=\begin{pmatrix}x(\alpha+x\beta)&tz\\z&\beta+x\alpha\end{pmatrix}.
\end{align*}
For the diagonal entries we first compute the scalar difference
\begin{align*}
 a(\alpha+x\beta)-c(\beta+x\alpha)
 &=h\alpha\bigl[(1-sxp)-x(x-sp)\bigr]\\
 &\quad+h\beta\bigl[x(1-sxp)-(x-sp)\bigr]\\
 &=h\alpha(1-x^2)+h\beta sp(1-x^2)
 =-zh(\alpha+sp\beta).
\end{align*}
Let $C=d(A_K[\theta_j]S_{j-1}-S_jB_K[\theta_j])$ and index its
entries by $0,1$.  Taking the four row-by-column products gives
\begin{align*}
 C_{00}
 &=ax(\alpha+x\beta)+xDz-x(\beta+x\alpha)c-tzxT\\*
 &=xz\bigl[D-(\alpha+sp\beta)h-tT\bigr]=0,\\
 C_{01}
 &=atz+xD(\beta+x\alpha)-x(\beta+x\alpha)D-tza=0,\\
 C_{10}
 &=Tx(\alpha+x\beta)+cz-zc-(\alpha+x\beta)xT=0,\\
 C_{11}
 &=Ttz+c(\beta+x\alpha)-zD-(\alpha+x\beta)a\\*
 &=-z\bigl[D-(\alpha+sp\beta)h-tT\bigr]=0.
\end{align*}
Thus \eqref{eq:fb-local-intertwining} is exactly the recurrence
relation, since $d=1-sx>0$.  No division by a boundary weight or
determinant occurred, so vanishing coefficients and singular $S_j$
are allowed.  The matrices $S_j$ need not be nonnegative.

For a physical column set $K_j=m_j(\mu)$,
$A_j=A_{K_j}[\theta_j]$ and $B_j=B_{K_j}[\theta_j]$.
The local identity sums the shared binary index:
\[
 \sum_{c=0}^1(A_j)_{u,c}(S_{j-1})_{c,v}
 =\sum_{d=0}^1(S_j)_{u,d}(B_j)_{d,v},
\]
where $u=\delta_{j+1}$, $v=1-\varepsilon_j$ are fixed,
$c=\delta_j$ and $d=1-\varepsilon_{j+1}$.
Figure~\ref{fig:Littlewood-matrix-proof} illustrates the identity
and its iteration; its lower lines show ordered matrix products.

\begin{figure}[htbp]
\centering
\begin{tikzpicture}[x=.95cm,y=.85cm,font=\scriptsize,
 amat/.style={draw=lred,line width=1pt,fill=white,minimum width=.82cm,minimum height=.55cm},
 bmat/.style={draw=lgray,line width=1pt,fill=white,minimum width=.82cm,minimum height=.55cm},
 smat/.style={draw=black,line width=.6pt,fill=white,minimum width=.82cm,minimum height=.55cm}]
\node at (6.25,1.1) {One column: sum the repeated binary index};
\draw (-.15,0)--(5.25,0);
\node[amat] at (1.15,0) {$A_j$};
\node[smat] at (3.85,0) {$S_{j-1}$};
\node[above] at (.15,0) {$u$};
\node[above] at (2.5,0) {$c\in\{0,1\}$};
\node[above] at (5,0) {$v$};
\node at (6.25,0) {$=$};
\draw (7.25,0)--(12.65,0);
\node[smat] at (8.55,0) {$S_j$};
\node[bmat] at (11.25,0) {$B_j$};
\node[above] at (7.55,0) {$u$};
\node[above] at (9.9,0) {$d\in\{0,1\}$};
\node[above] at (12.4,0) {$v$};
\node at (6.25,-1.0)
 {$\displaystyle\sum_c (A_j)_{u,c}(S_{j-1})_{c,v}
   =\sum_d(S_j)_{u,d}(B_j)_{d,v}$};
\node at (6.25,-2.0) {Apply the identity at columns $1,2,\ldots,R$};
\draw (.3,-2.85)--(12.2,-2.85);
\node[fill=white,inner sep=3pt] at (.3,-2.85) {$w_R$};
\node[amat] at (2.35,-2.85) {$A_R$};
\node[amat] at (4.45,-2.85) {$A_{R-1}$};
\node[fill=white,inner sep=3pt] at (6.55,-2.85) {$\cdots$};
\node[amat] at (8.45,-2.85) {$A_1$};
\node[smat] at (10.35,-2.85) {$S_0$};
\node[fill=white,inner sep=3pt] at (12.2,-2.85) {$\mathbf1$};
\node at (6.25,-3.5) {$=$};
\draw (.3,-4.2)--(12.2,-4.2);
\node[fill=white,inner sep=3pt] at (.3,-4.2) {$w_R$};
\node[smat] at (2.35,-4.2) {$S_R$};
\node[bmat] at (4.45,-4.2) {$B_R$};
\node[bmat] at (6.55,-4.2) {$B_{R-1}$};
\node[fill=white,inner sep=3pt] at (8.45,-4.2) {$\cdots$};
\node[bmat] at (10.35,-4.2) {$B_1$};
\node[fill=white,inner sep=3pt] at (12.2,-4.2) {$\mathbf1$};
\end{tikzpicture}
\caption{The matrix calculation proving the skew Littlewood identity.
Here $A_j=A_{m_j(\mu)}[\theta_j]$ and
$B_j=B_{m_j(\mu)}[\theta_j]$.  At the top, the external indices
$u=\delta_{j+1}$ and $v=1-\varepsilon_j$ are fixed; the internal
indices are $c=\delta_j$ on the left and
$d=1-\varepsilon_{j+1}$ on the right.  The lower two lines are ordered
matrix products, with every joined binary index summed.  The vector
$w_R$ accounts for columns beyond $R=\mu_1$, and $\mathbf1$ sums the
occupation at the left end of the row.  Evaluating $w_RS_R$ and
$S_0\mathbf1$ gives the scalar factor $\Phi_{t,\zeta}(x)$.}
\label{fig:Littlewood-matrix-proof}
\end{figure}

\smallskip
\noindent\emph{Step 4: finite partition sums and the columns at infinity.}
Set $R=\mu_1$, with $R=0$ for $\mu=\varnothing$, and put
$K_j=m_j(\mu)$, $A_j=A_{K_j}[\theta_j]$,
$B_j=B_{K_j}[\theta_j]$.  We use the row vectors
$e_0=(1,0)$, $e_1=(0,1)$, and the column vector
$\mathbf1=(1,1)^{\mathsf T}$.
First impose a cutoff $\lambda_1\leq L$, $L\geq R$.
The far-right occupation is $\delta_{L+1}=0$, while the left
occupation $\delta_1$ is summed over $\{0,1\}$.  It follows that
\[
 \sum_{\lambda:\,\lambda_1\leq L}W_\lambda f_{\lambda/\mu}(x)
 =e_0A_0[\theta_L]\cdots A_0[\theta_{R+1}]
       A_R\cdots A_1\mathbf1.
\]
The order is decreasing because row indices are outgoing
occupations.  The final vector is $\mathbf1$: the boundary factor
$x^{\delta_1}$ was already distributed among the columns.

For a binary string ending in $\delta_{L+1}=0$, set
$n_j=K_j+\delta_j-\delta_{j+1}$.  Negative $n_j$ gives a zero
matrix entry.  Otherwise these multiplicities define a unique
partition $\lambda$ satisfying
\[
 \lambda'_j-\mu'_j
 =\sum_{k=j}^{L}(n_k-K_k)
 =\sum_{k=j}^{L}(\delta_k-\delta_{k+1})=\delta_j.
\]
Conversely every $\mu\prec\lambda$ with $\lambda_1\leq L$
determines this string, proving exact enumeration.

For $j>R$, $m_j(\mu)=0$, whereas $\theta_j$ still depends on
parity.  The values $\theta_0=1$, $\theta_{-1}=0$, $E_1,O_1$ give
\[
 A_0[\theta_j]=\begin{pmatrix}1&b_j\\0&r\end{pmatrix},
 \quad r=\frac{x-s}{1-sx},\quad
 b_e=\frac{x(U+sV)}{1-sx},\quad
 b_o=\frac{x(V+sU)}{1-sx},
\]
where $b_j=b_e$ or $b_o$ according to the parity of $j$.
We have $r>0$ and
\[
 1-r=\frac{(1-x)(1+s)}{1-sx}>0,
 \qquad
 1-r^2=\frac{(1-s^2)(1-x^2)}{(1-sx)^2}.
\]
In particular, $0<r<1$.

Multiplying the triangular matrices gives
\[
 e_0A_0[\theta_L]\cdots A_0[\theta_{R+1}]
 =\left(1,\sum_{j=R+1}^{L}b_jr^{j-R-1}\right).
\]
The second entry sums the possible locations of the single
change of occupation in these columns.  For even
$R$, the coefficients alternate between $b_o$ and $b_e$, giving
\[
 \sum_{k\geq0}(b_or^{2k}+b_er^{2k+1})
 =\frac{b_o+rb_e}{1-r^2}.
\]
The numerator simplifies as
\begin{align*}
 b_o+rb_e
 &=\frac{x\bigl[(V+sU)(1-sx)+(x-s)(U+sV)\bigr]}{(1-sx)^2}\\
 &=\frac{x(1-s^2)(V+xU)}{(1-sx)^2}.
\end{align*}
Dividing by $1-r^2$ gives $x(V+xU)/(1-x^2)$.  For odd $R$,
exchange $b_e,b_o$ and $U,V$.  Hence
\begin{equation}
 e_0A_0[\theta_L]\cdots A_0[\theta_{R+1}]
 \longrightarrow w_R=
 \begin{cases}
 (1,x(V+xU)/(1-x^2)),&R\text{ even},\\
 (1,x(U+xV)/(1-x^2)),&R\text{ odd}.
 \end{cases}
 \label{eq:fb-vacuum-tail}
\end{equation}

This limit also establishes absolute convergence of the partition sum.
Including the configuration with no change of occupation, the sum
of absolute values of the tail weights is bounded by
\[
 1+\sum_{j>R}|b_j|r^{j-R-1}
 \leq1+\frac{\max\{|b_e|,|b_o|\}}{1-r}<\infty.
\]
The remaining $R$ columns contribute a finite factor.  Their
entries and $b_e,b_o$ are bounded on compact sets of $(t,\zeta)$
with $\zeta\ne0$, proving local uniform convergence as well,
without requiring positivity.  Taking $L\to\infty$ gives
\[
 \sum_\lambda W_\lambda f_{\lambda/\mu}(x)
 =w_RA_R\cdots A_1\mathbf1.
\]

For the right side, $\kappa\prec\mu$ implies
$\kappa\subseteq\mu$, giving a finite sum.  The far-right
occupation $\varepsilon_{R+1}=0$ selects $e_1$ in the
complemented indexing.  By \eqref{eq:fb-transfer-B},
\[
 B_0[\theta_j]=\begin{pmatrix}
 r&(1-s^2)\theta_1/(1-sx)\\0&1
 \end{pmatrix},\qquad e_1B_0[\theta_j]=e_1.
\]
Columns beyond $R$ can thus be omitted.  The same conservation
argument as above, now with multiplicities
$K_j-\varepsilon_j+\varepsilon_{j+1}$, yields
\[
 \sum_\kappa W_\kappa g_{\mu/\kappa}(x)
 =e_1B_R\cdots B_1\mathbf1.
\]
For $R=0$, the products are identity matrices and the right sum
is $e_1\mathbf1=1$.

\smallskip
\noindent\emph{Step 5: evaluation of the end vectors.}
Set $\Gamma=x^2+Vx-t$.  We will verify
\begin{equation}
 S_0\mathbf1=\Gamma\mathbf1,\qquad
 w_RS_R=\Gamma\Phi_{t,\zeta}(x)e_1.
 \label{eq:fb-end-vectors}
\end{equation}
The first identity follows from the two row sums of
$S_0=S_{\rm even}$:
\begin{align*}
 x(V+xU)+t(x^2-1)&=xV+x^2(U+t)-t=\Gamma,\\
 x^2-1+U+xV&=x^2+xV-t=\Gamma,
\end{align*}
where $U+t=1$.
For even $R$, put $H=1-x^2$ and use
$w_R=(1,x(V+xU)/H)$.  Its product with the first column of
$S_R$ is $x(V+xU)-x(V+xU)=0$; the second component is
\[
 -tH+\frac{x(V+xU)(U+xV)}{H}
 =\frac{-t(1-x^2)^2+x(V+xU)(U+xV)}{1-x^2}.
\]
To check its factorization, expand the numerator:
\begin{align*}
 &-t(1-x^2)^2+x(V+xU)(U+xV)\\
 &\hspace{1em}=-t+UVx+(U^2+V^2+2t)x^2+UVx^3-tx^4\\
 &\hspace{1em}=-t+(1-t)Vx+(1+t^2+V^2)x^2+(1-t)Vx^3-tx^4\\
 &\hspace{1em}=(x^2+Vx-t)(1+Vx-tx^2).
\end{align*}
Here $U=1-t$.  The second component is therefore
$\Gamma\Phi_{t,\zeta}(x)$.  For odd $R$, the same calculation
exchanges $U+xV$ and $V+xU$, proving \eqref{eq:fb-end-vectors}
for both parities.

\smallskip
\noindent\emph{Step 6: combining the columns and treating $\Gamma=0$.}
Iterating \eqref{eq:fb-local-intertwining} and using
\eqref{eq:fb-end-vectors} gives
\[
\begin{aligned}
 \Gamma w_RA_R\cdots A_1\mathbf1
 &=w_RA_R\cdots A_1S_0\mathbf1\\
 &=w_RS_RB_R\cdots B_1\mathbf1\\
 &=\Gamma\Phi_{t,\zeta}(x)e_1B_R\cdots B_1\mathbf1.
\end{aligned}
\]
If $\Gamma\ne0$, cancellation proves
\eqref{eq:fb-skew-one-row}.  The exceptional case cannot be
discarded: substitution of $V$ gives
\[
 \Gamma=x^2+(\zeta^{-1}-t\zeta)x-t
 =(x-t\zeta)(x+\zeta^{-1}),
\]
which vanishes, for example, at $\zeta=1$, $t=x\in(0,1)$,
where the inequalities in Proposition~\ref{prop:fb-positivity}
are both strict.

For fixed $q,s,x$, the finite coefficient formulas and the
evaluated tail \eqref{eq:fb-vacuum-tail} are polynomials in $t$
and Laurent polynomials in $\zeta$.  Hence the finite matrix
expressions for both partition sums are continuous for
$\zeta\ne0$.  The set $\Gamma\ne0$ is dense there: at fixed
$\zeta\ne-1/x$, $\Gamma=x^2+x/\zeta-t(1+x\zeta)$ has at most
one zero in $t$, and points with $\zeta=-1/x$ can first be
approximated by other $\zeta$.  Continuity therefore extends
\eqref{eq:fb-skew-one-row} to $\Gamma=0$.
\end{proof}

\smallskip
\noindent\emph{A direct check when $\mu=\varnothing$.}
Only $\varnothing$ and $(k)$, $k\geq1$, contribute.  With
$r=(x-s)/(1-sx)$ and $C=x(1-s^2)/(1-sx)$, the row weights give
$f_{(k)}(x)=Cr^{k-1}$.  Since $CO_1=b_o$ and $CE_1=b_e$,
\[
 \sum_\lambda W_\lambda f_\lambda(x)
 =1+\frac{C(O_1+rE_1)}{1-r^2}
 =1+\frac{x(V+xU)}{1-x^2}=\Phi_{t,\zeta}(x).
\]
This checks the parity convention and boundary normalization.

\smallskip
\noindent\emph{Zero spin and the Hall--Littlewood identity.}
Let $H_m(z;q)=\sum_{k=0}^m\genfrac{[}{]}{0pt}{}{m}{k}_qz^k$
be the Rogers--Szeg\H{o} polynomial.  Setting $s=0$ in the finite
formulas and reversing $k\mapsto m-k$ gives
\begin{equation}
 E_m\big|_{s=0}=H_m(-t;q),\qquad
 O_m\big|_{s=0}=\zeta^{-m}H_m(-t\zeta^2;q).
 \label{eq:fb-zero-spin}
\end{equation}
In particular the odd coefficient is directly defined at $t=0$.
The same index reversal proves
$H_m(z;q)=z^mH_m(z^{-1};q)$ for $z\ne0$.  For $t\ne0$ this gives
\[
 \zeta^{-m}H_m(-t\zeta^2;q)
 =(-t\zeta)^mH_m\!\left(-\frac1{t\zeta^2};q\right).
\]
To compare with \cite[Theorem~1.1]{Warnaar2007}, take its
Hall--Littlewood parameter to be $q$ and its boundary parameters
to be $a=-t\zeta$, $b=\zeta^{-1}$.  Then $ab=-t$,
$b/a=-1/(t\zeta^2)$, and
$(1+ax)(1+bx)/(1-x^2)=\Phi_{t,\zeta}(x)$.
The even part-size coefficient there is $H_m(ab;q)$, while the
odd part-size coefficient is $a^mH_m(b/a;q)$; these are exactly
\eqref{eq:fb-zero-spin}.  The polynomial expressions also identify
the case $t=0$ without dividing by $a$.

Warnaar's cited theorem is non-skew.  Using the known specialization
$f|_{s=0}=P$, $g|_{s=0}=Q$ in the normalization of
\cite[Section~3]{BufetovMucciconiPetrov1905}, our theorem gives
its corresponding one-variable skew form
\[
 \sum_\lambda W_\lambda\big|_{s=0}\,P_{\lambda/\mu}(x;q)
 =\Phi_{t,\zeta}(x)
 \sum_\kappa W_\kappa\big|_{s=0}\,Q_{\mu/\kappa}(x;q).
\]
The skew Cauchy identity and the extension to several variables
are discussed in Section~\ref{ssec:scls}.
The boundary parameters agree with those in
\cite[Section~2.3]{He2024Boundary} under $\nu=\zeta$, with
$q,t$ unchanged.  Thus at $s=0$ our boundary weights are exactly the
Rogers--Szeg\H{o} weights in He's half-space Hall--Littlewood measures.

\smallskip
\noindent\emph{The specialization $(t,\zeta)=(1,1)$.}
Here $U=V=0$, so both recurrences become
\[
 (1-s^2q^m)E_{m+1}=(1-q^m)E_{m-1},
 \qquad O_m=E_m.
\]
Their initial values give $E_{2r+1}=0$ and
\[
 E_{2r}=\frac{1-q^{2r-1}}{1-s^2q^{2r-1}}E_{2r-2}
 =\prod_{k=1}^r\frac{1-q^{2k-1}}{1-s^2q^{2k-1}}.
\]
Thus $W_\lambda(1,1)$ vanishes unless every multiplicity is even,
equivalently every part of $\lambda'$ is even, by
$m_j(\lambda)=\lambda'_j-\lambda'_{j+1}$ and
$\lambda'_j=\sum_{k\geq j}m_k(\lambda)$.
Define its nonzero coefficient by
\[
 b^{\rm el}_\lambda
 =\prod_{j\geq1}\prod_{k=1}^{m_j(\lambda)/2}
       \frac{1-q^{2k-1}}{1-s^2q^{2k-1}}
 \quad\text{when every $m_j(\lambda)$ is even}.
\]
Since $\Phi_{1,1}(x)=(1-x^2)/(1-x^2)=1$, the theorem specializes to
\[
 \sum_{\lambda:\,\lambda'\text{ has only even parts}}
 b^{\rm el}_\lambda f_{\lambda/\mu}(x)
 =\sum_{\kappa:\,\kappa'\text{ has only even parts}}
 b^{\rm el}_\kappa g_{\mu/\kappa}(x).
\]
This is the skew Littlewood identity with the restriction that
the conjugate partition has only even parts.  Each factor in
$b^{\rm el}_\lambda$ is positive.  Although $t=1$ is excluded
from Proposition~\ref{prop:fb-positivity}, the explicit formula
therefore gives nonnegative weights at this specialization as well.

\smallskip
\noindent\emph{The specialization $(t,\zeta)=(0,1)$.}
Now $U=V=1$.  The common recurrence is first order:
$(1-s^2q^m)E_{m+1}=(1+sq^m)E_m$, with $E_0=1$ and $O_m=E_m$.
Iterating gives
\[
 E_m=O_m=\prod_{r=0}^{m-1}\frac{1+sq^r}{1-s^2q^r}
 =\frac{(-s;q)_m}{(s^2;q)_m},\qquad
 \Phi_{0,1}(x)=\frac{1+x}{1-x^2}=\frac1{1-x}.
\]
We spell out the normalization in the comparison with
\cite[Theorem~1.1]{FischerGangl2026}.  Take their column spins
to be $s_0=0$, $s_j=s$ for $j\geq1$, and their spectral variables
to be $A=(a_1,\ldots,a_n)$.  Each one-variable row increases length
by at most one, so the chain definition gives
$f_\lambda(A)=g_\lambda(A)=0$ when $\ell(\lambda)>n$.
Their formal-series convention is
$a_i=(s+y_i)/(1+sy_i)$; equivalently
$y_i=(a_i-s)/(1-sa_i)$.  Write $F^{\rm ns}$ for their nonstable
functions.  For $\ell(\lambda)\leq n$, pad $\lambda$ by
$m_0=n-\ell(\lambda)$ zero parts.
The stable specialization in
\cite[Section~3.3, equation~(3.7)]{BufetovMucciconiPetrov1905} is
\[
 F^{\rm ns}_{\lambda\cup0^{m_0}}(A\mid0,s,s,\ldots)
 =(q;q)_{m_0}g_\lambda(A).
\]
Indeed, at column zero a continuing arrow has weight
$L_{a_i,0}(I,1;I,1)=a_i$, the stable boundary factor, while an
upward turn has weight $L_{a_i,0}(I,1;I+1,0)=1-q^{I+1}$.
The $m_0$ turns contribute $(q;q)_{m_0}$ independently of their
row positions, and the remaining weights are those of $g_\lambda$.
The zero-part coefficient in the cited Littlewood sum is
$(-s_0;q)_{m_0}/(q;q)_{m_0}=1/(q;q)_{m_0}$, cancelling this
factor.  For the positive parts, \eqref{eq:f-g-gauge} gives
\[
 \mathsf c(\lambda)W_\lambda(0,1)
 =\prod_{j\geq1}
 \frac{(s^2;q)_{m_j(\lambda)}}{(q;q)_{m_j(\lambda)}}
 \frac{(-s;q)_{m_j(\lambda)}}{(s^2;q)_{m_j(\lambda)}}
 =\prod_{j\geq1}\frac{(-s;q)_{m_j(\lambda)}}{(q;q)_{m_j(\lambda)}}.
\]
Since $f_\lambda=\mathsf c(\lambda)g_\lambda$, the resulting
summand is $W_\lambda(0,1)f_\lambda(A)$.  The non-skew
multivariable consequence proved in Section~\ref{ssec:scls} is
therefore the same identity:
\[
 \sum_\lambda W_\lambda(0,1)f_\lambda(A)
 =\prod_{i=1}^n\frac1{1-a_i}
   \prod_{1\leq i<j\leq n}\frac{1-qa_ia_j}{1-a_ia_j}.
\]
This comparison concerns the non-skew specialization of the present
identity.  For $0<a_i<1$, convergence follows from our identities
as explained in Section~\ref{ssec:scls}.

\subsection{The skew Cauchy--Littlewood structure}
\label{ssec:scls}

We extend the skew Cauchy structure of
\cite[Section~2.1]{BufetovMucciconiPetrov1905} by including a
skew Littlewood identity.  The resulting \emph{skew Cauchy--Littlewood
structure} consists of the following six properties of $f,g$ and
the boundary weights $W_\lambda(t,\zeta)$.
Throughout, assume $0<q<1$, $-1<s\leq0$, and spectral parameters
in $(0,1)$.  For the boundary parameters, assume
$0\leq t<1$, $\zeta>0$ and \eqref{eq:fb-positive-domain}, or
$(t,\zeta)=(1,1)$.

\begin{enumerate}[label=(\arabic*)]
\item \emph{Symmetry.}
The functions $f_{\lambda/\mu}$ and $g_{\lambda/\mu}$ are symmetric
rational functions of their spectral parameters.

\item \emph{Interlacing.}
For $0<x<1$,
\[
 f_{\lambda/\mu}(x)\neq0
 \quad\Longleftrightarrow\quad\mu\prec\lambda
 \quad\Longleftrightarrow\quad g_{\lambda/\mu}(x)\neq0.
\]
For $A=(a_1,\ldots,a_N)$, a multivariable function is nonzero
exactly when there is a chain
$\mu=\lambda^{(0)}\prec\cdots\prec\lambda^{(N)}=\lambda$.
In particular, $f_\lambda(A)=g_\lambda(A)=0$ if $\ell(\lambda)>N$.

\item \emph{Branching.}
For finite alphabets $A,B$, write $(A,B)$ for their concatenation.
For $h=f$ or $g$,
\[
 h_{\lambda/\mu}(A,B)
 =\sum_\nu h_{\lambda/\nu}(B)h_{\nu/\mu}(A),
 \qquad h_{\lambda/\mu}(\varnothing)=\mathbf1_{\lambda=\mu}.
\]
The sum is finite because $\mu\subseteq\nu\subseteq\lambda$.

\item \emph{Skew Cauchy identity.}
For any partitions $\lambda,\mu$,
\begin{equation}
 \Pi(x;y)\sum_\kappa f_{\mu/\kappa}(x)g_{\lambda/\kappa}(y)
 =\sum_\nu f_{\nu/\lambda}(x)g_{\nu/\mu}(y),
 \qquad \Pi(x;y)=\frac{1-qxy}{1-xy}.
 \label{eq:ssci}
\end{equation}
The left sum is finite and the right sum converges absolutely
in the stated parameter range.

\item \emph{Skew Littlewood identity.}
Theorem~\ref{thm:fb-skew-Littlewood} gives, for every partition $\mu$,
\[
 \sum_\lambda W_\lambda(t,\zeta)f_{\lambda/\mu}(x)
 =\Phi_{t,\zeta}(x)
   \sum_\kappa W_\kappa(t,\zeta)g_{\mu/\kappa}(x),
\]
where $\Phi_{t,\zeta}$ is defined in
\eqref{eq:fb-boundary-factor}.  The left sum converges absolutely
and the right sum is finite.

\item \emph{Nonnegativity.}
The functions $f_{\lambda/\mu}(A)$, $g_{\lambda/\mu}(A)$ and
the weights $W_\lambda(t,\zeta)$ are nonnegative.
\end{enumerate}

Properties (1)--(4) and (6), restricted to $f,g$, form the skew
Cauchy structure; see \cite[Sections~3.2--3.5]{BufetovMucciconiPetrov1905},
in particular Theorem~3.4 for \eqref{eq:ssci}.
Its convergence condition holds because
$0<(x-s)/(1-sx)<1$ and $0<(y-s)/(1-sy)<1$.
Property (5) supplies the additional boundary relation.
We also use the known stability $h_{\lambda/\mu}(A,0)=h_{\lambda/\mu}(A)$
for $h=f,g$, and the zero-spin specialization
$f_{\lambda/\mu}(A)|_{s=0}=P_{\lambda/\mu}(A;q)$,
$g_{\lambda/\mu}(A)|_{s=0}=Q_{\lambda/\mu}(A;q)$; the latter follows from
the row weights and branching, consistently with
\cite[Remark~5.1]{BorodinWheeler1701}.

The boundary function used in Section~\ref{ssec:contributions} is
\begin{equation}
 G^{t,\zeta}_\kappa(x)
 :=\sum_{\rho\prec\kappa} W_\rho(t,\zeta)g_{\kappa/\rho}(x)
 =\Phi_{t,\zeta}(x)^{-1}
   \sum_\eta W_\eta(t,\zeta)f_{\eta/\kappa}(x).
 \label{G}
\end{equation}
Its defining sum is finite and nonnegative.  The second expression
is property (5); division is valid because the same identity at
$\mu=\varnothing$ gives
$\Phi_{t,\zeta}(x)=\sum_\lambda W_\lambda f_\lambda(x)\geq1$.
Combining properties (4) and (5) yields the generalized skew
Littlewood identity
\begin{equation}
 \sum_\nu f_{\nu/\mu}(x)G^{t,\zeta}_\nu(y)
 =\Pi(x;y)\Phi_{t,\zeta}(x)
   \sum_\kappa G^{t,\zeta}_\kappa(x)g_{\mu/\kappa}(y).
 \label{eq:gene-sli}
\end{equation}
Indeed, suppressing the boundary parameters, we have
\[
\begin{aligned}
 \sum_\nu f_{\nu/\mu}(x)G_\nu(y)
 &=\sum_\rho W_\rho
     \sum_\nu f_{\nu/\mu}(x)g_{\nu/\rho}(y)\\
 &=\Pi(x;y)\sum_\kappa g_{\mu/\kappa}(y)
     \sum_\rho W_\rho f_{\rho/\kappa}(x)\\
 &=\Pi(x;y)\Phi(x)
     \sum_\kappa G_\kappa(x)g_{\mu/\kappa}(y).
\end{aligned}
\]
Tonelli's theorem permits the rearrangements; the final finite sum
over $\kappa\prec\mu$ proves convergence.
Identities \eqref{eq:ssci} and \eqref{eq:gene-sli} will be used for
the bulk and boundary transitions, respectively.

For $A=(a_1,\ldots,a_N)$, recall the normalizing constant
\begin{equation}
 Z_{t,\zeta}(A)
 =\prod_{i=1}^N\Phi_{t,\zeta}(a_i)
  \prod_{1\leq i<j\leq N}\Pi(a_i;a_j).
 \label{eq:fb-normalizer}
\end{equation}
Iterating property (5) using branching and the Cauchy identity gives
\begin{equation}
 \sum_\lambda W_\lambda(t,\zeta)f_{\lambda/\mu}(A)
 =Z_{t,\zeta}(A)
   \sum_\kappa W_\kappa(t,\zeta)g_{\mu/\kappa}(A).
 \label{eq:fb-multivariable-identity}
\end{equation}
For completeness, write $S_\mu(A)=\sum_\lambda W_\lambda f_{\lambda/\mu}(A)$
and $\Pi(A;x)=\prod_i\Pi(a_i;x)$, suppressing $t,\zeta$.
Branching, property (5), and the multivariable Cauchy identity
\cite[equation~(2.4)]{BufetovMucciconiPetrov1905} give
\[
 S_\mu(A,x)
 =\Phi(x)\Pi(A;x)\sum_\eta g_{\mu/\eta}(x)S_\eta(A).
\]
Starting with $S_\mu(\varnothing)=W_\mu$, induction and branching
for $g$ prove \eqref{eq:fb-multivariable-identity}, since
$Z(A,x)=Z(A)\Phi(x)\Pi(A;x)$ and $Z(\varnothing)=1$.
All rearrangements use nonnegativity; finiteness follows from the
final sum over $\kappa\subseteq\mu$.

Taking $\mu=\varnothing$ leaves only $\kappa=\varnothing$ on the
right, so $\sum_\lambda W_\lambda f_\lambda(A)=Z_{t,\zeta}(A)$.
Consequently,
\begin{equation}
 \mathbb P^A_{s;t,\zeta}(\lambda)
 =\frac{W_\lambda(t,\zeta)f_\lambda(A)}{Z_{t,\zeta}(A)}
 \label{eq:fb-multivariable-measure}
\end{equation}
is a probability measure.  By branching, it is the distribution of
$\lambda^{(N)}$ in the ascending process of
Section~\ref{ssec:contributions}.
At $(t,\zeta)=(0,1)$,
$\Phi_{0,1}(x)=(1-x)^{-1}$, giving the product formula
at the end of Section~\ref{sec:boundary-family}; at
$(t,\zeta)=(1,1)$, $\Phi_{1,1}(x)=1$.
The normalizing constant is independent of $s$; the corresponding
statement for joint lengths requires the separate argument in
Section~\ref{ssec:joint-lengths}.
\section{Half-space random field for stable spin Hall--Littlewood symmetric functions}
\label{sec:HSRF}

Throughout this section, $q\in(0,1)$, $s\in(-1,0]$, and $x_r\in(0,1)$
for $r\ge0$. We allow either $0\le t<1$, $\zeta>0$ satisfying
\eqref{eq:fb-positive-domain}, or $(t,\zeta)=(1,1)$. In the latter
case the weight $W_\lambda(t,\zeta)$ vanishes unless every part of
$\lambda'$ is even. All conditional probabilities below are defined
for conditioning events of positive probability.

We construct the field of Theorem~\ref{thm:intro-field} using
stochastic Yang--Baxter moves and the boundary weights $W_\lambda$.
The construction applies throughout the range above; autonomous
projections require the regimes of Section~\ref{ssec:boundary-to}.
Section~\ref{ssec:fb-local-projections} derives their scalar
identifications and two-layer continuous-time limit.

\subsection{The half-space random field}
\label{ssec:hsybrf}

Let $\mathbb H=\{(i,j)\in\mathbb Z^2:0\le i\le j\}$ and fix integers
$0\le n\le N$. An up-left path
$\omega=(\omega_0,\ldots,\omega_N)$ from $(n,n)$ to $(0,N)$
has steps $-\mathbf e_1$ or $\mathbf e_2$: there are $n$ left
steps and $N-n$ up steps. Each vertex carries a partition.
A horizontal edge between columns $a$ and $a+1$ has parameter
$x_a$, and a vertical edge between rows $b-1$ and $b$ has parameter
$x_b$. A left step contributes $f$, with the departure partition
as its outer index; an up step contributes $g$, with the arrival
partition as its outer index.

The initial diagonal segment in Figure~\ref{fig:path} represents
$G^{t,\zeta}_{\lambda^{(n,n)}}(x_n)$, defined in \eqref{G}.
Its lower endpoint carries no partition in the path measure;
in particular, the partition summed inside $G$ is not identified
with $\lambda^{(n-1,n-1)}$. Thus the diagonal factor is $G_\lambda$,
not $W_\lambda$. For $n=0$ the segment is formal and contributes
$G^{t,\zeta}_\varnothing(x_0)=1$.

\begin{figure}
\begin{center}
\begin{tikzpicture}[scale=.67]
\draw[dotted,->] (0,0) -- (7.8,7.8);
\draw[dotted,->] (0,0) -- (0,7.8);
\foreach \x in {0,1, ..., 7} 
{
\draw[dotted] (0,\x) -- (\x,\x) -- (\x,7.8 );
}
\begin{scope}
\draw[red, line width=2] (3,3) -- (4,4);
\draw[red, line width=2] (4,4) -- (4,5);
\draw[red, line width=2] (4,5) -- (2,5);
\draw[red, line width=2] (2,5) -- (2,6);
\draw[red, line width=2] (2,6) -- (1,6);
\draw[red, line width=2] (1,6) -- (1,7);
\draw[red, line width=2] (1,7) -- (0,7);
\fill (0,7) circle(0.1);
\fill (1,7) circle(0.1);
\fill (1,6) circle(0.1);
\fill (2,6) circle(0.1);
\fill (2,5) circle(0.1);
\fill (3,5) circle(0.1);
\fill (4,5) circle(0.1);
\fill (4,4) circle(0.1);
\draw (-0.3, 0.5) node{$x_1$};
\draw (-0.3, 1.5) node{$x_2$};
\draw (-0.3, 2.5) node{$x_3$};
\draw (-0.3, 3.5) node{$x_4$};
\draw (-0.3, 4.5) node{$x_5$};
\draw (-0.3, 5.5) node{$x_6$};
\draw (-0.3, 6.5) node{$x_7$};
\draw (0.5, 7.8) node{$x_0$};
\draw (1.5, 7.8) node{$x_1$};
\draw (2.5, 7.8) node{$x_2$};
\draw (3.5, 7.8) node{$x_3$};
\draw (4.5, 7.8) node{$x_4$};
\draw (5.5, 7.8) node{$x_5$};
\draw (6.5, 7.8) node{$x_6$};
\draw (0.2, 7.2) node{$\omega_7$};
\draw (1.2, 7.2) node{$\omega_6$};
\draw (0.8, 5.8) node{$\omega_5$};
\draw (2.2, 6.2) node{$\omega_4$};
\draw (1.8, 4.8) node{$\omega_3$};
\draw (3.2, 5.2) node{$\omega_2$};
\draw (4.2, 5.2) node{$\omega_1$};
\draw (3.8, 4.2) node{$\omega_0$};
\draw (0.5, 6.8) node{$f$};
\draw (1.2, 6.5) node{$g$};
\draw (1.5, 5.8) node{$f$};
\draw (2.2, 5.5) node{$g$};
\draw (2.5, 4.8) node{$f$};
\draw (3.5, 4.8) node{$f$};
\draw (4.2, 4.5) node{$g$};
\draw (3.4, 3.6) node{$G^{t,\zeta}$};
\draw (0.8, 0.3) node{$x_1$};
\draw (1.8, 1.3) node{$x_2$};
\draw (2.8, 2.3) node{$x_3$};
\draw (3.8, 3.3) node{$x_4$};
\draw (4.8, 4.3) node{$x_5$};
\draw (5.8, 5.3) node{$x_6$};
\draw (6.8, 6.3) node{$x_7$};
\end{scope}
\end{tikzpicture}
\end{center}
\caption{An up-left path from $(4,4)$ to $(0,7)$, with an initial
diagonal segment. A partition is assigned to each black vertex.
The symbols on the red edges specify the factors in
\eqref{eq:process}: $f$ for a leftward step, $g$ for an upward
step, and $G^{t,\zeta}$ at the initial diagonal vertex.
The labels along the three sides of the half-quadrant indicate the
spectral parameters of the corresponding edges.}
\label{fig:path}
\end{figure}

\begin{defn}
\label{defn:hsybrf}
A \emph{half-space stable spin Hall--Littlewood random field} is a
family of random partitions
$\boldsymbol\lambda=(\lambda^{(i,j)}:(i,j)\in\mathbb H)$ such that,
almost surely, $\lambda^{(0,j)}=\varnothing$ and
$\lambda^{(i,j)}\prec\lambda^{(i,j+1)}$,
$\lambda^{(i,j)}\prec\lambda^{(i+1,j)}$ whenever the vertices belong
to $\mathbb H$. For every path $\omega$ above, its partitions have
joint probability mass function
\begin{equation}
\label{eq:process}
 \frac{G^{t,\zeta}_{\lambda^{\omega_0}}(x_n)}{Z_\omega}
 \prod_{\omega_{k+1}=\omega_k+\mathbf e_2}
 g_{\lambda^{\omega_{k+1}}/\lambda^{\omega_k}}(x_{j_{k+1}})
 \prod_{\omega_{k+1}=\omega_k-\mathbf e_1}
 f_{\lambda^{\omega_k}/\lambda^{\omega_{k+1}}}(x_{i_k-1}),
\end{equation}
where $\omega_k=(i_k,j_k)$ and $\lambda^{\omega_N}=\varnothing$.
Here $Z_\omega$ is the sum of the numerator over all interlacing
path partitions; its finiteness and positivity are proved below.
We call \eqref{eq:process} the half-space stable spin
Hall--Littlewood process along $\omega$.
\end{defn}

For $n\ge1$, let $h_a$ be the height of the unique left step from
column $a+1$ to column $a$, for $0\le a<n$. Then
$N\ge h_0\ge\cdots\ge h_{n-1}\ge n$, and
\begin{equation}
\label{eq:path-normalizer-product}
 Z_\omega=
 \prod_{a=0}^{n-1}\Phi_{t,\zeta}(x_a)
 \prod_{a=0}^{n-1}\prod_{b=a+1}^{h_a}\Pi(x_a;x_b).
\end{equation}
Indeed, use the skew Cauchy identity \eqref{eq:ssci} to lower the
first left step from height $h_{n-1}$ to $n$, one square at a time
as in Figure~\ref{fig:contraction}(a). At height $b$, the partition
function before this change equals $\Pi(x_{n-1};x_b)$ times the
partition function afterwards. The boundary change in
Figure~\ref{fig:contraction}(b), using \eqref{eq:gene-sli}, then
moves the starting vertex to $(n-1,n-1)$ and contributes
$\Phi_{t,\zeta}(x_{n-1})\Pi(x_{n-1};x_n)$. The total factor at
this stage is therefore
$\Phi_{t,\zeta}(x_{n-1})\prod_{b=n}^{h_{n-1}}\Pi(x_{n-1};x_b)$.
The remaining crossing heights are unchanged. Repeating reduces
the path to the vertical axis, whose weight is one, and proves
\eqref{eq:path-normalizer-product}; for $n=0$ both products are empty.
Nonnegativity permits these summations by Tonelli's theorem.
The finite product proves convergence, and the all-empty
configuration gives $Z_\omega\ge1$.

For two paths differing by one change in Figure~\ref{fig:contraction},
fix the common partitions and sum over the changed corner.
The relevant identity multiplies both this summed weight and
$Z_\omega$ by the same scalar. After normalization the scalar
cancels, so the joint distributions on the common vertices agree.

\begin{figure}
	\begin{center}
		\begin{tikzpicture}[scale=.78]
		\begin{scope}
		\draw (-3,-.5) node{(a)};
		\draw[red, line width=2] (-1,0) -- (0,0) node[midway, anchor=south]{$x_i$};
		\draw[red, line width=2] (0,0) -- (0,-1) node[midway, anchor=west]{$x_j$};
		\fill (0,0) circle(0.1);
        \node[above right] at (.1,.1) {$\nu$};
        \fill (-1,0) circle(.06);
        \node[above] at (-1,.15) {$\mu$};
		\draw[ultra thick, gray, dotted] (-1.5,0) -- (-1,0);
		\draw[ultra thick, gray, dotted] (0,-1) -- (.5,-1);
        \fill (0,-1) circle(.06);
        \node[below] at (0,-1.1) {$\lambda$};
		\draw[thick, gray] (-1.5,0) ellipse (.5 and .3);
		\draw[thick, gray] (.5,-1) ellipse (.5 and .3);
		\draw (-.2,-.2)  -- ++(0.4,0) -- ++(0,0.4) -- ++(-0.4,0) -- cycle; 
		\draw (3,-.5) node{$=$}; 
		\draw[red, line width=2] (5,-1) -- (6,-1) node[midway, anchor=north]{$x_i$};
		\draw[red, line width=2] (5,0) -- (5,-1) node[midway, anchor=east]{$x_j$};
		\fill (5,-1) circle(0.1);
        \node[below left] at (4.9,-1.1) {$\kappa$};
        \fill (5,0) circle(.06);
        \node[above] at (5,.15) {$\mu$};
        \fill (6,-1) circle(.06);
        \node[below] at (6,-1.1) {$\lambda$};
		\draw[ultra thick, gray, dotted] (4.5,0) -- (5,0);
		\draw[ultra thick, gray, dotted] (6,-1) -- (6.5,-1);
		\draw[thick, gray] (4.5,0) ellipse (.5 and .3);
		\draw[thick, gray] (6.5,-1) ellipse (.5 and .3);
		\draw (4.8,-1.2)  -- ++(0.4,0) -- ++(0,0.4) -- ++(-0.4,0) -- cycle; 
		\draw (8.5,-.5) node{$\times\ \Pi(x_i;x_j)$};
		
		\begin{scope}[yshift=-3cm]
		\draw (-3,-.5) node{(b)};
		\draw[red, line width=2] (-1,-1) -- (0,0) node[midway, anchor=north west]{$x_{i+1}$};
		\draw[red, line width=2] (-1,0) -- (0,0) node[midway, anchor=south]{$x_i$};
		\fill (0,0) circle(0.1);
        \node[above right] at (.1,.1) {$\nu$};
        \fill (-1,0) circle(.06);
        \node[above] at (-1,.15) {$\mu$};
		\draw[ultra thick, gray, dotted] (-1.5,0) -- (-1,0);
		\draw[thick, gray] (-1.5,0) ellipse (.5 and .3);
		\draw (-.2,-.2)  -- ++(0.4,0) -- ++(0,0.4) -- ++(-0.4,0) -- cycle; 
		\draw (3,-.5) node{$=$}; 
		\draw[red, line width=2] (6,-1) -- (6,0) node[midway, anchor=east]{$x_{i+1}$};
		\draw[red, line width=2] (6,-1) -- (5,-2) node[midway, anchor=north west]{$x_i$};
		\fill (6,-1) circle(0.1);
        \node[right] at (6.25,-1) {$\kappa$};
        \fill (6,0) circle(.06);
        \node[above] at (6,.15) {$\mu$};
		\draw[ultra thick, gray, dotted] (5.5,0) -- (6,0);
		\draw[thick, gray] (5.5,0) ellipse (.5 and .3);
		\draw (5.8,-1.2)  -- ++(0.4,0) -- ++(0,0.4) -- ++(-0.4,0) -- cycle; 
     \draw (8.6,-.5) node{$\times\ \Pi(x_i;x_{i+1})\Phi_{t,\zeta}(x_i)$};
		\end{scope}
		\end{scope}
		\end{tikzpicture}
	\end{center}
\caption{Local changes of the path: (a) the skew Cauchy identity
\eqref{eq:ssci}; (b) the generalized skew Littlewood identity
\eqref{eq:gene-sli}. The boxed partitions $\nu$ and $\kappa$ are
summed over; the endpoint partitions $\lambda,\mu$ in (a), and
$\mu$ in (b), are fixed. Gray ellipses represent unchanged factors.
Each diagonal segment contributes $G^{t,\zeta}$; the displayed
scalar factors give the change in the partition function.}
\label{fig:contraction}
\end{figure}

At $s=0$ and $(t,\zeta)=(1,1)$, we have $f=P$, $g=Q$;
after reflecting the coordinates and reversing the path orientation,
\eqref{eq:process} is the Hall--Littlewood specialization of the
half-space Macdonald process in
\cite[Section~2.3]{BarraquandBorodinCorwin1802}.
The preceding identities normalize the path laws and establish
their compatibility under local changes. They do not specify
joint distributions at vertices that cannot share a path, such
as $(1,1)$ and $(2,2)$. Theorem~\ref{thm:half-space-field-sampling}
constructs a field by choosing compatible Markov transition
operators. Different choices can give different fields with the
same path laws; the projected models in Section~\ref{ssec:contributions}
use the particular kernels of Sections~\ref{ssec:bulk-to}
and~\ref{ssec:boundary-to}.

\subsection{Sampling by Markov transition operators}
\label{ssec:shsrf}

We now formulate the local balance relations that turn the path
identities into a sampling procedure. The construction in this
subsection applies to any kernels satisfying these relations.

At a bulk square, let $\varkappa,\lambda,\mu,\nu$ denote the
southwest, southeast, northwest, and northeast partitions,
respectively; see Figure~\ref{operators}. The forward kernel
$\Ufwd_{x,y}(\varkappa\to\nu\mid\lambda,\mu)$ samples $\nu$
from the other three partitions, while the backward kernel
$\Ubwd_{x,y}(\nu\to\varkappa\mid\lambda,\mu)$ samples
$\varkappa$ from $\nu,\lambda,\mu$. Both are nonnegative and
vanish unless $\varkappa\prec\lambda$, $\varkappa\prec\mu$,
$\lambda\prec\nu$, and $\mu\prec\nu$.
At the diagonal, $\varkappa$ is the old diagonal partition,
$\mu$ lies directly above it, and $\nu$ is the new diagonal
partition. The nonnegative kernels
$U^{\partial,+}_{x,y}(\varkappa\to\nu\mid\mu)$ and
$U^{\partial,-}_{x,y}(\nu\to\varkappa\mid\mu)$ are supported on
$\varkappa\prec\mu\prec\nu$. We require
\begin{align}
\label{bulk-to}
 &\Pi(x;y) f_{\lambda/\varkappa}(x)g_{\mu/\varkappa}(y)
 \Ufwd_{x,y}(\varkappa\to\nu\mid\lambda,\mu)
 =f_{\nu/\mu}(x)g_{\nu/\lambda}(y)
 \Ubwd_{x,y}(\nu\to\varkappa\mid\lambda,\mu),\\
\label{boundary-to}
 &\Pi(x;y)\Phi_{t,\zeta}(x)G^{t,\zeta}_\varkappa(x)
 g_{\mu/\varkappa}(y)
 U^{\partial,+}_{x,y}(\varkappa\to\nu\mid\mu)
 =f_{\nu/\mu}(x)G^{t,\zeta}_\nu(y)
 U^{\partial,-}_{x,y}(\nu\to\varkappa\mid\mu),
\end{align}
and the stochasticity conditions
\begin{align}
\label{normal-con1}
 \sum_\nu\Ufwd_{x,y}(\varkappa\to\nu\mid\lambda,\mu)&=1,
 &\sum_\varkappa\Ubwd_{x,y}(\nu\to\varkappa\mid\lambda,\mu)&=1,\\
\label{normal-con2}
 \sum_\nu U^{\partial,+}_{x,y}(\varkappa\to\nu\mid\mu)&=1,
 &\sum_\varkappa U^{\partial,-}_{x,y}(\nu\to\varkappa\mid\mu)&=1.
\end{align}
These row sums are imposed on interlacing inputs for which the
factor multiplying the corresponding kernel in \eqref{bulk-to}
or \eqref{boundary-to} is positive. On admissible zero-weight inputs,
any stochastic extension preserving the support conditions and
these identities may be used.

For fixed $\lambda,\mu$ with positive total weight, normalize
$f_{\lambda/\varkappa}(x)g_{\mu/\varkappa}(y)$ and
$f_{\nu/\mu}(x)g_{\nu/\lambda}(y)$ to probability distributions
$p_\varkappa$ and $r_\nu$. The skew Cauchy identity shows that their
normalizers differ by $\Pi(x;y)$, so \eqref{bulk-to} is precisely
$p_\varkappa\Ufwd(\varkappa\to\nu)=
r_\nu\Ubwd(\nu\to\varkappa)$.
Any coupling of $p$ and $r$, including their product measure,
therefore defines compatible forward and backward conditional
kernels. The same argument applies to \eqref{boundary-to}, using
\eqref{eq:gene-sli} and the weights
$G^{t,\zeta}_\varkappa(x)g_{\mu/\varkappa}(y)$ and
$f_{\nu/\mu}(x)G^{t,\zeta}_\nu(y)$.

\begin{figure}
\begin{center}

\begin{tikzpicture}[scale=.78]
\usetikzlibrary{shapes}
\usetikzlibrary{snakes}
\begin{scope}
%
\node (v1) at (0,1) {$\mu$};
\node (m) at (0,0) {$\varkappa$};
\node (k1) at (1,0) {$\lambda$};
\node[gray, circle, draw] (p1) at (1,1) {$?$};
\draw[red, line width=2] (v1) -- (m);
\node at (-0.2,0.5) {$y$};
\draw[red, line width=2] (m) -- (k1);
\node at (0.5,-0.2) {$x$};
\draw[dotted, gray] (v1) -- (p1);
\draw[dotted, gray] (p1) -- (k1);
\draw[thick, ->, >=stealth'] (3,0.75) --(5,0.75) node[midway, anchor=south] {$\Ufwd_{x,y}(\varkappa \rightarrow \nu|\lambda,\mu)$};
\draw[thick, <-, >=stealth'] (3,0.25) --(5,0.25) node[midway, anchor=north] {$\Ubwd_{x,y}(\nu \rightarrow \varkappa|\lambda,\mu)$};
\node (v2) at (7,1) {$\mu$};
\node (p) at (8,1) {$\nu$};
\node (k2) at (8,0) {$\lambda$};
\node[gray, circle, draw] (m2) at (7,0) {$?$};
\draw[red, line width=2] (v2) -- (p);
\node at (7.5,1.2) {$x$};
\draw[red, line width=2] (p) -- (k2);
\node at (8.2,0.5) {$y$};
\draw[dotted, gray] (v2) -- (m2);
\draw[dotted, gray] (m2) -- (k2);
\begin{scope}[yshift=-2.8cm]
%
\node (m) at (0,0) {$\varkappa$};
\node (bas) at (-1,-1){};
\node (k1) at (0,1) {$\mu$};
\node[gray, circle, draw] (p) at (1,1) {$?$};
\draw[red, line width=2] (bas) -- (m);
\node at (-0.35,-0.65) {$x$};
\draw[red, line width=2] (m) -- (k1);
\node at (-0.2,0.5) {$y$};
\draw[dotted, gray] (m) -- (p);
\draw[dotted, gray] (p) -- (k1);
\draw[thick, ->, >=stealth'] (3,0.2) --(5,0.2) node[midway, anchor=south] {$U^{\partial,+}_{x,y}(\varkappa \rightarrow \nu|\mu)$};
\draw[thick, <-, >=stealth'] (3,-0.3) --(5,-0.3) node[midway, anchor=north] {$U^{\partial,-}_{x,y}(\nu \rightarrow \varkappa|\mu)$};
\node (p) at (8,1) {$\nu$};
\node (bas) at (6,-1){};
\node (k1) at (7,1) {$\mu$};
\node[gray, circle, draw] (m2) at (7,0) {$?$};
\draw[red, line width=2] (m2) -- (p);
\node at (7.65,0.35) {$y$};
\draw[red, line width=2] (p) -- (k1);
\node at (7.5,1.2) {$x$};
\draw[dotted, gray] (m2) -- (k1);
\draw[dotted, gray] (bas) -- (m2);
\end{scope}
\end{scope} 
\end{tikzpicture}
\end{center}
\caption{The partition configurations used by the transition
operators. Top: with $\lambda$ and $\mu$ fixed, the forward bulk
operator samples the northeast partition $\nu$ from the southwest
partition $\varkappa$; the backward operator samples $\varkappa$
from $\nu$. Bottom: the analogous update along the diagonal, with
$\mu$ fixed. Red edges show the path before or after the update,
and a gray circled question mark marks the partition that is sampled.
The edge labels specify the arguments of the functions in
\eqref{bulk-to} and \eqref{boundary-to}.}
\label{operators}
\end{figure}

\begin{thm}[Sampling the field]
\label{thm:half-space-field-sampling}
Fix kernels satisfying \eqref{bulk-to}--\eqref{normal-con2} with
the support conventions above. Starting with
$\lambda^{(0,j)}=\varnothing$, sample vertices in increasing order
of $i+j$, using independent auxiliary randomness and
\begin{align*}
 \lambda^{(i,j)}&\sim
 \Ufwd_{x_{i-1},x_j}\bigl(\lambda^{(i-1,j-1)}\to\cdot\mid
               \lambda^{(i,j-1)},\lambda^{(i-1,j)}\bigr),
 &&1\le i<j,\\
 \lambda^{(i,i)}&\sim
 U^{\partial,+}_{x_{i-1},x_i}\bigl(\lambda^{(i-1,i-1)}\to\cdot\mid
               \lambda^{(i-1,i)}\bigr),&&i\ge1.
\end{align*}
The resulting field satisfies Definition~\ref{defn:hsybrf}.
\end{thm}
\begin{proof}
Assign independent uniform random variables to the non-axis vertices
and use them to implement the kernels. Every input precedes its
output in the dependency order, so the resulting partitions are
independent of the chosen order respecting these dependencies.

To obtain a fixed path $\omega$, add squares to the empty vertical
path in such an order. At a bulk update, \eqref{bulk-to} and backward
stochasticity give
\[
 \sum_\varkappa f_{\lambda/\varkappa}(x)g_{\mu/\varkappa}(y)
 \Ufwd_{x,y}(\varkappa\to\nu\mid\lambda,\mu)
 =\frac{f_{\nu/\mu}(x)g_{\nu/\lambda}(y)}{\Pi(x;y)}.
\]
All other path factors are unchanged, and the new normalizer is
$\Pi(x;y)Z_{\rm old}$. The boundary calculation is identical, with
factor $\Pi(x;y)\Phi_{t,\zeta}(x)$, by \eqref{boundary-to}.
Induction therefore establishes \eqref{eq:process} on each exposed
path. This uses its joint path law, without conditioning on the
partitions left behind.

The finite triangles $T_M=\{(i,j):0\le i\le j\le M\}$ contain every
predecessor of their vertices. Using the same assigned random
variables makes their constructions agree on overlaps; their union
is the required infinite field. The kernel supports give the
interlacings, and every prescribed path lies in some $T_M$.
\end{proof}

Thus, once the bulk and boundary kernels are specified below,
Theorem~\ref{thm:half-space-field-sampling} gives the existence
assertion of Theorem~\ref{thm:intro-field}. The additional autonomy
and consistency assertions are proved in
Theorem~\ref{thm:fb-local-projection}.

\subsection{Construction of the bulk transition operators}
\label{ssec:bulk-to}

We now specify the bulk kernels used in
Theorem~\ref{thm:intro-field} and all the projected models of
Section~\ref{ssec:contributions}. We use the stochastic
bijectivization of the Yang--Baxter equation from
\cite[Section~2]{BufetovPetrov1712} and
\cite[Sections~6.1--6.3]{BufetovMucciconiPetrov1905}:
coupling the normalized weights on its two sides gives compatible
forward and backward transitions. The construction below fixes
one such coupling and verifies the bulk balance relation
\eqref{bulk-to}.

Fix admissible input partitions $\varkappa,\lambda,\mu$ as in
Figure~\ref{operators}. For $h\ge1$, let
$I_h=m_h(\lambda)$, $J_h=m_h(\mu)$, and
$K_h=m_h(\varkappa)$, where
$m_h(\lambda)=\lambda'_h-\lambda'_{h+1}$ counts parts equal to $h$.
Here $h$ indexes columns of the vertex model defining $f,g$,
not vertices of the random field. For $h\ge0$, the horizontal
occupations are
\begin{align}
\label{eq:bulk-carrier-dictionary}
 k_h&=\mu'_{h+1}-\varkappa'_{h+1},&
 \ell_h&=\lambda'_{h+1}-\varkappa'_{h+1},&
 i_h&=\nu'_{h+1}-\lambda'_{h+1},&
 j_h&=\nu'_{h+1}-\mu'_{h+1}.
\end{align}
Interlacing makes these occupations binary. The first two are fixed
by the inputs; sampling the last two determines $\nu'_{h+1}$.

The dual crossing weight $R^*_{xy}(i,j;k,\ell)$ has arguments
ordered as lower left, upper left, upper right, lower right; see
Figure~\ref{fig:R-vert*}. Its occupation rule is
$i-j=k-\ell$. The arrows indicate line orientations, with this
signed convention for the occupations. The weights satisfy
$\sum_{k,\ell}R^*_{xy}(i,j;k,\ell)=\Pi(x;y)$ and require
normalization to define transition probabilities.

\begin{figure}[htbp]
\centering
\begin{tabular}{|c||c|c|c|c|c|c|}
\hline
\quad
\tikz{0.5}{
	\draw[lred,line width=1.5pt,->] (-1,-1) -- (1,1);
	\draw[lred,line width=1.5pt,->] (-1,1) -- (1,-1);
	\node at (-1.1,1.1) {\tiny $j$};\node at (1.1,-1.1) {\tiny $\ell$};
	\node at (-1.1,-1.1) {\tiny $i$};\node at (1.1,1.1) {\tiny $k$};
}
\quad
&
\tikz{0.5}{
	\draw[lred,line width=1.5pt,->] (-1,-1) -- (1,1);
	\draw[lred,line width=1.5pt,->] (-1,1) -- (1,-1);
	\node at (-1.1,1.1) {\tiny $0$};\node at (1.1,-1.1) {\tiny $0$};
	\node at (-1.1,-1.1) {\tiny $0$};\node at (1.1,1.1) {\tiny $0$};
}
&
\tikz{0.5}{
	\draw[lred,line width=1.5pt,->] (-1,-1) -- (1,1);
	\draw[lred,line width=1.5pt,->] (-1,1) -- (1,-1);
	\node at (-1.1,1.1) {\tiny $0$};\node at (1.1,-1.1) {\tiny $0$};
	\node at (-1.1,-1.1) {\tiny $1$};\node at (1.1,1.1) {\tiny $1$};
}
&
\tikz{0.5}{
	\draw[lred,line width=1.5pt,->] (-1,-1) -- (1,1);
	\draw[lred,line width=1.5pt,->] (-1,1) -- (1,-1);
	\node at (-1.1,1.1) {\tiny $1$};\node at (1.1,-1.1) {\tiny $0$};
	\node at (-1.1,-1.1) {\tiny $1$};\node at (1.1,1.1) {\tiny $0$};
}
& 
\tikz{0.5}{
	\draw[lred,line width=1.5pt,->] (-1,-1) -- (1,1);
	\draw[lred,line width=1.5pt,->] (-1,1) -- (1,-1);
	\node at (-1.1,1.1) {\tiny $0$};\node at (1.1,-1.1) {\tiny $1$};
	\node at (-1.1,-1.1) {\tiny $0$};\node at (1.1,1.1) {\tiny $1$};
}
&
\tikz{0.5}{
	\draw[lred,line width=1.5pt,->] (-1,-1) -- (1,1);
	\draw[lred,line width=1.5pt,->] (-1,1) -- (1,-1);
	\node at (-1.1,1.1) {\tiny $1$};\node at (1.1,-1.1) {\tiny $1$};
	\node at (-1.1,-1.1) {\tiny $0$};\node at (1.1,1.1) {\tiny $0$};
}
&
\tikz{0.5}{
	\draw[lred,line width=1.5pt,->] (-1,-1) -- (1,1);
	\draw[lred,line width=1.5pt,->] (-1,1) -- (1,-1);
	\node at (-1.1,1.1) {\tiny $1$};\node at (1.1,-1.1) {\tiny $1$};
	\node at (-1.1,-1.1) {\tiny $1$};\node at (1.1,1.1) {\tiny $1$};
}
\\\hline
\quad
$R^*(i,j;k,\ell)$
\quad
& 
$1$
& 
$\dfrac{1-qxy}{1-xy}$
& 
$\dfrac{1-q}{1-xy}$
& 
$\dfrac{(1-q)xy}{1-xy}$
& 
$\dfrac{1-qxy}{1-xy}$
&
$q$
\\\hline
\end{tabular}
\caption{The six nonzero weights of the dual crossing $R^*_{xy}$.
The first column specifies the order of the binary occupations;
for fixed $(i,j)$, the sum over $(k,\ell)$ is $\Pi(x;y)$.}
\label{fig:R-vert*}
\end{figure}

At column $h\ge1$, fix $I=I_h$, $J=J_h$, $k=k_h$, $\ell=\ell_h$
and the previously sampled $p=i_{h-1}$, $v=j_{h-1}$.
The two configurations in Figure~\ref{fig:bulk-local-YB} have weights
\begin{align}
 A(k_-,\ell_-,K)
 &=R^*_{xy}(p,v;k_-,\ell_-)
   L_{y,s}(K,k_-;J,k)M^*_{x,s}(I,\ell_-;K,\ell),
 \label{eq:bulk-local-A}\\
 B(i,j,M)
 &=L_{y,s}(I,p;M,i)M^*_{x,s}(M,v;J,j)
   R^*_{xy}(i,j;k,\ell).
 \label{eq:bulk-local-B}
\end{align}
The dual Yang--Baxter equation
\cite[Proposition~A.2]{BufetovMucciconiPetrov1905} reads
\begin{equation}
\label{eq:bulk-local-YB}
 Z_h:=\sum_{k_-,\ell_-,K} A(k_-,\ell_-,K)
     =\sum_{i,j,M}B(i,j,M).
\end{equation}
The horizontal occupations are binary and $K,M\ge0$.
Conservation gives $K=J+k-k_-=I+\ell-\ell_-$ and
$M=I+p-i=J+v-j$, so both sums are finite.
In the cited identity, substitute $y$ for its spectral parameter $u$
and $x$ for its spectral parameter $v$, matching our $L$ and $M^*$ rows.

\begin{figure}[htbp]
\centering
$\tikz{0.6}{
\draw[lred,line width=1.5pt,->] (-3,2) -- (-1,0);
\draw[lred,line width=1.5pt,->] (-3,0) -- (-1,2);
\draw[lred,line width=1.5pt,->] (-1,0) -- (1,0);
\draw[lred,line width=4pt,->] (0,1) -- (0,-1);
\draw[lgray,line width=1.5pt,->] (-1,2) -- (1,2);
\draw[lgray,line width=4pt,->] (0,1) -- (0,3);
\draw[black,dotted,->] (0,2) -- (1,3);
\draw[black,dotted,->] (0,0) -- (1,-1);
\node[above] at (-1,2) {\tiny $k_{h-1}$};
\node[above] at (0,3) {\tiny $J_h$};
\node at (0,1) {\tiny $K_h$};
\node[below] at (-1,0) {\tiny $\ell_{h-1}$};
\node[below] at (0,-1) {\tiny $I_h$};
\node[left] at (-3,0) {\tiny $i_{h-1}$};
\node[left] at (-3,2) {\tiny $j_{h-1}$};
\node[right] at (1,-1) {\tiny $x$};
\node[right] at (1,0) {\tiny $\ell_{h}$};
\node[right] at (1,2) {\tiny $k_{h}$};
\node[right] at (1,3) {\tiny $y$};
}\qquad\longleftrightarrow\qquad\tikz{0.6}{
\draw[lred,line width=1.5pt,->] (1,2) -- (3,0);
\draw[lred,line width=1.5pt,->] (1,0) -- (3,2);
\draw[lgray,line width=1.5pt,->] (-1,0) -- (1,0);
\draw[lgray,line width=4pt,->] (0,-1) -- (0,1);
\draw[lred,line width=1.5pt,->] (-1,2) -- (1,2);
\draw[lred,line width=4pt,->] (0,3) -- (0,1);
\draw[black,dotted,->] (0,2) -- (-1,3);
\draw[black,dotted,->] (0,0) -- (-1,-1);
\node[above] at (1,2) {\tiny $j_h$};
\node[above] at (0,3) {\tiny $J_h$};
\node at (0,1) {\tiny $M_{h}$};
\node[below] at (1,0) {\tiny $i_h$};
\node[below] at (0,-1) {\tiny $I_h$};
\node[left] at (-1,-1) {\tiny $y$};
\node[left] at (-1,0) {\tiny $i_{h-1}$};
\node[left] at (-1,2) {\tiny $j_{h-1}$};
\node[left] at (-1,3) {\tiny $x$};
\node[right] at (3,0) {\tiny $\ell_h$};
\node[right] at (3,2) {\tiny $k_h$};
}$
\caption{The two sides of \eqref{eq:bulk-local-YB}, with six exterior
occupations fixed. Gray vertices have weight $L_{y,s}$ and red vertices
have weight $M^*_{x,s}$; dotted arrows mark their spectral parameters.
The internal occupations are $(k_{h-1},\ell_{h-1},K_h)$ on the left
and $(i_h,j_h,M_h)$ on the right. The sums of the respective products
of three weights agree.}
\label{fig:bulk-local-YB}
\end{figure}

For $a=(k_-,\ell_-,K)$ and $b=(i,j,M)$, choose the coupling
$A(a)B(b)/Z_h^2$ as in the proof of
\cite[Proposition~6.4]{BufetovMucciconiPetrov1905}.
Thus independence is conditional on the six fixed exterior
occupations. On positive-weight configurations its conditional laws
$\mathbf p_h^+(a,b)=B(b)/Z_h$ and
$\mathbf p_h^-(b,a)=A(a)/Z_h$ sum to one and satisfy
\begin{equation}
\label{eq:local-bulk-reversibility}
 A(a)\mathbf p_h^+(a,b)=B(b)\mathbf p_h^-(b,a).
\end{equation}
We use this choice throughout the paper.

Explicitly, assign $(i,j)\in\{0,1\}^2$ weight zero unless
$I+p-i=J+v-j\ge0$, and otherwise set
\begin{equation}
\label{eq:bulk-four-weight-kernel}
 \begin{aligned}
 \mathsf B_h(i,j)&=
 L_{y,s}(I,p;I+p-i,i)
 M^*_{x,s}(I+p-i,v;J,j)R^*_{xy}(i,j;k,\ell),\\
 \mathbb P(i_h=i,j_h=j)&=
 \frac{\mathsf B_h(i,j)}{\sum_{u,w\in\{0,1\}}\mathsf B_h(u,w)}.
 \end{aligned}
\end{equation}
At most two weights are nonzero, since $i-j=k-\ell$;
for $k\ne\ell$, the output is deterministically $(k,\ell)$.
At each encountered step the actual incoming configuration contributes
a positive $A$-term, so the denominator $Z_h$ is positive.
The analogous observation using $B$ applies to the backward update.

At column zero, the factors $y^{k_0}x^{\ell_0}$ in
\eqref{eq:row-definition} and \eqref{eq:dual-row} give
\begin{equation}
\label{eq:bulk-zero-column-kernel}
 \mathbf p_0^+(i,j\mid k,\ell)
 =\frac{y^i x^j R^*_{xy}(i,j;k,\ell)}
 {\Pi(x;y)y^k x^\ell}.
\end{equation}
Indeed, $\sum_{i,j}y^ix^jR^*_{xy}(i,j;k,\ell)
=\Pi(x;y)y^kx^\ell$ by Figure~\ref{fig:R-vert*}.
For $(k,\ell)=(0,0)$, the probabilities of $(i,j)=(0,0),(1,1)$
are $(1-xy)/(1-qxy)$ and $(1-q)xy/(1-qxy)$;
for $(k,\ell)=(1,1)$, they are $(1-q)/(1-qxy)$ and
$q(1-xy)/(1-qxy)$. Unequal pairs remain unchanged.
The old internal configuration is unique, so set $\mathbf p_0^-=1$
on its support. Since
$\ell(\nu)=\ell(\lambda)+i_0=\ell(\mu)+j_0$,
the initial update determines the output length independently of $s$.

The forward kernel samples column zero and then columns
$h=1,2,\ldots$, determining $\nu'_1,\nu'_2,\ldots$ in order.
The backward kernel samples in the opposite direction, starting
beyond the largest parts of its prescribed partitions with all
occupations zero. The resulting partition kernels are
\begin{equation}
\label{eq:bulk-product-kernels}
 \begin{aligned}
 \Ufwd_{x,y}(\varkappa\to\nu\mid\lambda,\mu)
 &=\mathbf p_0^+(i_0,j_0\mid k_0,\ell_0)
   \prod_{h\ge1}\mathbf p_h^+(a_h,b_h),\\
 \Ubwd_{x,y}(\nu\to\varkappa\mid\lambda,\mu)
 &=\prod_{h\ge1}\mathbf p_h^-(b_h,a_h).
 \end{aligned}
\end{equation}
For fixed input and output partitions, the occupation dictionary
determines every local configuration uniquely.

\begin{prop}[Stochastic bulk kernels]
\label{prop:bulk-kernels}
The products \eqref{eq:bulk-product-kernels} define stochastic kernels
satisfying \eqref{bulk-to}. The forward construction produces a finite
partition $\nu$ almost surely.
\end{prop}
\begin{proof}
Conservation gives
$\nu'_h-\nu'_{h+1}=I_h+i_{h-1}-i_h=M_h\ge0$;
the binary occupations ensure the required interlacings.
For $h>H=\max(\lambda_1,\mu_1)$, with the largest part of
$\varnothing$ taken as zero, $I=J=k=\ell=0$.
An incoming pair $(0,0)$ remains zero. For an incoming pair $(1,1)$,
continuation and termination have respective weights
\[
 \mathsf B_h(1,1)=
 \frac{(1-q)(x-s)(y-s)}{(1-xy)(1-sx)(1-sy)},\qquad
 \mathsf B_h(0,0)=\frac{(1-q)(1-s^2)}{(1-sx)(1-sy)}.
\]
Their sum is $(1-q)/(1-xy)$. Thus continuation has probability
$r_s(x,y)=(x-s)(y-s)/((1-sx)(1-sy))$, with
\begin{equation}
\label{eq:vacuum-termination}
 1-r_s(x,y)=\frac{(1-s^2)(1-xy)}{(1-sx)(1-sy)}>0.
\end{equation}
Since $r_s(x,y)^m\to0$, the output is finite almost surely,
proving forward stochasticity. Backward sampling has finitely many
nontrivial steps; conservation similarly gives
$\varkappa'_h-\varkappa'_{h+1}=K_h\ge0$ and the required
interlacings. Each step is normalized, proving backward stochasticity.

Write $R_h=R^*_{xy}(i_h,j_h;k_h,\ell_h)$.
The crossing factors in \eqref{eq:local-bulk-reversibility} are
$R_{h-1}$ on the left and $R_h$ on the right. At column zero,
\[
 \Pi(x;y)y^{k_0}x^{\ell_0}\mathbf p_0^+
 =y^{i_0}x^{j_0}R_0\mathbf p_0^-,\qquad \mathbf p_0^-=1.
\]
Multiplying these identities over columns cancels the intermediate
crossings; the final empty crossing has weight one. By the row
definitions, the remaining products give exactly \eqref{bulk-to}.
\end{proof}

\subsection{Construction of the boundary transition operators}
\label{ssec:boundary-to}

The boundary update first samples an auxiliary partition $\eta$
and then applies the bulk kernel. We verify boundary balance
throughout the nonnegative range of Section~\ref{sec:HSRF}, on
inputs with positive normalizers, and classify the autonomous
regimes for these prescribed kernels.

Use $\varkappa,\mu,\nu$ for the boundary partitions as in
Section~\ref{ssec:shsrf}. With $G^{t,\zeta}$ defined in \eqref{G}, set
\begin{align}
 Q_x^+(\varkappa,\eta)
 &=\frac{W_\eta(t,\zeta)f_{\eta/\varkappa}(x)}
 {\Phi_{t,\zeta}(x)G^{t,\zeta}_\varkappa(x)},
 \label{eq:fb-local-completion}\\
 Q_y^-(\nu,\eta)
 &=\frac{W_\eta(t,\zeta)g_{\nu/\eta}(y)}
 {G^{t,\zeta}_\nu(y)}.
 \label{eq:fb-local-reverse-completion}
\end{align}
These laws are supported on $\varkappa\prec\eta$ and
$\eta\prec\nu$, respectively. The skew Littlewood identity
normalizes $Q_x^+$, and the definition of $G$ normalizes $Q_y^-$.
After sampling $\eta$, apply the corresponding bulk kernel:
\begin{align}
 U^{\partial,+}_{x,y}(\varkappa\to\nu\mid\mu)
 &=\sum_\eta Q_x^+(\varkappa,\eta)
   \Ufwd_{x,y}(\varkappa\to\nu\mid\eta,\mu),\nonumber\\
 U^{\partial,-}_{x,y}(\nu\to\varkappa\mid\mu)
 &=\sum_\eta Q_y^-(\nu,\eta)
   \Ubwd_{x,y}(\nu\to\varkappa\mid\eta,\mu).
 \label{eq:fb-local-boundary-kernels}
\end{align}
Both kernels are stochastic by summing the two normalized steps;
their support satisfies $\varkappa\prec\mu\prec\nu$.
Figure~\ref{fig:boundary-partition-sampling} illustrates the forward
construction. Multiplying \eqref{bulk-to}, with southeast partition
$\eta$, by $W_\eta(t,\zeta)$ and summing gives
\begin{align}
 &\Pi(x;y)\Phi_{t,\zeta}(x)G^{t,\zeta}_\varkappa(x)
 g_{\mu/\varkappa}(y)
 U^{\partial,+}_{x,y}(\varkappa\to\nu\mid\mu)\nonumber\\
 &\hspace{12mm}=f_{\nu/\mu}(x)G^{t,\zeta}_\nu(y)
 U^{\partial,-}_{x,y}(\nu\to\varkappa\mid\mu).
 \label{eq:fb-local-boundary-balance}
\end{align}
Thus \eqref{boundary-to} holds. Proposition~\ref{prop:bulk-kernels}
and Theorem~\ref{thm:half-space-field-sampling} now establish the
field throughout the stated nonnegative parameter range.

\begin{figure}[t]
\centering
\begin{tikzpicture}[>=stealth,scale=.9,font=\small]
\begin{scope}
 \fill[lred!35] (0,0)--(2.8,0)--(2.8,2.3)--cycle;
 \draw[dashed,gray] (-.25,-.2)--(3.05,2.5);
 \draw[gray] (0,0)--(0,2.3)--(2.8,2.3);
 \draw[gray] (0,0)--(2.8,0)--(2.8,2.3);
 \draw[->,very thick,red] (.2,0)--(2.6,0)
   node[midway,below=4pt] {$Q_x^+(\varkappa,\cdot)$};
 \draw[->,thick] (0,.2)--(0,2.1);
 \draw[->,thick] (.2,2.3)--(2.6,2.3);
 \draw[->,thick] (2.8,.2)--(2.8,2.1);
 \fill (0,0) circle(2pt) node[left=3pt] {$\varkappa$};
 \fill (0,2.3) circle(2pt) node[left=3pt] {$\mu$};
 \fill[red] (2.8,0) circle(2pt) node[right=3pt] {$\eta$};
 \fill (2.8,2.3) circle(2pt) node[right=3pt] {$\nu$};
 \node[fill=white,inner sep=3pt] at (1.35,1.3)
   {$\Ufwd_{x,y}$};
 \node[align=center] at (1.4,3.0) {Boundary update};
\end{scope}
\begin{scope}[xshift=6.2cm]
 \node at (2.0,3.0) {Sampling $\eta$ when $t\zeta=a$};
 \draw[gray,line width=3pt] (.9,.4)--(.9,2.15);
 \draw[gray,line width=3pt] (3.1,.4)--(3.1,2.15);
 \draw[->,thick] (-.45,1.3)--(.55,1.3)
  node[midway,above] {$h_0$};
 \draw[->,thick] (1.25,1.3)--(2.75,1.3)
  node[midway,above] {$h_1$};
 \draw[->,thick] (3.45,1.3)--(4.45,1.3)
  node[midway,above] {$h_2$};
 \node[draw,fill=lred!40,minimum width=.75cm,minimum height=.7cm]
  at (.9,1.3) {$T^{(1)}_{K_1}$};
 \node[draw,fill=lred!40,minimum width=.75cm,minimum height=.7cm]
  at (3.1,1.3) {$T^{(2)}_{K_2}$};
 \node[above] at (.9,2.15) {$m_1(\varkappa)$};
 \node[above] at (3.1,2.15) {$m_2(\varkappa)$};
 \node[below] at (.9,.4) {$m_1(\eta)$};
 \node[below] at (3.1,.4) {$m_2(\eta)$};
 \node at (4.75,1.3) {$\cdots$};
 \node[align=center] at (2.0,-.45)
 {$m_j(\eta)=K_j+h_{j-1}-h_j$};
\end{scope}
\end{tikzpicture}
\caption{Boundary sampling. Left: given $\varkappa,\mu$, first sample
$\eta$ using $Q_x^+$, then sample $\nu$ using the bulk kernel.
Right: under \eqref{eq:fb-local-tuning}, sample the binary occupations
$h_{j-1}=\eta'_j-\varkappa'_j$ by \eqref{eq:fb-local-carrier},
with $K_j=m_j(\varkappa)$ and
$h_0\sim\operatorname{Bernoulli}(t/(1+t))$.}
\label{fig:boundary-partition-sampling}
\end{figure}

For general parameters, the auxiliary law $Q_x^+$ uses the full
input partition. To determine when the resulting boundary update
closes on the retained columns, recall
$\pi_R(\lambda)=(\lambda'_1,\ldots,\lambda'_R)$ for $R\ge1$.
This projection is \emph{autonomous} if the distribution of
$\pi_R(\nu)$ depends on the inputs only through
$\pi_R(\varkappa),\pi_R(\mu)$, on all inputs with positive
normalizing constants. The following classification is specific
to \eqref{eq:fb-local-boundary-kernels} with the bulk coupling
fixed in Section~\ref{ssec:bulk-to}.

\begin{prop}[Classification of autonomous finite-column projections]
\label{prop:fb-autonomy-classification}
Assume the parameter conditions of Section~\ref{sec:HSRF}, and fix
$x,y\in(0,1)$. For the forward kernel
\eqref{eq:fb-local-boundary-kernels}, using the bulk operators of
Section~\ref{ssec:bulk-to}, the projection $\pi_R$ is autonomous
for a fixed $R\ge1$ if and only if
\[
 (t,\zeta)=(1,1)\qquad\text{or}\qquad x=t\zeta.
\]
In either case it is autonomous for every finite $R$.
\end{prop}
\begin{proof}
For necessity assume $0\le t<1$, and put $U=1-t$,
$V=\zeta^{-1}-t\zeta$. The nonnegativity conditions give $U>0$,
$V\ge0$. Set
\[
 \rho=\frac{x-s}{1-sx}\in(0,1),\qquad
 b=\frac{x(V+xU)}{1-x^2}>0,\qquad
 c=\frac{x(U+xV)}{1-x^2}>0.
\]
The empty-tail vectors to the right of even and odd columns are
$(1,b)$ and $(1,c)$, respectively, by \eqref{eq:fb-vacuum-tail}.
The zero-multiplicity matrices in \eqref{eq:fb-transfer-A} are
\[
 A_0[E]=\begin{pmatrix}1&c-\rho b\\0&\rho\end{pmatrix},\qquad
 A_0[O]=\begin{pmatrix}1&b-\rho c\\0&\rho\end{pmatrix}.
\]
Multiplication of a row vector by these matrices sums the outgoing
occupation. It preserves its zeroth component and multiplies the
deviation from the alternating ratio $b,c$ by $\rho$.

Choose $\varkappa=\mu=(M)$ with even $M\ge R$. All these inputs
have projection $(1,\ldots,1)$, and their only nonzero multiplicity
is $m_M(\varkappa)=1$. After summing the tail and column $M$, put
$B=(1,b)A_1[E]$ and $z=B_1/B_0$. These quantities do not depend on
$M$, and
\[
 B_0=\frac{E_1(1-sxq)+b(1-q)}{1-sx}>0.
\]
The preceding zero-multiplicity matrices preserve $B_0$, so the
auxiliary normalizer is positive. Also,
$g_{(M)/(M)}(y)=(1-syq)/(1-sy)>0$; hence all these conditioning
inputs are admissible, including at positivity endpoints.

Write $h_{j-1}=\eta'_j-\varkappa'_j$. If
$\beta_M=Q_x^+((M),\{\eta:\ell(\eta)=2\})$, multiplication
through columns $M-1,\ldots,1$ yields
\[
 \frac{\beta_M}{1-\beta_M}
 =b+\rho^{M-1}(z-c).
\]
The matrices already contain the factors $x^{h_{j-1}-h_j}$, whose
product is the initial row factor $x^{h_0}$.
For the following bulk update, $(k_0,\ell_0)=(0,h_0)$.
By \eqref{eq:bulk-zero-column-kernel}, with
$C=(1-xy)/(1-qxy)>0$,
\[
 \mathbb P\{\ell(\nu)=2\mid\varkappa=\mu=(M)\}
 =1-C+C\beta_M.
\]
Since the output length is retained by $\pi_R$, autonomy forces
$\beta_M=\beta_{M+2}$. Thus
$\rho^{M-1}(1-\rho^2)(z-c)=0$, and $B_1=cB_0$.

Now $c=xU+xb$ and $b=xV+xc$. The recurrence
\eqref{eq:fb-even-recurrence} at multiplicity one gives
$(1-s^2q)E_2=(U+sqV)E_1+t(1-q)$. Consequently,
\[
\begin{aligned}
 (1-sx)(B_1-cB_0)
 &=x(1-s^2q)E_2+b(x-sq)E_1
   -c\bigl[(1-sxq)E_1+b(1-q)\bigr]\\
 &=(xt-bc)(1-q).
\end{aligned}
\]
Hence $bc=xt$, equivalently
\[
 0=x(V+xU)(U+xV)-t(1-x^2)^2
 =\frac{(x-t\zeta)(1-t\zeta x)(x+\zeta)(1+\zeta x)}{\zeta^2}.
\]
If $t>0$, $V\ge0$ implies $t\zeta\le\sqrt t<1$; if $t=0$,
$t\zeta=0$. All factors except $x-t\zeta$ are therefore strictly
positive, proving necessity.

For sufficiency, at $(t,\zeta)=(1,1)$,
Proposition~\ref{prop:fb-deterministic-completion} gives
$\eta'_j=2\lceil\varkappa'_j/2\rceil$.
When $x=t\zeta$, Proposition~\ref{prop:fb-local-factorization}
samples $h_0$ independently of $\varkappa$ and each $h_j$ from
$h_{j-1}$ and $m_j(\varkappa)$. The first $R$ auxiliary coordinates
therefore require only $m_j(\varkappa)$ for $j<R$, which are
determined by $\pi_R(\varkappa)$. In both regimes, the subsequent
bulk columns $0,\ldots,R-1$ determine $\pi_R(\nu)$ using only
$\pi_R(\varkappa),\pi_R(\eta),\pi_R(\mu)$. The remaining
normalized updates sum to one by almost-sure termination. Thus
$\pi_R$ is autonomous for every finite $R$.
\end{proof}

\begin{prop}[Deterministic auxiliary partition at $(t,\zeta)=(1,1)$]
\label{prop:fb-deterministic-completion}
Let $(t,\zeta)=(1,1)$ and $x,y\in(0,1)$. Appending zero parts, set
\begin{equation}
 C^+(\varkappa)=(\varkappa_1,\varkappa_1,\varkappa_3,\varkappa_3,\ldots),
 \qquad
 C^-(\nu)=(\nu_2,\nu_2,\nu_4,\nu_4,\ldots).
 \label{eq:even-completions}
\end{equation}
To sample $Q_x^+(\varkappa,\cdot)$, set $\eta=C^+(\varkappa)$;
to sample $Q_y^-(\nu,\cdot)$, set $\eta=C^-(\nu)$. Equivalently,
\begin{equation}
 Q_x^+(\varkappa,\eta)=\mathbf1_{\{\eta=C^+(\varkappa)\}},\qquad
 Q_y^-(\nu,\eta)=\mathbf1_{\{\eta=C^-(\nu)\}}.
 \label{eq:even-completion-kernels}
\end{equation}
\end{prop}
\begin{proof}
Here $\Phi_{1,1}(x)=1$, and $W_\eta(1,1)>0$ exactly when
$\eta_1=\eta_2$, $\eta_3=\eta_4$, and so on, equivalently when
every conjugate coordinate is even.
If $\varkappa\prec\eta$ and $\eta_{2r-1}=\eta_{2r}$, then
$\eta_{2r-1}\ge\varkappa_{2r-1}\ge\eta_{2r}$ forces both parts
to equal $\varkappa_{2r-1}$. Likewise, $\eta\prec\nu$ forces
each equal pair to equal $\nu_{2r}$. Each resulting partition has
positive weight in its auxiliary law. Thus the normalizers are
positive and each law is concentrated on the stated partition.
\end{proof}
Only the auxiliary step is deterministic; the subsequent partition
is sampled using the bulk kernel in \eqref{eq:fb-local-boundary-kernels}.

For the second case in Proposition~\ref{prop:fb-autonomy-classification},
write $x=a=t\zeta\in(0,1)$. The next proposition derives the
auxiliary sampler and the same closed nonnegativity interval as in
\eqref{eq:intro-local-curve}, including its endpoints.

\begin{prop}[Column transition probabilities when $t\zeta=a$]
\label{prop:fb-local-factorization}
Suppose $0<a<1$, $-1<s\le0$, $0<t<1$, and
\begin{equation}
 t\zeta=a,\qquad
 r_s(a)\le\zeta\le r_s(a)^{-1},\qquad
 r_s(a)=\frac{a-s}{1-sa}.
 \label{eq:fb-local-tuning}
\end{equation}
The weights $W_\rho(t,\zeta)$ are nonnegative and
$W_\rho(qt,\zeta)>0$ for every partition $\rho$. Moreover,
\begin{equation}
 G^{t,\zeta}_\varkappa(a)=W_\varkappa(qt,\zeta),\qquad
 Q_a^+(\varkappa,\eta)=
 \frac{W_\eta(t,\zeta)f_{\eta/\varkappa}(a)}
 {(1+t)W_\varkappa(qt,\zeta)}.
 \label{eq:fb-local-parameter-shift}
\end{equation}
Let $\theta_j$ denote the sequence $E$ for even $j$ and $O$ for odd $j$,
at parameters $(t,\zeta)$. Define row vectors indexed by $0,1$ and scalars by
\begin{equation}
 w_j=\begin{cases}(1,t),&j\text{ even},\\(1,a),&j\text{ odd},\end{cases}
 \qquad \gamma_K^{(j)}=\theta_j(K;qt,\zeta).
 \label{eq:fb-local-vectors}
\end{equation}
The following algorithm samples $Q_a^+(\varkappa,\cdot)$.
First sample $h_0\sim\operatorname{Bernoulli}(t/(1+t))$.
Then, successively for $j=1,2,\ldots$, put $K=m_j(\varkappa)$
and sample $h_j\in\{0,1\}$ using
\begin{equation}
 \mathbb P(h_j=l\mid h_{j-1}=r)
 =T_K^{(j)}(r,l):=
 \frac{(w_j)_l A_K[\theta_j](l,r)}
 {\gamma_K^{(j)}(w_{j-1})_r},\qquad r,l\in\{0,1\}.
 \label{eq:fb-local-carrier}
\end{equation}
Here $A_K[\theta_j]$ is evaluated at $x=a$, and its first index
is the outgoing occupation. Stop at the first $J\ge\varkappa_1$
with $h_J=0$, taking $\varkappa_1=0$ for the empty partition,
and set $h_j=0$ for $j>J$. Return the partition determined by
\begin{equation}
 \eta'_j=\varkappa'_j+h_{j-1},\qquad
 m_j(\eta)=m_j(\varkappa)+h_{j-1}-h_j.
 \label{eq:fb-local-output}
\end{equation}
The stopping time is finite almost surely.
\end{prop}
\begin{proof}
With $\sigma=-s$ and $t=a/\zeta$, the nonnegativity conditions
\eqref{eq:fb-positive-domain} become
$(a+\sigma)/(1+\sigma a)\le\zeta\le(1+\sigma a)/(a+\sigma)$,
which are exactly \eqref{eq:fb-local-tuning}.

For $A_K[\theta_j]$ in \eqref{eq:fb-transfer-A}, evaluated at $x=a$,
we claim
\begin{equation}
 w_jA_K[\theta_j]=\gamma_K^{(j)}w_{j-1},\qquad
 \gamma_K^{(j)}=
 \frac{\theta_j(K)(1-saq^K)+(w_j)_1\theta_j(K-1)(1-q^K)}{1-sa}.
 \label{eq:fb-local-factorization}
\end{equation}
To verify both components, abbreviate $p=q^K$,
$b=(w_j)_1$, $c=(w_{j-1})_1$, and $\theta_K=\theta_j(K)$.
The zeroth component gives the displayed scalar expression.
For the first component, substituting the boundary recurrence gives
\[
\begin{aligned}
 &a(1-s^2p)\theta_{K+1}+b(a-sp)\theta_K
 -c\bigl[(1-sap)\theta_K+b(1-p)\theta_{K-1}\bigr]\\
 &\qquad=
 \begin{cases}
 sp(a/\zeta-t)\theta_K,&j\text{ even},\\
 (a/\zeta-t)\theta_K,&j\text{ odd},
 \end{cases}
 =0,
\end{aligned}
\]
where $bc=at$ and $t\zeta=a$ were used.

It remains to identify the scalar with $\theta_j(K;qt,\zeta)$.
Write
\[
 F_j(z)=\sum_{K\ge0}\frac{(s^2;q)_K}{(q;q)_K}\theta_j(K)z^K,
\]
and let $\widetilde F_j$ replace $\theta_j(K)$ by the scalar
expression in \eqref{eq:fb-local-factorization}.
After multiplication by $1-sa$, shifting the term containing
$\theta_j(K-1)$ gives $bz[F_j(z)-s^2F_j(qz)]$, hence
\[
 (1-sa)\widetilde F_j(z)
 =(1+bz)F_j(z)-s(a+sbz)F_j(qz).
\]
Using \eqref{eq:fb-even-generating}--\eqref{eq:fb-odd-generating}
and $t\zeta=a$, we obtain
\[
 \frac{\widetilde F_j(z)}{F_j(z)}=
 \begin{cases}
 \displaystyle\frac{1+tz}{1-sa}
 \left(1-\frac{sa(1-z)}{1-saz}\right)
 =\frac{1+tz}{1-saz},&j\text{ even},\\[6pt]
 \displaystyle\frac{1+az}{1-sa}
 \left(1-\frac{sa(1-z/\zeta)}{1-stz}\right)
 =\frac{1+az}{1-stz},&j\text{ odd}.
 \end{cases}
\]
These are exactly the ratios of the generating products at
$(qt,\zeta)$ and $(t,\zeta)$. Coefficient comparison proves
\eqref{eq:fb-local-factorization}. Since $t>0$, the nonnegative
boundary recurrences admit no two consecutive zero terms: such a
pair would propagate backwards to $\theta_j(0)=0$, a contradiction.
The scalar expression in \eqref{eq:fb-local-factorization} therefore
gives $\gamma_K^{(j)}>0$ for $K\ge1$, and $\gamma_0^{(j)}=1$.
This proves positivity of $W_\rho(qt,\zeta)$ and of all sampling
denominators, including at the endpoints of
\eqref{eq:fb-local-tuning}.

The entries of $A_K$ are nonnegative. Summing over the outgoing
occupation gives
$\sum_lT_K^{(j)}(r,l)=(w_jA_K)_r/
(\gamma_K^{(j)}(w_{j-1})_r)=1$ by
\eqref{eq:fb-local-factorization}.
Negative multiplicities have probability zero because $A_0(1,0)=0$.
Beyond the largest part of $\varkappa$, state zero is absorbing,
while continuation from state one has probability
$\zeta r_s(a)$ at odd columns and $r_s(a)/\zeta$ at even columns.
Both are at most one, and their two-column product is
$r_s(a)^2<1$. Thus the output is a finite partition almost surely,
also when one continuation probability equals one.

For an eventually zero sequence $(h_j)$, the vector factors in
\eqref{eq:fb-local-carrier} telescope against
$\mathbb P(h_0=r)=(w_0)_r/(1+t)$, leaving
\[
 \frac{\prod_{j\ge1}A_{m_j(\varkappa)}[\theta_j](h_j,h_{j-1})}
 {(1+t)\prod_{j\ge1}\gamma_{m_j(\varkappa)}^{(j)}}
 =\frac{W_\eta(t,\zeta)f_{\eta/\varkappa}(a)}
 {(1+t)W_\varkappa(qt,\zeta)}.
\]
Here the factors $a^{h_{j-1}-h_j}$ already included in $A_K$
multiply to $a^{h_0}$, the initial factor in the row definition.
The displayed probabilities sum to one by termination. Comparing
their normalizer with the skew Littlewood identity and
$\Phi_{t,\zeta}(a)=1+t$ proves \eqref{eq:fb-local-parameter-shift}.
\end{proof}

At the diagonal vertex $(i,i)$ the local parameter $x$ is
$x_{i-1}$. Thus, with fixed boundary parameters, imposing the
second autonomous regime at every diagonal vertex gives
$x_r=a=t\zeta$ for all $r\ge0$. The specialization
$(t,\zeta)=(1,1)$ allows arbitrary $x_r\in(0,1)$. These are exactly
the two regimes stated in Theorem~\ref{thm:intro-field}; the
consistency of their finite-column fields is proved next.

\subsection{Finite-column projections}
\label{ssec:fb-local-projections}

We prove the projection statement of
Theorem~\ref{thm:intro-field}, the scalar identifications in
Corollary~\ref{cor:intro-scalar}, and the continuous-time limit
underlying Theorem~\ref{thm:intro-continuous-time}. Throughout this
subsection $0<q<1$ and $-1<s\le0$. Unless stated otherwise, the
spectral parameters are homogeneous and equal to $a\in(0,1)$.
The two boundary regimes are
\[
 \text{(i)}\quad(t,\zeta)=(1,1),\qquad
 \text{(ii)}\quad t\zeta=a,\quad
 0<t<1,\quad r_s(a)\le\zeta\le r_s(a)^{-1},
 \qquad r_s(a)=\frac{a-s}{1-sa}.
\]
We always use the bulk kernels of Section~\ref{ssec:bulk-to} and
the corresponding boundary kernels of Section~\ref{ssec:boundary-to}.

\begin{thm}[Consistent finite-column projections]
\label{thm:fb-local-projection}
In either regime, the fields
$\mathbf H_R(i,j)=\pi_R(\lambda^{(i,j)})$, $R\ge1$, have autonomous
local transition probabilities. Their laws are consistent under
deleting the last coordinates. Regime~\textup{(i)} also permits
inhomogeneous spectral parameters $0<x_r<1$.
\end{thm}
\begin{proof}
At the boundary this is the sufficiency statement of
Proposition~\ref{prop:fb-autonomy-classification}. At a bulk site,
write $\varkappa,\eta,\mu,\nu$ for the southwest, south, west and
northeast partitions. The binary occupations are
\begin{equation}
 k_h=\mu'_{h+1}-\varkappa'_{h+1},\quad
 \ell_h=\eta'_{h+1}-\varkappa'_{h+1},\quad
 i_h=\nu'_{h+1}-\eta'_{h+1},\quad
 j_h=\nu'_{h+1}-\mu'_{h+1}.
 \label{eq:fb-projection-carriers}
\end{equation}
Sampling columns $0,\ldots,R-1$ uses only the first $R$ input
coordinates. All later sampling probabilities sum to one.
Stopping this same procedure earlier proves consistency.
\end{proof}

Autonomy concerns the joint vector of retained coordinates at local
updates. It does not imply that a higher coordinate alone, or the
sequence of diagonal vectors alone, is a Markov chain.

\smallskip\noindent\textbf{The scalar stochastic vertex models.}
Write $H(i,j)=(\lambda^{(i,j)})'_1$. At a bulk site, the south,
west, north and east arrow occupations are
\begin{equation}
 (\iota_{\rm S},\iota_{\rm W},\iota_{\rm N},\iota_{\rm E})
 =(\ell_0,1-k_0,j_0,1-i_0).
 \label{eq:height-arrow-dictionary}
\end{equation}
They obey arrow conservation. Moreover, $H(0,j)=0$ and $H(i,j)$
counts the north arrows above $(1,j),\ldots,(i,j)$. With
$c=(1-a^2)/(1-qa^2)$, their bulk probabilities are
\begin{equation}
\begin{array}{c|c|c}
 (\iota_{\rm W},\iota_{\rm S})&(\iota_{\rm E},\iota_{\rm N})
 &\text{probability}\\ \hline
 (0,0)&(0,0)&1\\
 (1,1)&(1,1)&1\\
 (1,0)&(1,0)&c\\
 (1,0)&(0,1)&1-c\\
 (0,1)&(0,1)&qc\\
 (0,1)&(1,0)&1-qc
\end{array}
 \label{eq:fb-scalar-bulk-table}
\end{equation}
The external west arrows are one.

First recall the half-space stochastic six-vertex model of
Barraquand, Borodin, Corwin and Wheeler
\cite[Section~4.1]{BarraquandBorodinCorwinWheeler1704}.
In the homogeneous case it has the bulk probabilities
\eqref{eq:fb-scalar-bulk-table}, external west arrows equal to one,
and the diagonal rule $\iota_{\rm N}=1-\iota_{\rm W}$.
Write $H_{\rm BBCW}$ for its north-edge height function.

He extends this diagonal rule to a two-parameter boundary family
\cite[Section~1.3]{He2024Boundary}. Denote the corresponding height
function by $H_{\rm He}^{b,\zeta}$, with bulk parameters $(q,a)$.
Conditional on west input zero or one, its diagonal north-arrow
probabilities are, respectively, $bp_{b,\zeta}(a)$ and
$1-p_{b,\zeta}(a)$, where
\[
 p_{b,\zeta}(a)=\frac{1-a^2}{(1-\zeta ba)(1+a/\zeta)}.
\]
At $(b,\zeta)=(1,1)$, one has $p_{1,1}(a)=1$, so this is the
BBCW model. The vertex configurations are illustrated in
Figure~\ref{fig:intro-scalar}. We now derive the boundary rule of
our scalar projection and explain the coordinate changes in
Corollary~\ref{cor:intro-scalar}.

\begin{prop}[Regime \textup{(ii)}: the scalar field]
\label{prop:fb-local-scalar-law}
In regime~\textup{(ii)}, $H$ is exactly $H_{\rm He}^{qt,\zeta}$.
Its diagonal transition table is
\begin{equation}
\begin{array}{c|c|c}
 \iota_{\rm W}&\iota_{\rm N}&\text{probability}\\ \hline
 0&0&1-qt\,p'\\
 0&1&qt\,p'\\
 1&0&p'\\
 1&1&1-p'
\end{array}
 \qquad p'=\frac{c}{1+t}
 =\frac{1-a^2}{(1-\zeta qt\,a)(1+a/\zeta)}.
 \label{eq:fb-scalar-boundary-table}
\end{equation}
For $A_n=(a,\ldots,a)$ with $n$ entries, the partition at the
diagonal has law
\begin{equation}
 \mathbb P(\lambda^{(n,n)}=\rho)
 =\frac{W_\rho(qt,\zeta)f_\rho(A_n)}{Z_{qt,\zeta}(A_n)}.
 \label{eq:fb-local-diagonal-law}
\end{equation}
\end{prop}
\begin{proof}
The auxiliary south arrow $h_0$ is a fresh Bernoulli variable with
parameter $\beta=t/(1+t)$. Averaging the bulk table gives north-arrow
probabilities $\beta qc$ and $1-(1-\beta)c$, which are the displayed
ones. Since $t\zeta=a$, these agree with He's parameters
$(b,\zeta)=(qt,\zeta)$. The diagonal partition mass is proportional
to $G^{t,\zeta}_\rho(a)f_\rho(A_n)$; now use
\eqref{eq:fb-local-parameter-shift} and the Littlewood normalization.
\end{proof}

Thus the bulk and boundary probabilities use the configurations in
Figure~\ref{fig:intro-scalar}, with $b=qt$ and
$p_{b,\zeta}(a)=p'=c/(1+t)$.

\begin{prop}[Regime \textup{(ii)}: deletion of the diagonal]
\label{prop:He-row-shift}
The following equalities hold jointly over $1\le i\le j$:
\begin{equation}
 \{H(i,j)\}\stackrel{\rm d}=
 \{H_{\rm He}^{t,\zeta}(i,j+1)\}\stackrel{\rm d}=
 \{H_{\rm He}^{qt,\zeta}(i,j)\}.
 \label{eq:He-joint-row-shift}
\end{equation}
\end{prop}
\begin{proof}
For boundary parameters $(t,\zeta)$ with $t\zeta=a$,
$p_{t,\zeta}(a)=1/(1+t)$. Both possible west inputs therefore
give the same north-arrow probability $t/(1+t)$. Delete the diagonal
vertices and relabel $(i,J)$ as $(i,J-1)$ for $J>i$.
This deletion is illustrated in Figure~\ref{fig:scalar-diagonal-deletion}.
The old vertex $(i,i+1)$ becomes a boundary vertex whose south
input is this independent Bernoulli arrow. Its west input is sampled
strictly to the left and is independent of that arrow. Averaging
therefore gives \eqref{eq:fb-scalar-boundary-table}. All other bulk
weights and external arrows are preserved. This proves the joint
identities by coupling the remaining random choices.
\end{proof}

\begin{prop}[Regime \textup{(i)}: the BBCW model after deletion]
\label{prop:even-boundary-BBCW}
Write $h_{\rm ev}(i,j)=(\lambda^{(i,j)})'_1$ in regime~\textup{(i)}.
Allow spectral parameters $0<x_r<1$. Let $H_{\rm BBCW}$ have the
parameters $t_{\rm BBCW}=q$, $a_r=x_{r-1}$ and the north-edge
counting convention above. There is a coupling such that
\begin{equation}
 h_{\rm ev}(i,j)=H_{\rm BBCW}(i,j+1),\qquad 1\le i\le j,
 \label{eq:even-BBCW-row-shift}
\end{equation}
and, with $h_{\rm ev}(0,0)=0$,
\begin{equation}
 H_{\rm BBCW}(n+1,n+1)
 =h_{\rm ev}(n,n)+(h_{\rm ev}(n,n)\bmod2).
 \label{eq:BBCW-diagonal-rounding}
\end{equation}
\end{prop}
\begin{proof}
Our auxiliary arrow is $h_0=\varkappa'_1\bmod2$.
Let $D_i$ be the north arrow of the BBCW corner $(i,i)$.
Its west arrow is $1-D_i$. Row conservation gives
\[
 H_{\rm BBCW}(i-1,i)=H_{\rm BBCW}(i-1,i-1)+D_i,
 \qquad
 H_{\rm BBCW}(i,i)=H_{\rm BBCW}(i-1,i-1)+2D_i.
\]
The diagonal heights are therefore even, and
$D_i=H_{\rm BBCW}(i-1,i)\bmod2$. Deleting the corners gives
precisely our auxiliary parity rule. The remaining bulk parameters
agree since $a_i a_{j+1}=x_{i-1}x_j$, proving the coupling.
The last equation follows by adding $D_{n+1}$ to
$H_{\rm BBCW}(n,n+1)$.
\end{proof}

\begin{figure}[t]
\centering
\begin{tikzpicture}[>=stealth,font=\small,scale=.87]
\begin{scope}
 \node at (2.15,5.2) {Before deleting the diagonal};
 \foreach \j in {1,2,3,4}{
  \draw[gray,->] (.25,\j)--({min(\j,3.6)},\j);
 }
 \foreach \i in {1,2,3}{
  \draw[gray,->] (\i,\i)--(\i,4.65);
  \fill[lgray] (\i,\i) circle(4pt);
  \draw[black] ({\i-.12},{\i-.12})--({\i+.12},{\i+.12});
  \draw[black] ({\i-.12},{\i+.12})--({\i+.12},{\i-.12});
  \draw[red,thick,->] (\i,{\i+.15})--(\i,{\i+.85});
  \fill[red] (\i,{\i+1}) circle(3pt);
 }
 \fill (1,3) circle(2pt);
 \fill (1,4) circle(2pt);
 \fill (2,4) circle(2pt);
 \draw[dashed,gray] (.65,.65)--(3.4,3.4);
 \node[below right] at (2,2) {$(i,i)$};
 \node[above right] at (2,3) {$(i,i+1)$};
 \node[left,red] at (2,2.5) {$D_i$};
 \node[below] at (2.1,.25) {Gray vertices are removed};
\end{scope}
\draw[->,thick] (4.7,2.8)--(6.4,2.8)
 node[midway,above] {$J\mapsto J-1$};
\begin{scope}[xshift=7.0cm,yshift=1.0cm]
 \node at (1.7,4.2) {After relabeling};
 \foreach \j in {1,2,3}{
  \draw[gray,->] (.25,\j)--(\j,\j);
 }
 \foreach \i in {1,2,3}{
  \draw[gray,->] (\i,\i)--(\i,3.65);
  \fill[red] (\i,\i) circle(3pt);
 }
 \fill (1,2) circle(2pt);
 \fill (1,3) circle(2pt);
 \fill (2,3) circle(2pt);
 \draw[dashed,gray] (.65,.65)--(3.4,3.4);
 \node[below right] at (2,2) {$(i,i)$};
 \node[below,align=center] at (1.8,-.75)
 {The red bulk vertices become\\the new diagonal vertices};
\end{scope}
\end{tikzpicture}
\caption{Deleting the diagonal vertices and relabeling
$(i,J)$ as $(i,J-1)$ for $J>i$. The old vertex $(i,i+1)$ becomes
the new diagonal vertex $(i,i)$. Its south input $D_i$ is the north
output of the removed vertex $(i,i)$. For BBCW,
$D_i=H_{\rm BBCW}(i-1,i)\bmod2$, where $H_{\rm BBCW}$ is its
north-edge height function; for the model of He with
$t\zeta=a$, the $D_i$ are independent Bernoulli variables with
parameter $t/(1+t)$. Thus the same geometric deletion produces two
different boundary transition rules. The gray lines indicate lattice
edges, not a sampled arrow configuration.}
\label{fig:scalar-diagonal-deletion}
\end{figure}

In regime~\textup{(i)}, deletion leaves a boundary rule depending
on the height parity. In regime~\textup{(ii)}, deletion changes the
He parameter $b=t$ to $b=qt$ and leaves a boundary rule depending
only on the incoming arrow. The latter realizes the subfamily
$b\zeta=qa$, rather than arbitrary He boundary parameters.
Both scalar fields are independent of the spin $s$.

\smallskip\noindent\textbf{The first two columns.}
Set $\mathbf H_2=(H_1,H_2)$, where
$H_r(i,j)=(\lambda^{(i,j)})'_r$. Its site values satisfy
$H_1\ge H_2\ge0$, and neighboring differences in each coordinate
belong to $\{0,1\}$.

Figure~\ref{fig:fb-two-column-model} shows the projection of a
Young diagram and both height fields in one finite realization.

\begin{figure}[t]
\centering
\begin{tikzpicture}[>=stealth,font=\small,x=1cm,y=1cm]
 \node at (1.1,4.45) {Retain the first two columns};
 \node at (1.1,3.93) {$\lambda^{(4,4)}=(4,3,1)$};
 \begin{scope}[shift={(.14,3.5)},x=.51cm,y=-.51cm]
  \foreach \row/\length in {0/4,1/3,2/1}{
   \foreach \col in {1,...,\length}{
    \ifnum\col=1
     \fill[black!75] ({\col-1},\row) rectangle (\col,{\row+1});
    \else\ifnum\col=2
     \fill[red!65] ({\col-1},\row) rectangle (\col,{\row+1});
    \else
     \fill[gray!20] ({\col-1},\row) rectangle (\col,{\row+1});
    \fi\fi
    \draw[white,line width=.65pt] ({\col-1},\row) rectangle (\col,{\row+1});
   }
  }
 \end{scope}
 \draw[->,thick] (1.15,1.72)--(1.15,1.02)
  node[midway,right] {$\pi_2$};
 \node at (1.15,.65) {$\bigl(\underbrace{\vphantom{3}3}_{H_1},
                                  \underbrace{\textcolor{red}{2}}_{\textcolor{red}{H_2}}\bigr)$};
 \node[gray,align=center,font=\footnotesize] at (1.15,-.12)
  {Gray columns are discarded};
 \draw[dotted,gray] (3.05,-.3)--(3.05,4.7);
 \node at (7.85,4.97) {Two height fields from one half-space realization};
 \foreach \origin/\layer/\layercolor in {3.85/1/black,8.8/2/red}{
  \begin{scope}[shift={(\origin,.3)},x=.8cm,y=.8cm]
   \node[\layercolor] at (2,5.17) {$H_{\layer}(i,j)=(\lambda^{(i,j)})'_{\layer}$};
   \foreach \j in {1,2,3,4}{
    \draw[gray!45] (0,\j)--(\j,\j);
    \node[left,font=\scriptsize,gray] at (-.19,\j) {$\j$};
   }
   \foreach \i in {1,2,3,4}{
    \draw[gray!45] (\i,\i)--(\i,4);
    \draw[gray] (\i,-.05)--(\i,.05);
    \node[below,font=\scriptsize,gray] at (\i,-.08) {$\i$};
   }
   \draw[gray,->] (0,0)--(4.5,0) node[right,black] {$i$};
   \draw[gray,->] (0,0)--(0,4.6) node[above,black] {$j$};
   \draw[gray,dashed] (.35,.35)--(4.28,4.28);
   \node[gray,rotate=45,font=\scriptsize] at (2.83,2.38) {$i=j$};
   \foreach \j in {0,1,2,3,4}{
    \node[circle,fill=white,inner sep=1.6pt,text=gray] at (0,\j) {$0$};
   }
   \foreach \i/\j/\first/\second in
    {1/1/0/0,1/2/1/0,2/2/1/1,
     1/3/1/1,2/3/2/1,3/3/2/2,
     1/4/1/1,2/4/2/1,3/4/3/2,4/4/3/2}{
    \ifnum\layer=1\def\heightvalue{\first}\else\def\heightvalue{\second}\fi
    \node[circle,fill=white,inner sep=2pt,text=\layercolor]
     at (\i,\j) {$\heightvalue$};
   }
   \draw[\layercolor,thick] (4,4) circle(.19);
   \node[above right,font=\scriptsize] at (4.15,4.03) {$(4,4)$};
  \end{scope}
 }
\end{tikzpicture}
\caption{An actual two-column projection and an admissible finite
realization in regime~\textup{(i)}. Left: the black and red columns
of $(4,3,1)$ have lengths $3$ and $2$; all later columns are removed.
Right: the numbers are the values of the two coupled height functions
on $0\le i\le j\le4$, with their common diagonal shown dashed.
The circled values are the projections of the displayed Young diagram.
At every site $H_1\ge H_2$, and each horizontal or vertical increment
of either height is zero or one. Both height fields arise from the same
partition-valued realization.}
\label{fig:fb-two-column-model}
\end{figure}

\begin{prop}[Sampling the two-column projection]
\label{prop:fb-two-column-sampling}
At a bulk site retain the first two columns of the southwest,
south and west partitions $\varkappa,\eta,\mu$. Put
$k_r=\mu'_{r+1}-\varkappa'_{r+1}$ and
$\ell_r=\eta'_{r+1}-\varkappa'_{r+1}$, $r=0,1$.
First sample $(i_0,j_0)$ by \eqref{eq:bulk-zero-column-kernel} and
set $\nu'_1=\eta'_1+i_0=\mu'_1+j_0$. Then put
$I=\eta'_1-\eta'_2$, $J=\mu'_1-\mu'_2$, and sample
$(i_1,j_1)$ with probabilities proportional to
\begin{equation}
 B_{i,j}=L_{a,s}(I,i_0;M,i)M^*_{a,s}(M,j_0;J,j)
 R^*_{a^2}(i,j;k_1,\ell_1),\qquad M=I+i_0-i,
 \label{eq:fb-two-column-bulk-weights}
\end{equation}
for $(i,j)\in\{0,1\}^2$, setting the weight to zero unless
$M=J+j_0-j\ge0$. Set $\nu'_2=\eta'_2+i_1=\mu'_2+j_1$.

At a boundary site the input partitions are $\varkappa,\mu$.
Sample the retained coordinates of the auxiliary partition $\eta$
as follows, and then apply the same bulk update.
In regime~\textup{(i)}, set
$\eta'_r=2\lceil\varkappa'_r/2\rceil$, $r=1,2$.
In regime~\textup{(ii)}, put $D=\varkappa'_1-\varkappa'_2$, sample
$h_0\sim\operatorname{Bernoulli}(t/(1+t))$, and sample $h_1$ using
\begin{equation}
 \begin{aligned}
 \mathbb P(h_1=1\mid h_0=0)
 &=\frac{a(1-q^D)O_{D-1}(t,\zeta)}{(1-sa)O_D(qt,\zeta)},\\
 \mathbb P(h_1=1\mid h_0=1)
 &=\frac{\zeta(a-sq^D)O_D(t,\zeta)}{(1-sa)O_D(qt,\zeta)}.
 \end{aligned}
 \label{eq:fb-two-column-boundary}
\end{equation}
Set $\eta'_1=\varkappa'_1+h_0$ and
$\eta'_2=\varkappa'_2+h_1$.
\end{prop}
\begin{proof}
These are columns zero and one of the specified kernels.
Conservation leaves at most two positive bulk candidates, and
$\nu'_1-\nu'_2=M\ge0$ preserves order. The boundary formulas are the
first two steps of the algorithms in Section~\ref{ssec:boundary-to}.
At $D=0$ the first probability in \eqref{eq:fb-two-column-boundary}
is zero; hence the auxiliary state is also ordered.
\end{proof}

For example, with $\varkappa=\eta=\mu=\varnothing$, the retained
bulk output has distribution
\[
\begin{array}{c|ccc}
 \pi_2(\nu)&(0,0)&(1,0)&(1,1)\\ \hline
 \text{probability}&c&(1-c)(1-r_s(a)^2)&(1-c)r_s(a)^2.
\end{array}
\]
Indeed, conditional on $\nu'_1=1$, the two weights are
$(1-q)(1-s^2)/(1-sa)^2$ and
$(1-q)(a-s)^2/((1-sa)^2(1-a^2))$. The identity
$(1-s^2)(1-a^2)+(a-s)^2=(1-sa)^2$ gives
\begin{equation}
 \mathbb P(\nu'_2=1\mid\nu'_1=1,\,
 \varkappa=\eta=\mu=\varnothing)=r_s(a)^2.
 \label{eq:fb-two-column-bulk-example}
\end{equation}
This explicitly retains the spin. A marginal check in regime~\textup{(ii)}
is obtained at $n=1$. The one-variable function is supported on
one-row partitions; thus $H_2=0$ precisely for $\varnothing$ and $(1)$.
With $Z=\Phi_{qt,\zeta}(a)$, $U_*=1-qt$ and $V_*=\zeta^{-1}-qa$,
subtracting these two masses gives
\begin{equation}
 \mathbb P((\lambda^{(1,1)})'_2=1)
 =1-Z^{-1}-\frac{a(V_*+sU_*)}{(1-sa)Z}.
 \label{eq:fb-local-spin-example}
\end{equation}

For $R>2$, use the first $R$ boundary sampling steps followed by
bulk columns $0,\ldots,R-1$. These are coupled height fields in the
Yang--Baxter construction of
\cite{BufetovPetrov1712,BufetovMucciconiPetrov1905}, with the boundary
transitions specified here. At $s=0$ the kernels have their
Hall--Littlewood specializations. Equations
\eqref{eq:fb-two-column-bulk-example}--\eqref{eq:fb-local-spin-example}
show why the scalar identifications do not determine the higher
coordinates. The bulk construction has the full-space precedents
just cited; the boundary rules above and the limit below specify
the resulting half-space coupling.

\smallskip\noindent\textbf{A continuous-time limit of the two-column field.}
Keep $q,s$ fixed, put $a_\varepsilon=1-(1-q)\varepsilon/2$, and
use either regime~\textup{(i)} or the following choice in
regime~\textup{(ii)}:
\begin{equation}
 a_\varepsilon=1-\frac{1-q}{2}\varepsilon,\qquad
 \zeta=1,\qquad t_\varepsilon=a_\varepsilon,
 \qquad n_\varepsilon=\lfloor T/\varepsilon\rfloor.
 \label{eq:fb-scalar-ASEP-scaling}
\end{equation}
The latter satisfies the positivity inequalities for small
$\varepsilon>0$. For a row at time $n$, define particle occupations
on the positive integers by
\begin{equation}
 \xi_r^{(n)}(x)=1-H_r(n-x+1,n)+H_r(n-x,n),\quad 1\le x\le n,
 \qquad \xi_r^{(n)}(x)=0\quad(x>n),\quad r=1,2.
 \label{eq:fb-two-column-particles}
\end{equation}
These particles are holes in the north-arrow configuration, in
coordinates measured from the diagonal. Write
\begin{equation}
 \begin{gathered}
 N_r=\sum_{x\ge1}\xi_r(x),\qquad
 D_x=\sum_{y\ge x}(\xi_2(y)-\xi_1(y)),\\
 \mathcal X=\{(\xi_1,\xi_2):N_1+N_2<\infty,\ D_x\ge0\ (x\ge1)\}.
 \end{gathered}
 \label{eq:fb-two-column-state}
\end{equation}
Here $\xi_r(x)\in\{0,1\}$. On row $n$, telescoping gives
$D_x=H_1(n-x+1,n)-H_2(n-x+1,n)$ for $1\le x\le n$.
The height ordering gives $D_x\ge0$;
it does not require $\xi_1(x)\le\xi_2(x)$ at each site. In particular,
$N_r^{(n)}=n-H_r(n,n)$ and $D_1=N_2-N_1$.
For row $n=4$ in Figure~\ref{fig:fb-two-column-model}, the particle
positions are $\{1\}$ in layer 1 and $\{1,3\}$ in layer 2.

The rates of this process are specified in
Section~\ref{ssec:contributions}: use the bulk table
\eqref{eq:fb-two-column-bulk-rates} and the boundary factor
\eqref{eq:fb-two-column-boundary-rho}, with $d=D_{x+1}$ at bond
$(x,x+1)$ and $(D,k)=(D_1,\xi_1(1))$ at the boundary.
In particular, a first-layer right jump or entry forces the
corresponding second-layer move when its tail gap is zero;
second-layer left jumps and exits are blocked at a zero gap.
We now write the generator and derive every rate from the discrete
kernels. Figure~\ref{fig:intro-two-layer-moves} displays the right
jumps and entries, while Figure~\ref{fig:fb-two-column-complementary-moves}
displays the left jumps, exits and blocked moves.

\begin{figure}[tbp]
\centering
\begingroup
\newcommand{\fbsite}[4]{%
 \ifnum#3=2
  \node[text=#4,inner sep=0pt] at (#1,#2) {$\ast$};
 \else
  \draw[draw=#4,fill=white] (#1,#2) circle(2.7pt);
  \ifnum#3=1\fill[#4] (#1,#2) circle(2.7pt);\fi
 \fi}
\newcommand{\fbstate}[5]{%
 \begin{scope}[xshift=#1cm]
  \draw[gray!50] (0,.65)--(1.3,.65);
  \draw[gray!50] (0,-.25)--(1.3,-.25);
  \fbsite{.3}{.65}{#2}{black}\fbsite{1}{.65}{#3}{black}
  \fbsite{.3}{-.25}{#4}{red}\fbsite{1}{-.25}{#5}{red}
  \node[below,font=\scriptsize] at (.3,-.55) {$x$};
  \node[below,font=\scriptsize] at (1,-.55) {$x+1$};
 \end{scope}}
\newcommand{\fbboundarystate}[3]{%
 \begin{scope}[xshift=#1cm]
  \draw[gray!50] (.1,.65)--(1.2,.65);
  \draw[gray!50] (.1,-.25)--(1.2,-.25);
  \fbsite{.65}{.65}{#2}{black}\fbsite{.65}{-.25}{#3}{red}
  \node[below,font=\scriptsize] at (.65,-.55) {site $1$};
 \end{scope}}
\newcommand{\fbmove}[3]{%
 \draw[->,thick] (1.6,.2)--(3.1,.2)
  node[midway,above=4pt] {$#1$}
  node[midway,below=4pt] {$#2$};
 \ifnum#3=1
  \draw[red,thick] (2.23,.08)--(2.47,.32);
  \draw[red,thick] (2.23,.32)--(2.47,.08);
 \fi}
\begin{tikzpicture}[>=stealth,font=\small,x=1cm,y=1cm]
\begin{scope}[shift={(0,0)}]
 \node at (2.5,1.25) {(a) Layer-1 left jump};
 \fbstate{0}{0}{1}{2}{2}
 \fbstate{3.6}{1}{0}{2}{2}
 \fbmove{q}{d\ge0}{0}
\end{scope}
\begin{scope}[shift={(6.25,0)}]
 \node at (2.5,1.25) {(b) Layer-1 exit};
 \fbboundarystate{0}{1}{2}
 \fbboundarystate{3.6}{0}{2}
 \fbmove{\tfrac q2}{D\ge0}{0}
\end{scope}
\begin{scope}[shift={(0,-2.75)}]
 \node at (2.5,1.25) {(c) Layer-2 left jump};
 \fbstate{0}{2}{2}{0}{1}
 \fbstate{3.6}{2}{2}{1}{0}
 \fbmove{v_d(\omega)}{d>0}{0}
\end{scope}
\begin{scope}[shift={(6.25,-2.75)}]
 \node at (2.5,1.25) {(d) Layer-2 exit};
 \fbboundarystate{0}{2}{1}
 \fbboundarystate{3.6}{2}{0}
 \fbmove{\tfrac{q}{2\rho_{D-1}(k)}}{D>0}{0}
\end{scope}
\begin{scope}[shift={(0,-5.5)}]
 \node at (2.5,1.25) {(e) Blocked layer-2 left jump};
 \fbstate{0}{0}{1}{0}{1}
 \fbstate{3.6}{0}{1}{1}{0}
 \fbmove{0}{d=0}{1}
\end{scope}
\begin{scope}[shift={(6.25,-5.5)}]
 \node at (2.5,1.25) {(f) Blocked layer-2 exit};
 \fbboundarystate{0}{1}{1}
 \fbboundarystate{3.6}{1}{0}
 \fbmove{0}{D=0}{1}
\end{scope}
\end{tikzpicture}
\endgroup
\caption{Left jumps, exits and blocked moves, complementing the
right jumps and entries in Figure~\ref{fig:intro-two-layer-moves}.
The upper black row is layer 1 and the lower red row is layer 2;
filled and open circles denote particles and vacancies.
Each star denotes an occupation that is unchanged by the move.
In the left panels, $d=D_{x+1}$ and
$\omega=(\xi_1(x),\xi_1(x+1))$; in the right panels,
$D=D_1$ and $k=\xi_1(1)$. Rates are given in
\eqref{eq:fb-two-column-bulk-rates} and
\eqref{eq:fb-two-column-boundary-rho}. Only admissible input states
are considered. The crossed arrows show proposed outputs outside
$\mathcal X$, which are forbidden. All unlisted sites are unchanged.}
\label{fig:fb-two-column-complementary-moves}
\end{figure}

\smallskip\noindent\textbf{The generator.}
For $\xi=(\xi_1,\xi_2)\in\mathcal X$, let $\sigma_{r,x}$ exchange
$\xi_r(x)$ and $\xi_r(x+1)$, and let $e_r^+$ and $e_r^-$ set
$\xi_r(1)$ to one and zero, respectively. Each map leaves all other
occupations unchanged and is interpreted as the identity if its
proposed output lies outside $\mathcal X$. Define
\[
 \mathsf P_x\xi=
 \begin{cases}
  \sigma_{1,x}\sigma_{2,x}\xi,&D_{x+1}=0,\\
  \sigma_{1,x}\xi,&D_{x+1}>0,
 \end{cases}
 \qquad
 \mathsf P_\partial\xi=
 \begin{cases}
  e_1^+e_2^+\xi,&D_1=0,\\
  e_1^+\xi,&D_1>0.
 \end{cases}
\]
The rightmost map acts first. In an allowed layer-1 right jump or
entry, these compositions describe one simultaneous transition;
applying the layer-2 map first keeps the intermediate state in
$\mathcal X$. Put
\[
 \chi_{r,x}^+=\xi_r(x)(1-\xi_r(x+1)),\qquad
 \chi_{r,x}^-=\xi_r(x+1)(1-\xi_r(x)),\qquad
 \omega_x=(\xi_1(x),\xi_1(x+1)),
\]
and define the layer-2 boundary removal rate by
\[
 v_D^\partial(k)=
 \begin{cases}
  0,&D=0,\\
  q/(2\rho_{D-1}(k)),&D\ge1.
 \end{cases}
\]
For a bounded test function $\varphi:\mathcal X\to\mathbb R$, write
$\Delta_T\varphi(\xi)=\varphi(T\xi)-\varphi(\xi)$.
With $D=D_1$ and $k=\xi_1(1)$, the generator is
\begin{equation}
\begin{aligned}
 (\mathcal L\varphi)(\xi)
 ={}&\sum_{x\ge1}\Bigl[
 \chi_{1,x}^+\Delta_{\mathsf P_x}\varphi(\xi)
 +q\chi_{1,x}^-\Delta_{\sigma_{1,x}}\varphi(\xi)\\
 &\qquad\quad+
 \bigl(\chi_{2,x}^+u_{D_{x+1}}(\omega_x)
       +\chi_{2,x}^-v_{D_{x+1}}(\omega_x)\bigr)
       \Delta_{\sigma_{2,x}}\varphi(\xi)\Bigr]\\
 &+\frac{1-k}{2}\Delta_{\mathsf P_\partial}\varphi(\xi)
   +\frac{qk}{2}\Delta_{e_1^-}\varphi(\xi)\\
 &+\frac{1-\xi_2(1)}{2}\rho_D(k)\Delta_{e_2^+}\varphi(\xi)
   +\xi_2(1)v_D^\partial(k)\Delta_{e_2^-}\varphi(\xi).
\end{aligned}
\label{eq:fb-two-column-generator}
\end{equation}
The occupation products impose exclusion. The definitions
$v_0=0$ and $v_0^\partial=0$ avoid negative subscripts in the rates.
Only finitely many summands are nonzero for each $\xi\in\mathcal X$.

\begin{prop}[Continuous-time limit in both boundary regimes]
\label{prop:fb-two-column-continuous-time}
In either regime with the scaling above, the row process
$\bigl(\xi_1^{(\lfloor T/\varepsilon\rfloor)},
\xi_2^{(\lfloor T/\varepsilon\rfloor)}\bigr)$, started empty,
converges in distribution in the Skorokhod $J_1$ topology on paths
in $\mathcal X$, equipped with the discrete topology, over bounded
time intervals, to the nonexplosive Markov chain with generator
$\mathcal L$.
In particular, its first layer is open ASEP on $\mathbb Z_{\ge1}$,
with right and left jump rates $1,q$ and entry and exit rates
$1/2,q/2$. In both regimes,
\begin{equation}
 n_\varepsilon-H_1(n_\varepsilon,n_\varepsilon)
 \ \Longrightarrow\ N_1(T).
 \label{eq:fb-scalar-ASEP-current}
\end{equation}
\end{prop}
\begin{proof}
The row is sampled from large to small $x$ in the coordinates
\eqref{eq:fb-two-column-particles}. At a bulk bond, in the absence
of an earlier exceptional output, its inputs are
$k_r=\xi_{r+1}(x+1)$ and $\ell_r=1-\xi_{r+1}(x)$, $r=0,1$.
The usual output is $(i_r,j_r)=(1-\ell_r,1-k_r)$; it leaves the
particle configuration unchanged. After the following usual vertex,
the occupations at $(x,x+1)$ are $(i_r,1-j_r)$, which fixes the
directions of the jumps below. Since
$c=\varepsilon+O(\varepsilon^2)$, the exceptional first-column
outputs give right and left rates $1,q$.

To calculate the second-column rates, remove the crossing factor
from \eqref{eq:fb-two-column-bulk-weights} and denote the remaining
product by $C_{ij}$. If $k_1=\ell_1=0$ or $1$, respectively, the
exceptional probabilities are exactly
\begin{equation}
 \frac{(1-a^2)C_{00}}{(1-a^2)C_{00}+(1-q)C_{11}},\qquad
 \frac{q(1-a^2)C_{11}}{(1-q)a^2C_{00}+q(1-a^2)C_{11}}.
 \label{eq:fb-two-column-rare-probabilities}
\end{equation}
For unequal inputs the output is deterministic. Following a usual
first-column output, the usual second-column candidate has
multiplicity $M=d$ and positive coefficient. Consequently the two
rates in \eqref{eq:fb-two-column-rare-probabilities} are
$C_{00}/C_{11}$ and $qC_{11}/C_{00}$ evaluated at $a=1$.
Substitution of the four $L,M^*$ weights gives
\eqref{eq:fb-two-column-bulk-rates}. For instance, when both
layers have pair $10$, these coefficients are
$C_{00}=B_d/(1-s)^2$ and $C_{11}=A_d^2/(1-s)^2$.
If a first-layer right jump occurs, the candidate leaving layer 2
fixed has multiplicity $d-1$. It has positive coefficient for
$d>0$; at $d=0$ it is forbidden and the layer-2 right jump is
forced. A first-layer left jump instead gives multiplicity $d+1$
and forces no second jump. This proves every bulk rule, including
the simultaneous transition at contact.

At the boundary put $D=\varkappa'_1-\varkappa'_2$. In regime~\textup{(ii)}, the
recurrences for $O_m$ and \eqref{eq:fb-two-column-boundary} give,
for each fixed $D$,
\[
 \begin{gathered}
 \mathbb P(h_0=z)=\tfrac12+O(\varepsilon),\qquad z=0,1,\\
 h_1\equiv h_0+D\pmod2
 \quad\text{with probability }1-O(\varepsilon).
 \end{gathered}
\]
Indeed $O_{2m+1}(1,1)=0$, $O_{2m}(1,1)>0$, and the limiting
normalizers $O_D(q,1)$ are positive. For any admissible auxiliary
pair the usual bulk outputs leave the particles fixed, with
probability $1-O(\varepsilon)$. Thus errors of order $\varepsilon$
in the auxiliary distribution affect nonidentity transitions only
at order $\varepsilon^2$.

Write $k_r=\xi_{r+1}(1)$. The changes in particle counts are
$\Delta N_{r+1}=1-h_r-i_r$. A first-layer entry is possible when
$k_0=h_0=0$, with conditional rate $1$; an exit is possible when
$k_0=h_0=1$, with conditional rate $q$. Averaging the two phases
gives rates $1/2,q/2$. For a layer-2-only event one must have
$k_1=h_1=z$. In \eqref{eq:fb-two-column-rare-probabilities} set
\[
 I=D+h_0-z,\quad J=D+k_0-z,\quad
 i_0=1-h_0,\quad j_0=1-k_0,\quad
 h_0\equiv z+D\pmod2.
\]
The same coefficient ratios then give $\rho_D(k_0)$ for entry
and $q/\rho_{D-1}(k_0)$ for exit. Multiplication by the phase
probability $1/2$ gives the stated boundary rates. Following a
first-layer entry, keeping layer 2 fixed requires multiplicity
$D-1$; at $D=0$ the second entry is therefore forced.

In regime~\textup{(i)}, $h_0=\varkappa'_1\bmod2$ and
$h_1\equiv h_0+D\pmod2$ exactly. Since $\varkappa'_1=n-N_1$ before the
next row is sampled, this phase alternates whenever the particle
configuration stays fixed. On any finite set of configurations,
two consecutive row kernels therefore have expansion
$I+2\varepsilon\mathcal L+O(\varepsilon^2)$: a particle change
in the first row affects a second exceptional event only at order
$\varepsilon^2$. The average over the two phases is precisely the
boundary generator just computed. Regime~\textup{(ii)} has the
one-row expansion $I+\varepsilon\mathcal L+O(\varepsilon^2)$.

For completeness, these expansions give a process limit by
localization. The listed moves preserve every inequality $D_x\ge0$:
a move that would decrease a zero gap is blocked or accompanied
by the specified simultaneous move. All rates are bounded uniformly
in the gap, because
$A_d\ge1$ and $B_d\ge(1-q)(1-s^2)>0$. Boundary births have bounded
total rate, and only a number of bonds proportional to $N_1+N_2$
can change. The limiting chain is therefore nonexplosive.
For the discrete chains, let $R_n$ be the rightmost occupied site
in either layer, with $R_n=0$ for the empty state. Vertices strictly
beyond the particles are deterministic, and the right-to-left sweep
can increase $R_n$ by at most one. If $R_n>0$, tail ordering forces
$\xi_2(R_n)=1$. At the first potentially changing vertex, the
probability of increasing the rightmost occupied site is, according
as $\xi_1(R_n)=1$ or $0$,
\[
 c+(1-c)\bigl(1-r_s(a)^2\bigr),\qquad
 \frac{(1-a^2)(1-saq)}
 {(1-a^2)(1-saq)+(1-q)a(a-s)}.
\]
Both are $O(\varepsilon)$ under the stated scaling, with a bound
depending only on $q,s$. From the empty state, only a boundary
birth can increase $R_n$; the two boundary auxiliary laws and the
same bulk probabilities again give an $O(\varepsilon)$ bound.
Hence the conditional probability of $R_{n+1}>R_n$ is at most
$C\varepsilon$, uniformly in $n$ and the state. On
$n\le T_0/\varepsilon$, the maximum of $R_n$ is consequently
stochastically bounded by
$R_0+\operatorname{Binomial}(\lfloor T_0/\varepsilon\rfloor,
C\varepsilon)$, giving uniform compact containment.

Stopping at $R_n>M$ leaves a finite state space, with one cemetery
state. Only finitely many vertices are then active; their rational
probabilities have uniform expansions and two initiating events
in one row have probability $O(\varepsilon^2)$.
The one-row or two-row expansions prove convergence of these
stopped chains. In the deterministic case, before stopping, the
probability of two particle changes in any paired interval before
$T_0$ is $O_M(\varepsilon)$; otherwise unpairing shifts jump times
by at most $2\varepsilon$. Compact containment then permits
$M\to\infty$ and proves convergence in the Skorokhod $J_1$
topology on paths in the
countable state space $\mathcal X$, equipped with the discrete topology.
\end{proof}

The first-layer limit is the known critical open ASEP, consistent
with \cite[Proposition~5.2]{BarraquandBorodinCorwinWheeler1704} and
\cite[proof of Corollary~4.5]{He2024Boundary}. The deletion in
\eqref{eq:even-BBCW-row-shift} is relevant even in this limit:
setting the BBCW row to $n+1$ and its distance from the diagonal to
one gives exactly $n-h_{\rm ev}(n,n)$. Its diagonal rounding in
\eqref{eq:BBCW-diagonal-rounding} is of order one and cannot be
discarded in a fixed-time continuous-time limit.

The two-layer generator retains $s$. For example, from the empty
configuration a simultaneous entry into both layers has rate $1/2$,
whereas entry into layer 2 alone has rate
$(1-q)(1+s)/(2(1-s))$. This proves Theorem~\ref{thm:intro-continuous-time}: both discrete
boundary regimes have the same continuous-time two-layer limit. Its jump rates depend on
the tail gap and the boundary rates also depend on its parity;
the table gives an explicit coupled exclusion process, with open
ASEP as its first-layer projection. Higher-column continuous-time
limits and joint large-time fluctuations of the layers require
further analysis. Section~\ref{sec:boundary-KPZ} concerns the
first-coordinate KPZ fluctuations; convergence at fixed time above
does not interchange the continuous-time and large-time limits.

\section{Joint partition lengths and boundary fluctuations}
\label{sec:boundary-KPZ}

We prove Theorem~\ref{thm:intro-joint-length} and
Corollary~\ref{cor:intro-fluctuations}. Joint ascending lengths,
and hence lengths along each fixed horizontal row, have the
zero-spin Hall--Littlewood laws throughout the nonnegative boundary
range, including $(t,\zeta)=(1,1)$. These comparisons suffice for
diagonal fluctuation limits without requiring autonomy of the
whole field. We combine them with He's asymptotic results
\cite{He2024Boundary}; for the continuous-time model we use its
first-layer open ASEP projection.

\subsection{Joint distribution of partition lengths in an ascending process}
\label{ssec:joint-lengths}

Fix $N\ge1$, $0<q<1$, $-1<s\le0$ and
$A=(a_1,\ldots,a_N)\in(0,1)^N$. Assume either
$0\le t<1$, $\zeta>0$ and \eqref{eq:fb-positive-domain}, or
$(t,\zeta)=(1,1)$. The ascending process has probability mass function
\begin{equation}
\label{eq:fb-ascending-process}
\mathbb P^{A,\uparrow}_{s;t,\zeta}
(\lambda^{(1)},\ldots,\lambda^{(N)})
=\frac{W_{\lambda^{(N)}}(t,\zeta)}{Z_{t,\zeta}(A)}
 \prod_{i=1}^N f_{\lambda^{(i)}/\lambda^{(i-1)}}(a_i),
\qquad \lambda^{(0)}=\varnothing.
\end{equation}
Branching and \eqref{eq:fb-multivariable-measure} prove normalization
and identify its terminal marginal. Interlacing gives
$\ell(\lambda^{(i)})-\ell(\lambda^{(i-1)})\in\{0,1\}$, hence
$0\le\ell(\lambda^{(i)})\le i$.

To identify the joint length law, more than the spin independence
of $Z_{t,\zeta}$ is needed. For $A_n=(a_1,\ldots,a_n)$, define
\begin{equation}
\label{eq:tau-first-difference-operator}
\mathcal D_n=\sum_{i=1}^n
 \prod_{\substack{1\le j\le n\\j\ne i}}
 \frac{qa_i-a_j}{a_i-a_j}T_{0,a_i},
\qquad d_{n,l}=\frac{1-q^{n-l}}{1-q}\quad(0\le l\le n),
\end{equation}
where $T_{0,a_i}$ sets $a_i$ to zero. Initially the variables are
distinct. In \cite[Section~8]{BufetovMucciconiPetrov1905}, replace
the Hall--Littlewood parameter $t$ by our $q$; it is unrelated to
our boundary parameter $t$. The function $\mathsf F_\lambda$ there
is our $g_\lambda$. Since
$f_\lambda=\mathsf c(\lambda)\mathsf F_\lambda$ and the gauge
factor is independent of the spectral variables, the $r=1$ case of
\cite[Theorem~8.2]{BufetovMucciconiPetrov1905} gives
\begin{equation}
\label{eq:length-eigenrelation}
\mathcal D_nf_\lambda(A_n)
=d_{n,\ell(\lambda)}f_\lambda(A_n),\qquad \ell(\lambda)\le n.
\end{equation}
The eigenvalues are distinct:
$d_{n,l}-d_{n,l+1}=q^{n-l-1}>0$. Their spectral projectors are
\begin{equation}
\label{eq:fb-prefix-projector}
\mathcal E_{n,l}
=\prod_{\substack{0\le j\le n\\j\ne l}}
 \frac{\mathcal D_n-d_{n,j}}{d_{n,l}-d_{n,j}},
\qquad
\mathcal E_{n,l}f_\lambda(A_n)
=\mathbf1_{\{\ell(\lambda)=l\}}f_\lambda(A_n).
\end{equation}
Scalar terms multiply the identity operator; each projector acts
only on the first $n$ variables.

\begin{thm}[Equality of joint length distributions]
\label{thm:fb-joint-length}
Under the assumptions of \eqref{eq:fb-ascending-process}, the joint
distribution of
$(\ell(\lambda^{(1)}),\ldots,\ell(\lambda^{(N)}))$
is independent of $s$ among the admissible spin values.
It equals the length distribution of the half-space Hall--Littlewood
process obtained by setting $s=0$, with the same parameters
$(q,t,\zeta,A)$. This includes $(t,\zeta)=(1,1)$.

For $1\leq n_1<\cdots<n_k\leq N$ and $0\leq l_j\leq n_j$,
\begin{equation}
\label{eq:fb-joint-length-projectors}
\mathbb P^{A,\uparrow}_{s;t,\zeta}
\{\ell(\lambda^{(n_j)})=l_j,\ 1\leq j\leq k\}
=\frac{\mathcal E_{n_1,l_1}\cdots\mathcal E_{n_k,l_k}
Z_{t,\zeta}(A)}{Z_{t,\zeta}(A)}.
\end{equation}
The rightmost operator acts first, and division by $Z_{t,\zeta}(A)$
is performed after all operators have acted. The formula extends
continuously to coincident spectral parameters.
\end{thm}

\begin{proof}
Restrict the chain sum to $\lambda^{(N)}_1\le M$, obtaining the
finite unnormalized sum $Z^{[M]}(A)$. Suppress the boundary parameters.
For $n<m\le N$, put $B=(a_{n+1},\ldots,a_m)$ and
$C=(a_{m+1},\ldots,a_N)$. Branching gives
\[
Z^{[M]}(A)=\sum_{\substack{\lambda:\lambda_1\le M\\\ell(\lambda)\le N}}
 \sum_\mu W_\lambda f_{\lambda/\mu}(C)f_\mu(A_n,B).
\]
Applying $\mathcal E_{m,h}$ inserts
$\mathbf1_{\{\ell(\mu)=h\}}$. Expanding
$f_\mu(A_n,B)=\sum_\kappa f_{\mu/\kappa}(B)f_\kappa(A_n)$
and then applying $\mathcal E_{n,l}$ yields
\[
\begin{aligned}
\mathcal E_{n,l}\mathcal E_{m,h}Z^{[M]}(A)
={}&\sum_{\substack{\lambda:\lambda_1\le M\\\ell(\lambda)\le N}}\sum_{\mu,\kappa}
 W_\lambda f_{\lambda/\mu}(C)f_{\mu/\kappa}(B)f_\kappa(A_n)
 \\
 &\hspace{12mm}\times
 \mathbf1_{\{\ell(\mu)=h\}}\mathbf1_{\{\ell(\kappa)=l\}}.
\end{aligned}
\]
All partitions in these sums have length at most $N$.
Repeating in decreasing order of the selected positions inserts
every required indicator. Branching exposes a non-skew function
at each step; no commutation of different prefix operators is used.

We justify passage to the infinite sum on $|a_i|\le r<1$.
Write $C_N=N(N+1)/2$. At most $C_N$ distinct positive part sizes
occur among the partitions in a chain. In every other column up
to $\lambda^{(N)}_1$, all vertical occupations vanish and at least
one row has a continuing horizontal arrow: the sum of the horizontal
occupations is the positive terminal column height. Such an arrow
has weight bounded in absolute value by
\[
\left|\frac{a_i-s}{1-sa_i}\right|
\le\rho:=\frac{r+|s|}{1+r|s|}<1,
\]
and an empty vertex has weight one. The remaining vertex weights
are uniformly bounded, since their occupations are at most $N$.
The boundary coefficient has at most $N$ nontrivial factors among
$E_1,\ldots,E_N,O_1,\ldots,O_N$. Thus the absolute chain weight
is at most $C\rho^{\lambda^{(N)}_1-C_N}$.
There are at most $(m+1)^{C_N}$ chains with all parts at most $m$,
so a polynomial times geometric bound proves local uniform absolute
convergence, also after inserting indicators.

The operator $\mathcal D_n$ preserves holomorphy on functions
symmetric in the first $n$ variables. Indeed, after multiplication
by their Vandermonde determinant its numerator is holomorphic and
alternating, and hence divisible by that determinant. To check
continuity for locally uniform convergence, bound the defining sum
on product circles with distinct radii between $r$ and a larger
$R<1$, then use the maximum principle on the enclosed polydisc.
This bounds the output on the smaller polydisc by a constant times
the input supremum on the larger one. Applying the estimate with
successively nested radii treats every required operator polynomial,
including repeated or zero variables. The grouped sums have the
prefix symmetry needed at each successive application. We can
therefore let $M\to\infty$ and divide by $Z(A)$, proving
\eqref{eq:fb-joint-length-projectors}.

Both the projectors and $Z(A)$ are independent of $s$. Moreover,
the boundary parameters remain admissible at $s=0$, since
\eqref{eq:fb-positive-domain} implies $V\ge0$; positivity also
holds at $(1,1)$. This proves the asserted distributional equality
and, by taking every position, Theorem~\ref{thm:intro-joint-length}.
\end{proof}

At $s=0$ the functions become $P_{\lambda/\mu}(\,\cdot\,;q)$,
and the boundary coefficients are He's with $\nu=\zeta$
(Section~\ref{sec:boundary-family}). Hence
\cite[Theorem~2.5]{He2024Boundary} gives
\begin{equation}
\label{eq:joint-He-row}
(\ell(\lambda^{(i)}))_{i=1}^N
\stackrel{\mathrm d}=
(H_{\rm He}^{t,\zeta}(i,N))_{i=1}^N,
\end{equation}
with spectral parameters $A$ and bulk parameter $q$.
Initially impose $t\zeta^2<1$ and $a_i<\min\{1,\zeta\}$,
which satisfy the sufficient convergence conditions of
\cite[Lemma~2.2]{He2024Boundary}; $t=0$ is allowed there.
By \cite[Remark~2.9]{He2024Boundary}, the comparison is a rational
identity. Indeed, \eqref{eq:fb-joint-length-projectors} and the
finite vertex-model probabilities are rational functions, so the
identity extends to all $a_i\in(0,1)$. Our uniform convergence
argument identifies the continued partition sums with probabilities.
Positivity gives $t\zeta^2\le1$, hence $t\zeta<1$ for $t<1$,
so the vertex denominators do not vanish. Continuity includes
$t\zeta^2=1$ and, by taking $t\uparrow1$ with $\zeta=1$,
$(t,\zeta)=(1,1)$: the finite-multiplicity coefficients are uniformly
bounded on compact boundary-parameter sets, and the geometric
majorant above is uniform on such sets. At $(1,1)$ the boundary
rule is the BBCW rule of Section~\ref{ssec:fb-local-projections}.

To obtain a horizontal row of the random field, use the ascending
process with $n+1$ variables $(x_0,\ldots,x_n)$ and sum out its
last partition $\eta$. For the penultimate partition $\rho$,
the skew Littlewood identity gives
\[
\sum_\eta W_\eta(t,\zeta)f_{\eta/\rho}(x_n)
=\Phi_{t,\zeta}(x_n)G^{t,\zeta}_\rho(x_n).
\]
The first $n$ partitions thus have the row law \eqref{eq:process},
with normalizer $Z_{t,\zeta}(x_0,\ldots,x_n)/\Phi_{t,\zeta}(x_n)$.
For each fixed $n$ this proves
\[
\bigl(\ell(\lambda^{(i,n)})\bigr)_{i=1}^n
\stackrel{\mathrm d}=
\bigl(H_{\rm He}^{t,\zeta}(i,n+1)\bigr)_{i=1}^n,
\]
using spectral list $(x_0,\ldots,x_n)$ on the right.
The comparison holds throughout the theorem's boundary range,
without an autonomous projection. At $(1,1)$ it gives the
corresponding row of BBCW with its diagonal deleted.

In the homogeneous regime \eqref{eq:fb-local-tuning},
$G^{t,\zeta}_\rho(a)=W_\rho(qt,\zeta)$, so the row is also the
ascending process with $n$ variables equal to $a$ and boundary
parameters $(qt,\zeta)$. Consequently,
\[
\bigl(\ell(\lambda^{(i,n)})\bigr)_{i=1}^n
\stackrel{\mathrm d}=
\bigl(H_{\rm He}^{qt,\zeta}(i,n)\bigr)_{i=1}^n.
\]
Its terminal marginal is \eqref{eq:fb-local-diagonal-law}.
These identities compare one fixed row at a time, which suffices
for diagonal fluctuation limits. The entire-field identities in
Corollary~\ref{cor:intro-scalar} additionally use the local-transition
argument of Section~\ref{ssec:fb-local-projections} in the two
autonomous regimes.

\subsection{Boundary fluctuations and the GSE--GOE crossover}
\label{ssec:boundary-limits}

We prove Corollary~\ref{cor:intro-fluctuations}, beginning with
fixed-parameter limits and the separate case $(t,\zeta)=(1,1)$.
For the ascending process take $A_N=(a,\ldots,a)$, and for the
field take $x_r=a$ for every $r\ge0$, where $a\in(0,1)$.
Recall the centering and scale
\[
 v(a)=\frac{2a}{1+a},
 \qquad \sigma(a)=\frac{[a(1-a)]^{1/3}}{1+a}.
\]
We use the Tracy--Widom distribution functions $F_{\mathrm{GSE}}$
and $F_{\mathrm{GOE}}$, and the Baik--Rains crossover distribution
$F_{\mathrm{cross}}(z;\xi)$, in the normalization of
\cite[Definition~5.1]{He2024Boundary}. The effective boundary density
in He's model is $\varrho=\zeta/(1+\zeta)$; thus
$\zeta>1$, $\zeta=1$ and $\zeta<1$ correspond respectively to
$\varrho>1/2$, $\varrho=1/2$ and $\varrho<1/2$.
The boundary parameters remain fixed except in
Corollary~\ref{cor:fb-boundary-crossover}.

\begin{cor}[Boundary fluctuation limits]
\label{cor:fb-boundary-limits}
Fix $0<q<1$, $-1<s\le0$, $0\le t<1$, $\zeta>0$ satisfying
\eqref{eq:fb-positive-domain}, and $a\in(0,1)$.
In \eqref{eq:fb-ascending-process} with $A=(a,\ldots,a)$, write
$L_N=\ell(\lambda^{(N)})$. If $\zeta\ge1$, then, for every
$z\in\mathbb R$,
\begin{equation}
\label{eq:fb-TW-limits}
\lim_{N\to\infty}\mathbb P\left\{
\frac{L_N-v(a)N}{\sigma(a)N^{1/3}}\le z\right\}
=\begin{cases}F_{\mathrm{GSE}}(z),&\zeta>1,\\
F_{\mathrm{GOE}}(z),&\zeta=1.
\end{cases}
\end{equation}
If $0<\zeta<1$, then
$(L_N-m_{\mathrm G}N)/(\sigma_{\mathrm G}\sqrt N)$ converges in
distribution to a standard Gaussian, where
\begin{equation}
\label{eq:fb-Gaussian-constants}
m_{\mathrm G}=
\frac{2a^2+a(\zeta+\zeta^{-1})}{(1+a\zeta)(1+a\zeta^{-1})},
\qquad
\sigma_{\mathrm G}^2=
\frac{a(1-a^2)(\zeta^{-1}-\zeta)}
{(1+a\zeta)^2(1+a\zeta^{-1})^2}.
\end{equation}
The same limits hold with $L_N$ replaced by the diagonal length
$J_N=\ell(\lambda^{(N,N)})$ of any half-space random field with
the path laws of Section~\ref{ssec:hsybrf}, homogeneous parameters
$x_r=a$, and these boundary parameters. No autonomy assumption
is needed for this diagonal statement.
\end{cor}

\begin{proof}
Theorem~\ref{thm:fb-joint-length} and the Hall--Littlewood
identification reduce the ascending length to He's diagonal height
with boundary parameters $(b,\nu)=(t,\zeta)$.
Positivity gives $t\zeta^2\le1$, whence $t\zeta<1$ because $t<1$.
The limits therefore follow from
\cite[Theorem~1.4]{He2024Boundary}.

For the field, Section~\ref{ssec:joint-lengths} gives
$Y_n:=\ell(\lambda^{(n,n)})
\stackrel{\mathrm d}=H_{\rm He}^{t,\zeta}(n,n+1)$.
In the same six-vertex configuration, put
$X_{n+1}=H_{\rm He}^{t,\zeta}(n+1,n+1)$.
The additional north arrow gives a coupling with
$0\le X_{n+1}-Y_n\le1$.
Let $(m,c,\beta)$ denote $(v(a),\sigma(a),1/3)$ in the
Tracy--Widom cases and $(m_{\mathrm G},\sigma_{\mathrm G},1/2)$
in the Gaussian case. Then
\[
\frac{Y_n-mn}{cn^\beta}
=\left(\frac{n+1}{n}\right)^\beta
\frac{X_{n+1}-m(n+1)}{c(n+1)^\beta}
+\frac{m-(X_{n+1}-Y_n)}{cn^\beta}.
\]
The last term tends to zero uniformly, proving the field statement.
\end{proof}

This proves the $t<1$ cases of
Corollary~\ref{cor:intro-fluctuations}\textup{(i)}.
In particular, the first coordinate of every autonomous projection
$\mathbf H_R=(\lambda'_1,\ldots,\lambda'_R)$ under
\eqref{eq:fb-local-tuning} has these limits.
All three regimes occur: for fixed $a,s$,
\[
 a\le r_s(a)=\frac{a-s}{1-sa}<1,
 \qquad
 1-r_s(a)=\frac{(1-a)(1+s)}{1-sa}>0,
\]
so the admissible interval $(r_s(a),r_s(a)^{-1})$ contains points
on both sides of one. The Gaussian regime has $N^{1/2}$ fluctuations;
the GOE and GSE regimes have $N^{1/3}$ fluctuations.

The remaining case in Corollary~\ref{cor:intro-fluctuations}\textup{(i)}
is $(t,\zeta)=(1,1)$. We compare its BBCW diagonal law with the
Hall--Littlewood law at boundary parameters $(q,1)$ by a coupling
whose two lengths differ by at most one. Diagonal deletion then
adds at most one further arrow. This permits us to use He's GOE
theorem at $b=q<1$ directly.

\begin{cor}[GOE limit for the specialization $(t,\zeta)=(1,1)$]
\label{cor:even-boundary-GOE}
Let $(t,\zeta)=(1,1)$ and $x_r=a\in(0,1)$ for all $r\ge0$.
For every fixed $s\in(-1,0]$, the scalar projection
$h_{\rm ev}(i,j)=\ell(\lambda^{(i,j)})$ of
Proposition~\ref{prop:even-boundary-BBCW} satisfies
\begin{equation}
 \lim_{n\to\infty}\mathbb P\left\{
 \frac{h_{\rm ev}(n,n)-v(a)n}{\sigma(a)n^{1/3}}\le z\right\}
 =F_{\mathrm{GOE}}(z),\qquad z\in\mathbb R.
 \label{eq:even-boundary-GOE}
\end{equation}
This is also the diagonal first-coordinate limit for every
finite-column projection at this specialization.
\end{cor}

\begin{proof}
Set $A_M=(a,\ldots,a)$, $b_\mu=W_\mu(1,1)|_{s=0}$,
$Z_0=Z_{1,1}(A_M)$ and $C=((1-qa)/(1-a))^M$.
For ordinary Hall--Littlewood functions $P,Q$, consider
\[
 \mathbb P(\mu,\rho)
 =\frac{b_\mu P_\rho(A_M;q)Q_{\rho/\mu}(1;q)}{CZ_0}.
\]
Here $1$ denotes one ordinary variable. The skew Cauchy identity,
whose sums converge since $a<1$, gives
\[
 \sum_\rho P_\rho(A_M;q)Q_{\rho/\mu}(1;q)
 =C P_\mu(A_M;q).
\]
Thus these nonnegative weights sum to one, and their
$\mu$-marginal is $b_\mu P_\mu(A_M;q)/Z_0$.

At $(b,\nu)=(q,1)$, the specialization in
\cite[Lemma~2.4]{He2024Boundary} satisfies
\[
 p_k=\frac{1+(-1)^k-q^k-(-1)^k}{1-q^k}=1.
\]
Consequently
\[
 \sum_\mu b_\mu Q_{\rho/\mu}(1;q)=W_\rho(q,1)|_{s=0}.
\]
Since $\Phi_{1,1}(a)=1$ and
$\Phi_{q,1}(a)=(1-qa)/(1-a)$, we have
$CZ_0=Z_{q,1}(A_M)$. The $\rho$-marginal is therefore
\[
 \frac{W_\rho(q,1)|_{s=0}\,P_\rho(A_M;q)}{Z_{q,1}(A_M)}.
\]
By the Hall--Littlewood identifications, $E_M:=\ell(\mu)$ has
the BBCW diagonal law and $Y_M:=\ell(\rho)$ has the law of
$H_{\rm He}^{q,1}(M,M)$.
The support forces $\mu'_j$ to be even and
$\rho'_j-\mu'_j\in\{0,1\}$, so
\[
 \mu'_j=2\left\lfloor\frac{\rho'_j}{2}\right\rfloor,
 \qquad 0\le Y_M-E_M\le1.
\]
Gluing this coupling to the BBCW coupling over their common
diagonal marginal, and using \eqref{eq:BBCW-diagonal-rounding}
at $M=n+1$ gives $0\le Y_{n+1}-h_{\rm ev}(n,n)\le2$.
Hence
\[
 \frac{h_{\rm ev}(n,n)-v(a)n}{\sigma(a)n^{1/3}}
 =\left(\frac{n+1}{n}\right)^{1/3}
 \frac{Y_{n+1}-v(a)(n+1)}{\sigma(a)(n+1)^{1/3}}
 +O(n^{-1/3}),
\]
where the remainder is bounded by
$(2+v(a))/(\sigma(a)n^{1/3})$.
He's GOE theorem applies at $b=q<1$, proving the result.
\end{proof}

For Corollary~\ref{cor:intro-fluctuations}\textup{(ii)}, we apply
the open ASEP asymptotic theorem to the first-layer projection
of the continuous-time generator.

\begin{cor}[GOE limit for the first layer in continuous time]
\label{cor:fb-two-layer-GOE}
Consider the two-layer process of
Theorem~\ref{thm:intro-continuous-time}, obtained in
Proposition~\ref{prop:fb-two-column-continuous-time}, started from
the empty configuration, and put
$N_1(T)=\sum_{x\ge1}\xi_1(T,x)$.
For every fixed $0<q<1$, $-1<s\le0$ and $z\in\mathbb R$,
\begin{equation}
\label{eq:ASEP-known-GOE}
\lim_{T\to\infty}\mathbb P\left\{
\frac{(1-q)T/4-N_1(T)}{2^{-4/3}((1-q)T)^{1/3}}\le z\right\}
=F_{\mathrm{GOE}}(z).
\end{equation}
This applies to the common continuous-time limit of boundary
regimes \textup{(i)} and \textup{(ii)} in
Proposition~\ref{prop:fb-two-column-continuous-time}.
\end{cor}

\begin{proof}
Apply \eqref{eq:fb-two-column-generator} to
$\varphi(\xi_1,\xi_2)=\psi(\xi_1)$.
Moves of the second layer alone contribute zero, and simultaneous
moves act on $\psi$ as the corresponding first-layer move.
Thus $\xi_1$ is open ASEP with right and left rates $1,q$,
entry rate $\alpha=1/2$ and exit rate $\beta=q/2$.
In He's notation $b=\beta/\alpha=q$, and
\[
 \nu^{-1}
 =\frac{1-q+\beta-\alpha+
 \sqrt{(1-q+\beta-\alpha)^2+4\alpha\beta}}{2\alpha}
 =\frac{1-q}{2}+\frac{1+q}{2}=1.
\]
The effective density is therefore $1/2$.
Apply \cite[Theorem~1.1]{He2024Boundary} with
$\tau=(1-q)T$; this critical case was also proved in
\cite[Theorem~6.2]{BarraquandBorodinCorwinWheeler1704}.
The sign agrees with
$H_1(n,n)-v(a)n=(1-v(a))n-N_1^{(n)}$.
This uses the asymptotic theorem for ASEP directly and does not
interchange the continuous-time and large-time limits.
\end{proof}

Finally, we prove Corollary~\ref{cor:intro-fluctuations}\textup{(iii)}.
Here the boundary varies with $N$ while remaining in the autonomous
regime $t_N\zeta_N=a$. The scalar identification uses He's
parameters $(b_N,\nu_N)=(qt_N,\zeta_N)$. The proof below checks
both this tuning and the bounds needed to pass to the crossover law.

\begin{cor}[GSE--GOE crossover for the first column]
\label{cor:fb-boundary-crossover}
Fix $0<q<1$, $-1<s\le0$, $a\in(0,1)$ and $\xi>0$.
For each sufficiently large $N$, take the homogeneous field with
$x_r=a$ and boundary parameters
\begin{equation}
 \zeta_N^{-1}=1-\frac{\xi}{\sigma(a)N^{1/3}},
 \qquad t_N=\frac{a}{\zeta_N}.
 \label{eq:fb-crossover-scaling}
\end{equation}
Write $J_N=(\lambda^{(N,N)})'_1$ for its diagonal length.
Then, for every $z\in\mathbb R$,
\begin{equation}
 \lim_{N\to\infty}\mathbb P\left\{
 \frac{J_N-v(a)N}{\sigma(a)N^{1/3}}\le z\right\}
 =F_{\mathrm{cross}}(z;\xi).
 \label{eq:fb-crossover-limit}
\end{equation}
This is the first-coordinate limit for every fixed $R$ in
$\mathbf H_R$. The limiting family interpolates between GOE
as $\xi\downarrow0$ and GSE as $\xi\to\infty$.
\end{cor}

\begin{proof}
Since $r_s(a)<1<r_s(a)^{-1}$ and $\zeta_N\to1$,
\eqref{eq:fb-local-tuning} holds for all sufficiently large $N$.
Proposition~\ref{prop:fb-local-scalar-law} identifies $J_N$ with
$H_{\rm He}^{b_N,\nu_N}(N,N)$, where
\[
 b_N=qt_N=\frac{qa}{\zeta_N},\qquad \nu_N=\zeta_N,
 \qquad \gamma_2=-b_N\nu_N=-qa.
\]
Although $b_N$ varies, the kernel parameter $\gamma_2$ is fixed.
We check why the crossover argument of
\cite[Lemma~6.6 and Propositions~6.3 and~6.7]{He2024Boundary}
applies in these parameters.

Put $d_N=\sigma(a)N^{1/3}$ and $u_N=\nu_N^{-1}$.
In the scalar kernel of \cite[Corollary~4.4]{He2024Boundary},
the boundary-dependent factor is
\[
 g_N(w)=
 \frac{(u_N/w,\gamma_2/w;q)_\infty}
      {(u_Nw,\gamma_2qw;q)_\infty}.
\]
For $w=1+\alpha/d_N$, factoring the first term from the
$u_N$-products gives
\[
 g_N(w)=
 \frac{1-u_N/w}{1-u_Nw}
 \frac{(qu_N/w;q)_\infty}{(qu_Nw;q)_\infty}
 \frac{(\gamma_2/w;q)_\infty}{(\gamma_2qw;q)_\infty}
 \longrightarrow
 (1-\gamma_2)\frac{\xi+\alpha}{\xi-\alpha}.
\]
Here $(c,d;q)_\infty=(c;q)_\infty(d;q)_\infty$.
For the phase
$\mathcal A(w)=\log(1+aw)-\log(1+a/w)-v(a)\log w$,
\[
 \mathcal A(1)=\mathcal A'(1)=\mathcal A''(1)=0,
 \qquad \mathcal A'''(1)=2\sigma(a)^3,
\]
so
$N\mathcal A(1+\alpha/d_N)-d_Nz\log(1+\alpha/d_N)
\to\alpha^3/3-\alpha z$.
The two factors $1-\gamma_2$ cancel the kernel prefactor
$(1-\gamma_2)^{-2}$. Thus the limiting kernel is precisely the
crossover kernel with parameter $\xi$.

To justify passage to Fredholm Pfaffians, use the contours of
\cite[Lemmas~6.5--6.6]{He2024Boundary}, with inner arc of radius
$1+\delta/d_N$, where $0<\delta<\xi$, and a fixed outer radius
$1<R<\min\{(qa)^{-1},q^{-1/2},a^{-1}\}$ sufficiently close to
one for steepest descent. The pole $\nu_N$ stays outside these
contours. After extracting $(1-u_N/w)/(1-u_Nw)$, the remaining
factor
\[
 h_N(w)=\frac{(qu_N/w;q)_\infty}{(qu_Nw;q)_\infty}
         \frac{(\gamma_2/w;q)_\infty}
              {(\gamma_2qw;q)_\infty}
\]
and any fixed number of its derivatives are uniformly bounded on
a fixed annulus containing the contours. On the contours themselves,
their geometry gives
\[
 |1-u_Nw|\ge c(d_N^{-1}+|w-1|),\qquad
 |1-u_N/w|\le C(d_N^{-1}+|w-1|),
\]
so the extracted quotient is uniformly bounded there.
The nonboundary factors are precisely those of He's six-vertex
kernel, and $\gamma_2$, hence $D^{(\gamma_2)}$, is fixed.
The steepest-descent argument of
\cite[Lemma~6.6]{He2024Boundary} therefore gives Assumption~6.1
of that paper with constants independent of $N$, including the
estimates after difference operators and local spatial perturbations.
With its discrete size parameter $n=N$, Proposition~6.3 replaces
the singular kernel by the derivative kernel with Fredholm-Pfaffian
error $O(N^{-1/12})$. The exponential kernel bounds then provide the
summable Pfaffian-series majorant used in Proposition~6.7, so
dominated convergence gives $F_{\mathrm{cross}}(z;\xi)$.

It remains to remove the auxiliary shifts in that formula.
The shifted observable is $J_N+\chi_N+2S$, with independent shifts
$\chi_N\sim RS(q,b_N)$ and
\[
 \mathbb P(S=k)=
 \frac{q^{k^2}\vartheta^{2k}}
 {\sum_{j\in\mathbb Z}q^{j^2}\vartheta^{2j}},
 \qquad k\in\mathbb Z,
\]
where $\vartheta\in(0,1)$ is fixed.
Since $0<b_N<q$, this Rogers--Szeg\H{o} distribution is
nonnegative: for $r=b_N/q\in(0,1)$, the recurrence
$h_{k+1}(-r;q)=(1-r)h_k(-r;q)+r(1-q^k)h_{k-1}(-r;q)$
has nonnegative coefficients and initial values $1,1-r$.
By \cite[Lemma~5.7]{He2024Boundary}, its masses obey
\[
 0\le\mathbb P(\chi_N=k)
 \le \frac{(-q;q)_\infty}{(q;q)_\infty}(k+1)q^k,
 \qquad k\ge0.
\]
Indeed, expand its mass as
$(q;q)_\infty(-b_N;q)_\infty
\sum_{i+j=k}q^i(-b_N)^j/((q;q)_i(q;q)_j)$
and bound each term in absolute value.
This uniform summable bound gives
$(\chi_N+2S)/d_N\to0$ in probability and proves
\eqref{eq:fb-crossover-limit}.
\end{proof}

The parameter in \eqref{eq:fb-crossover-scaling} is the one in
the kernel calculation of \cite[Lemma~6.6]{He2024Boundary}.
In terms of the effective density it reads
\[
 \varrho_N=\frac{1}{1+\zeta_N^{-1}}
 =\frac12+\frac{\xi}{4\sigma(a)}N^{-1/3}+O(N^{-2/3}).
\]
Thus the crossover varies the boundary with the observation scale;
it cannot be obtained by substituting $\zeta=1$ into the Gaussian
limit. For the ascending process there is also a fixed-$t$ version.
Fix $0\le t<1$, keep $q,s,a$ fixed as above, and set
$\zeta=\zeta_N$. Its nonnegative parameter conditions hold for
all sufficiently large $N$: at $\zeta=1$ one has $U=V=1-t>0$
and both inequalities in \eqref{eq:fb-positive-domain} are strict
because $-s<1$. Then \eqref{eq:fb-crossover-limit} holds with
$J_N$ replaced by $L_N$, by Theorem~\ref{thm:fb-joint-length} and
the argument with fixed first boundary parameter in
\cite[Lemma~6.6 and Proposition~6.8]{He2024Boundary}.

The partition fluctuation limits concern the first column,
and the continuous-time fluctuation limit concerns the first particle layer.
The operator in \eqref{eq:fb-prefix-projector} does not determine
the higher conjugate coordinates, and their laws can depend on $s$
by \eqref{eq:fb-local-spin-example}.
No limit for the second column, the second-layer count $N_2(T)$,
or their joint fluctuations is asserted here.

\end{document}